\pdfoutput=1
\documentclass[12pt,a4paper]{article}
\usepackage{ifthen}
\newboolean{pdflatex}
\setboolean{pdflatex}{true}

\newboolean{articletitles}
\setboolean{articletitles}{true}

\newboolean{uprightparticles}
\setboolean{uprightparticles}{false}

\newboolean{inbibliography}
\setboolean{inbibliography}{false}

\newcommand{\mpr}        {\mbox{$m^\prime$}}
\newcommand{\thpr}       {\mbox{$\theta^\prime$}}

\usepackage{microtype}
\usepackage{lineno}
\usepackage{xspace}
\usepackage{caption}

\usepackage{graphicx}
\usepackage{rotating}
\usepackage{color}
\usepackage{colortbl}
\graphicspath{{./figs/}}

\usepackage{amsmath}
\usepackage{amssymb}
\usepackage{amsfonts}
\usepackage{upgreek}

\newcommand*\patchAmsMathEnvironmentForLineno[1]{%
\expandafter\let\csname old#1\expandafter\endcsname\csname #1\endcsname
\expandafter\let\csname oldend#1\expandafter\endcsname\csname
end#1\endcsname
 \renewenvironment{#1}%
   {\linenomath\csname old#1\endcsname}%
   {\csname oldend#1\endcsname\endlinenomath}%
}
\newcommand*\patchBothAmsMathEnvironmentsForLineno[1]{%
  \patchAmsMathEnvironmentForLineno{#1}%
  \patchAmsMathEnvironmentForLineno{#1*}%
}
\AtBeginDocument{%
\patchBothAmsMathEnvironmentsForLineno{equation}%
\patchBothAmsMathEnvironmentsForLineno{align}%
\patchBothAmsMathEnvironmentsForLineno{flalign}%
\patchBothAmsMathEnvironmentsForLineno{alignat}%
\patchBothAmsMathEnvironmentsForLineno{gather}%
\patchBothAmsMathEnvironmentsForLineno{multline}%
\patchBothAmsMathEnvironmentsForLineno{eqnarray}%
}

\usepackage{hyperref}   
\usepackage[all]{hypcap}

\def\lhcb {\mbox{LHCb}\xspace}

\def\babar  {\mbox{BaBar}\xspace}

\def\MagUp {\mbox{\em Mag\kern -0.05em Up}\xspace}

\ifthenelse{\boolean{uprightparticles}}%
{

 \def\Ppi         {\ensuremath{\uppi}\xspace}

 \def\PDelta      {\ensuremath{\Delta}\xspace}                 
 \def\PXi      {\ensuremath{\Xi}\xspace}                 
 \def\PLambda      {\ensuremath{\Lambda}\xspace}                 
 \def\PSigma      {\ensuremath{\Sigma}\xspace}                 
 \def\POmega      {\ensuremath{\Omega}\xspace}                 
 \def\PUpsilon      {\ensuremath{\Upsilon}\xspace}

 \def\PB      {\ensuremath{\mathrm{B}}\xspace}                 
                  
 \def\PD      {\ensuremath{\mathrm{D}}\xspace}

 \def\PK      {\ensuremath{\mathrm{K}}\xspace}

 \def\Pb      {\ensuremath{\mathrm{b}}\xspace}                 
 \def\Pc      {\ensuremath{\mathrm{c}}\xspace}                 
                  
 \def\Pe      {\ensuremath{\mathrm{e}}\xspace}

 \def\Pi      {\ensuremath{\mathrm{i}}\xspace}

 \def\Ps      {\ensuremath{\mathrm{s}}\xspace}

}
{

 \def\Ppi         {\ensuremath{\pi}\xspace}

 \mathchardef\PDelta="7101
 \mathchardef\PXi="7104
 \mathchardef\PLambda="7103
 \mathchardef\PSigma="7106
 \mathchardef\POmega="710A
 \mathchardef\PUpsilon="7107
                  
 \def\PB      {\ensuremath{B}\xspace}                 
                  
 \def\PD      {\ensuremath{D}\xspace}

 \def\PK      {\ensuremath{K}\xspace}

 \def\Pb      {\ensuremath{b}\xspace}                 
 \def\Pc      {\ensuremath{c}\xspace}                 
                  
 \def\Pe      {\ensuremath{e}\xspace}

 \def\Pi      {\ensuremath{i}\xspace}

 \def\Ps      {\ensuremath{s}\xspace}

}

\makeatletter
\ifcase \@ptsize \relax
  \newcommand{\miniscule}{\@setfontsize\miniscule{4}{5}}
\or
  \newcommand{\miniscule}{\@setfontsize\miniscule{5}{6}}
\or
  \newcommand{\miniscule}{\@setfontsize\miniscule{5}{6}}
\fi
\makeatother

\DeclareRobustCommand{\optbar}[1]{\shortstack{{\miniscule (\rule[.5ex]{1.25em}{.18mm})}
  \\ [-.7ex] $#1$}}

\def\epem       {{\ensuremath{\Pe^+\Pe^-}}\xspace}

\def\squark    {{\ensuremath{\Ps}}\xspace}

\def\cquark    {{\ensuremath{\Pc}}\xspace}

\def\bquark    {{\ensuremath{\Pb}}\xspace}

\def\pion   {{\ensuremath{\Ppi}}\xspace}
\def\piz    {{\ensuremath{\pion^0}}\xspace}

\def\pip    {{\ensuremath{\pion^+}}\xspace}
\def\pim    {{\ensuremath{\pion^-}}\xspace}

\def\kaon    {{\ensuremath{\PK}}\xspace}
  \def\Kbar    {{\kern 0.2em\overline{\kern -0.2em \PK}{}}\xspace}

\def\KorKbar    {\kern 0.18em\optbar{\kern -0.18em K}{}\xspace}

\def\Kp      {{\ensuremath{\kaon^+}}\xspace}
\def\Km      {{\ensuremath{\kaon^-}}\xspace}

\def\KS      {{\ensuremath{\kaon^0_{\rm\scriptscriptstyle S}}}\xspace}

  \def\Dbar    {{\kern 0.2em\overline{\kern -0.2em \PD}{}}\xspace}
\def\D       {{\ensuremath{\PD}}\xspace}

\def\DorDbar    {\kern 0.18em\optbar{\kern -0.18em D}{}\xspace}
\def\Dz      {{\ensuremath{\D^0}}\xspace}
\def\Dzb     {{\ensuremath{\Dbar{}^0}}\xspace}
\def\Dp      {{\ensuremath{\D^+}}\xspace}
\def\Dm      {{\ensuremath{\D^-}}\xspace}

\def\Dstarp  {{\ensuremath{\D^{*+}}}\xspace}

\def\Dsp     {{\ensuremath{\D^+_\squark}}\xspace}

\def\B       {{\ensuremath{\PB}}\xspace}
\def\Bbar    {{\ensuremath{\kern 0.18em\overline{\kern -0.18em \PB}{}}}\xspace}

\def\BorBbar    {\kern 0.18em\optbar{\kern -0.18em B}{}\xspace}
\def\Bz      {{\ensuremath{\B^0}}\xspace}
\def\Bzb     {{\ensuremath{\Bbar{}^0}}\xspace}
\def\Bu      {{\ensuremath{\B^+}}\xspace}
\def\Bub     {{\ensuremath{\B^-}}\xspace}
\def\Bp      {{\ensuremath{\Bu}}\xspace}
\def\Bm      {{\ensuremath{\Bub}}\xspace}

\def\Bd      {{\ensuremath{\B^0}}\xspace}
\def\Bs      {{\ensuremath{\B^0_\squark}}\xspace}
\def\Bsb     {{\ensuremath{\Bbar{}^0_\squark}}\xspace}

  \def\Y#1S{\ensuremath{\PUpsilon{(#1S)}}\xspace}

\def\Lz          {{\ensuremath{\PLambda}}\xspace}
\def\Lbar        {{\ensuremath{\kern 0.1em\overline{\kern -0.1em\PLambda}}}\xspace}
\def\LorLbar    {\kern 0.18em\optbar{\kern -0.18em \PLambda}{}\xspace}

\def\Lc      {{\ensuremath{\Lz^+_\cquark}}\xspace}

\def\to                 {\ensuremath{\rightarrow}\xspace}

\def\CP                {{\ensuremath{C\!P}}\xspace}

\def\AT#1     {\ensuremath{A_{\mathrm{T}}^{#1}}\xspace}

\def\C#1      {\ensuremath{\mathcal{C}_{#1}}\xspace}                       
\def\Cp#1     {\ensuremath{\mathcal{C}_{#1}^{'}}\xspace}                    
\def\Ceff#1   {\ensuremath{\mathcal{C}_{#1}^{\mathrm{(eff)}}}\xspace}        
\def\Cpeff#1  {\ensuremath{\mathcal{C}_{#1}^{'\mathrm{(eff)}}}\xspace}       
\def\Ope#1    {\ensuremath{\mathcal{O}_{#1}}\xspace}                       
\def\Opep#1   {\ensuremath{\mathcal{O}_{#1}^{'}}\xspace}

\newcommand{\tev}{\ifthenelse{\boolean{inbibliography}}{\ensuremath{~T\kern -0.05em eV}\xspace}{\ensuremath{\mathrm{\,Te\kern -0.1em V}}}\xspace}
\newcommand{\gev}{\ensuremath{\mathrm{\,Ge\kern -0.1em V}}\xspace}
\newcommand{\mev}{\ensuremath{\mathrm{\,Me\kern -0.1em V}}\xspace}
\newcommand{\kev}{\ensuremath{\mathrm{\,ke\kern -0.1em V}}\xspace}
\newcommand{\gevnsp}{\ensuremath{\mathrm{Ge\kern -0.1em V}}\xspace}
\newcommand{\mevnsp}{\ensuremath{\mathrm{Me\kern -0.1em V}}\xspace}
\newcommand{\kevnsp}{\ensuremath{\mathrm{ke\kern -0.1em V}}\xspace}
\newcommand{\ev}{\ensuremath{\mathrm{\,e\kern -0.1em V}}\xspace}
\newcommand{\gevc}{\ensuremath{{\mathrm{\,Ge\kern -0.1em V\!/}c}}\xspace}
\newcommand{\mevc}{\ensuremath{{\mathrm{\,Me\kern -0.1em V\!/}c}}\xspace}
\newcommand{\gevcc}{\ensuremath{{\mathrm{\,Ge\kern -0.1em V\!/}c^2}}\xspace}
\newcommand{\gevgevcccc}{\ensuremath{{\mathrm{\,Ge\kern -0.1em V^2\!/}c^4}}\xspace}
\newcommand{\mevcc}{\ensuremath{{\mathrm{\,Me\kern -0.1em V\!/}c^2}}\xspace}

\def\mm   {\ensuremath{\rm \,mm}\xspace}

\def\mum  {\ensuremath{{\,\upmu\rm m}}\xspace}

\def\fm   {\ensuremath{\rm \,fm}\xspace}

\def\invfb   {\ensuremath{\mbox{\,fb}^{-1}}\xspace}

\def\gsim{{~\raise.15em\hbox{$>$}\kern-.85em
          \lower.35em\hbox{$\sim$}~}\xspace}
\def\lsim{{~\raise.15em\hbox{$<$}\kern-.85em
          \lower.35em\hbox{$\sim$}~}\xspace}

\newcommand{\Real}{\ensuremath{\mathcal{R}e}\xspace}

\def\ptot       {\mbox{$p$}\xspace}
\def\pt         {\mbox{$p_{\rm T}$}\xspace}

\def\evtgen     {\mbox{\textsc{EvtGen}}\xspace}

\def\geant      {\mbox{\textsc{Geant4}}\xspace}

\def\photos     {\mbox{\textsc{Photos}}\xspace}

\def\pythia     {\mbox{\textsc{Pythia}}\xspace}

\def\tell1  {TELL1\xspace}
\def\ukl1   {UKL1\xspace}

\newcommand{\eg}{\mbox{\itshape e.g.}\xspace}
\newcommand{\ie}{\mbox{\itshape i.e.}\xspace}

\usepackage{cite} 
\usepackage{mciteplus}
\usepackage{multirow}

\begin{document}

\renewcommand{\thefootnote}{\fnsymbol{footnote}}
\setcounter{footnote}{1}

\begin{titlepage}
\pagenumbering{roman}

\vspace*{-1.5cm}
\centerline{\large EUROPEAN ORGANIZATION FOR NUCLEAR RESEARCH (CERN)}
\vspace*{1.5cm}
\hspace*{-0.5cm}
\begin{tabular*}{\linewidth}{lc@{\extracolsep{\fill}}r}
\ifthenelse{\boolean{pdflatex}}
{\vspace*{-2.7cm}\mbox{\!\!\!\includegraphics[width=.14\textwidth]{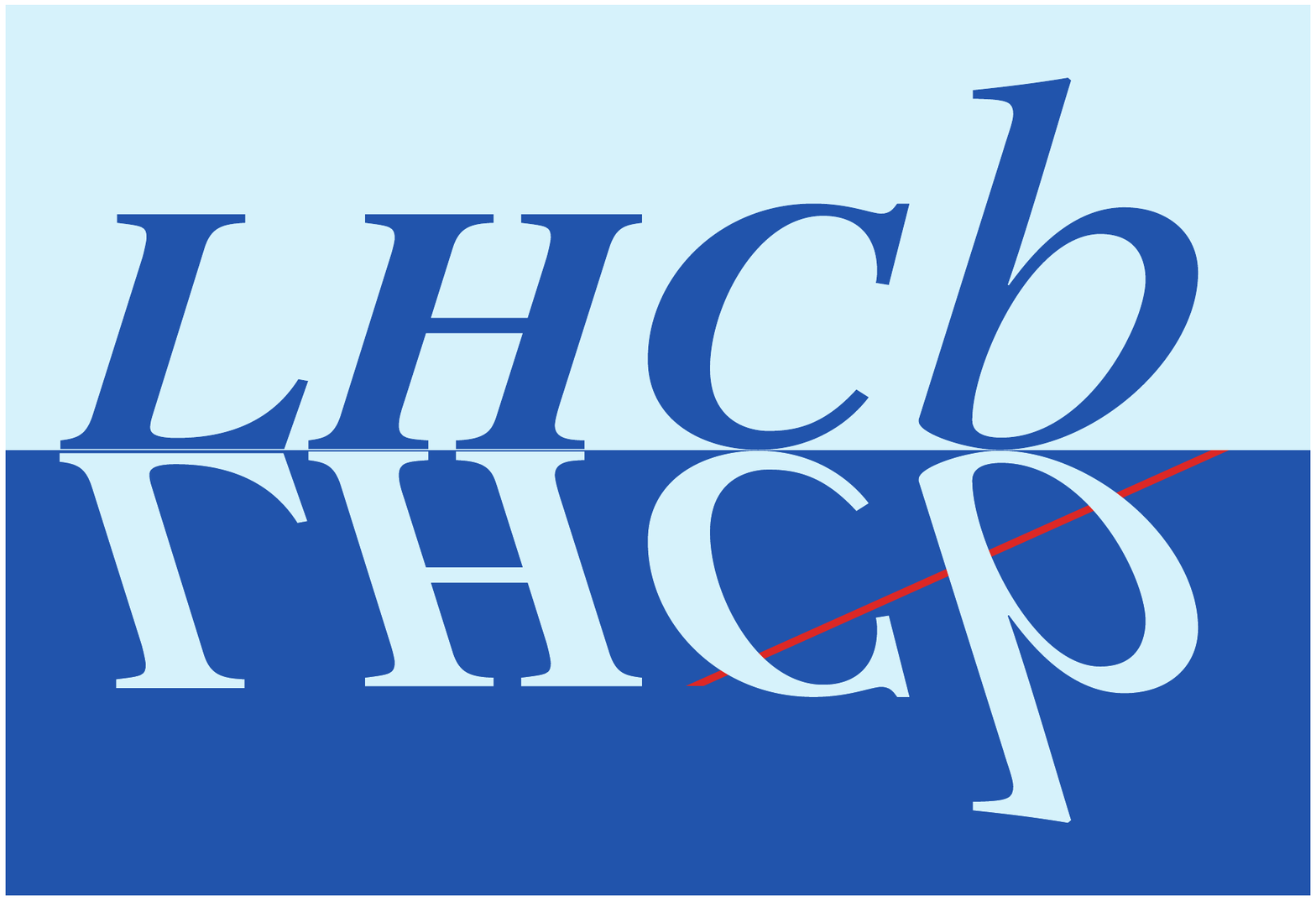}} & &}%
{\vspace*{-1.2cm}\mbox{\!\!\!\includegraphics[width=.12\textwidth]{lhcb-logo.eps}} & &}%
\\
 & & CERN-PH-EP-2015-051 \\  
 & & LHCb-PAPER-2015-007 \\  
 & & 21 June 2016 \\ 
 & & \\
\end{tabular*}

\vspace*{2.5cm}

{\bf\boldmath\huge
\begin{center}
  First observation and amplitude analysis of the $\Bm\to\Dp\Km\pim$ decay
\end{center}
}

\vspace*{1.5cm}

\begin{center}
The LHCb collaboration\footnote{Authors are listed at the end of this paper.}
\end{center}

\vspace{\fill}

\begin{abstract}
  \noindent
  The $\Bm\to\Dp\Km\pim$ decay is observed in a data sample corresponding to
  $3.0\invfb$ of $pp$ collision data recorded by the LHCb experiment during
  2011 and 2012.
  Its branching fraction is measured to be 
  ${\cal B}(\Bm\to\Dp\Km\pim) = (7.31 \pm 0.19 \pm 0.22 \pm 0.39) \times 10^{-5}$
  where the uncertainties are statistical, systematic and from the branching fraction of the normalisation channel $\Bm\to\Dp\pim\pim$, respectively.
  An amplitude analysis of the resonant structure of the $\Bm\to\Dp\Km\pim$
  decay is used to measure the contributions from quasi-two-body
  $\Bm\to \D_0^*(2400)^0\Km$, $\Bm\to \D_2^*(2460)^0\Km$, and 
  $\Bm\to \D_J^*(2760)^0\Km$ decays, as well as from nonresonant sources.
  The $\D_J^*(2760)^0$ resonance is  determined to have spin~1. 
\end{abstract}

\vspace*{1.5cm}

\begin{center}
  Submitted to Phys.~Rev.~D
\end{center}

\vspace{\fill}

{\footnotesize 
\centerline{\copyright~CERN on behalf of the \lhcb collaboration, licence \href{http://creativecommons.org/licenses/by/4.0/}{CC-BY-4.0}.}}
\vspace*{2mm}

\end{titlepage}


\newpage
\setcounter{page}{2}
\mbox{~}

\cleardoublepage

\renewcommand{\thefootnote}{\arabic{footnote}}
\setcounter{footnote}{0}


\pagestyle{plain} 
\setcounter{page}{1}
\pagenumbering{arabic}


\section{Introduction}
\label{sec:introduction}

Excited charmed mesons are of great theoretical and experimental interest as they allow detailed studies of QCD in an interesting energy regime.
Good progress has been achieved in identifying and measuring the parameters of
the orbitally excited states, notably from Dalitz plot (DP) analyses of three-body $B$ decays.
Relevant examples include the studies of $\Bm \to \Dp\pim\pim$~\cite{Abe:2003zm,Aubert:2009wg} and $\Bzb\to\Dz\pip\pim$~\cite{Kuzmin:2006mw} decays, which provide information on excited neutral and charged charmed mesons (collectively referred to as $D^{**}$ states), respectively.
First results on excited charm-strange mesons have also recently been obtained
with the DP analysis technique~\cite{LHCb-PAPER-2014-035,LHCb-PAPER-2014-036,Lees:2014abp}.
Studies of prompt charm resonance production in \epem and $pp$ collisions~\cite{delAmoSanchez:2010vq,LHCb-PAPER-2013-026} have revealed a number of additional high mass states.
Most of these higher mass states are not yet confirmed by independent analyses, and their spectroscopic identification is unclear.
Analyses of resonances produced directly from \epem and $pp$ collisions do not allow determination of the quantum numbers of the produced states, but can distinguish whether or not they have natural spin parity (\ie\ $J^P$ in the series $0^+, 1^-, 2^+, ...$).
The current experimental knowledge of the neutral $D^{**}$ states is summarised in Table~\ref{tab:PDG} (here and throughout the paper, natural units with $\hbar = c = 1$ are used).
The $D^*_0(2400)^0$, $D_1(2420)^0$, $D_1^\prime(2430)^0$ and $D^*_2(2460)^0$ mesons are generally understood to be the four orbitally excited (1P) states.
The experimental situation as well as the spectroscopic identification of the heavier states is less clear.

\begin{table}[!b]
\centering
\caption{\small
  Measured properties of neutral $D^{**}$ states.
  Where more than one uncertainty is given, the first is statistical and the others systematic.
}
\begin{tabular}{ccccc}
  \hline
  Resonance & Mass & Width & $J^P$ & Ref. \\
            & $(\mevnsp)$ & $(\mevnsp)$ & & \\
  \hline
  $D^*_0(2400)^0$ & $2318 \pm 29$ & $267 \pm 40$ & $0^+$ & \cite{PDG2014} \\
  $D_1(2420)^0$ & $2421.4 \pm 0.6$ & $27.4 \pm 2.5$ & $1^+$ & \cite{PDG2014} \\
  $D_1^\prime(2430)^0$ & $2427 \pm 26 \pm 20 \pm 15$ & $384\,^{+107}_{-75} \pm 24 \pm 70$ & $1^+$ & \cite{Abe:2003zm} \\
  $D^*_2(2460)^0$ & $2462.6 \pm 0.6$ & $49.0 \pm 1.3$ & $2^+$ & \cite{PDG2014} \\
  \hline
  $D^*(2600)$ & $2608.7 \pm 2.4 \pm 2.5$ & $93 \pm 6 \pm 13$ & natural & \cite{delAmoSanchez:2010vq} \\
  $D^*(2650)$ & $2649.2 \pm 3.5 \pm 3.5$ & $140 \pm 17 \pm 19$ & natural & \cite{LHCb-PAPER-2013-026} \\
  $D^*(2760)$ & $2763.3 \pm 2.3 \pm 2.3$ & $60.9 \pm 5.1 \pm 3.6$ & natural & \cite{delAmoSanchez:2010vq} \\
  $D^*(2760)$ & $2760.1 \pm 1.1 \pm 3.7$ & $74.4 \pm 3.4 \pm 19.1$ & natural & \cite{LHCb-PAPER-2013-026} \\
  \hline
\end{tabular}
\label{tab:PDG}
\end{table}

The $\Bm\to\Dp\Km\pim$ decay can be used to study neutral $D^{**}$ states. 
The $\Dp\Km\pim$ final state is expected to exhibit resonant structure only in the $\Dp\pim$ channel, and unlike the Cabibbo-favoured $\Dp\pim\pim$ final state does not contain any pair of identical particles.
This simplifies the analysis of the contributing excited charm states, since partial wave analysis can be used to help determine the resonances that contribute.

One further motivation to study $\Bm \to \Dp\Km\pim$ decays is related to the measurement of the angle $\gamma$ of the unitarity triangle defined as $\gamma \equiv \arg\left[ - V^{}_{ud}V_{ub}^*/(V^{}_{cd}V_{cb}^*) \right]$, where $V^{}_{xy}$ are elements of the Cabibbo-Kobayashi-Maskawa (CKM) quark mixing matrix~\cite{PhysRevLett.10.531,PTP.49.652}.
One of the most powerful methods to determine $\gamma$ uses $\Bm \to D \Km$ decays, with the neutral $D$ meson decaying to \CP eigenstates~\cite{Gronau:1990ra,Gronau:1991dp}.
The sensitivity to $\gamma$ arises due to the interference of amplitudes proportional to the CKM matrix elements $V^{}_{ub}$ and $V^{}_{cb}$, associated with $\Dzb$ and $\Dz$ production respectively.
However, a challenge for such methods is to determine the ratio of magnitudes of the two amplitudes, $r_B$, that must be known to extract $\gamma$.
This is usually handled by including \D meson decays to additional final states in the analysis.
By contrast, in $\Bm \to D^{**}\Km$ decays the efficiency-corrected ratio of yields of $\Bm \to D^{**} \Km \to \Dm \pip \Km$ and $\Bm \to D^{**} \Km \to \Dp \pim \Km$ decays gives $r_B^2$ directly~\cite{Sinha:2004ct}.
The decay $\Bm \to D^{**} \Km \to D \pi^0 \Km$ where the \D meson is reconstructed in \CP eigenstates can be used to search for \CP violation driven by $\gamma$. 
Measurement of the first two of these processes would therefore provide knowledge of $r_B$ in $\Bm \to D^{**}\Km$ decays, indicating whether or not a competitive measurement of $\gamma$ can be made with this approach.

In this paper, the $\Bm\to\Dp\Km\pim$ decay is studied for the first time, with the $\Dp$ meson reconstructed through the $\Km\pip\pip$ decay mode. 
The inclusion of charge conjugate processes is implied.
The topologically similar $\Bm \to \Dp\pim\pim$ decay is used as a control channel and for normalisation of the branching fraction measurement.
A large $\Bm\to\Dp\Km\pim$ signal yield is found, corresponding to a clear first observation of the decay, and allowing investigation of the DP structure of the decay.
The amplitude analysis allows studies of known resonances, searches for higher mass states and measurement of the properties, including the quantum numbers, of any resonances that are observed.
The analysis is based on a data sample corresponding to an integrated luminosity of $3.0  \,{\rm fb}^{-1}$ of $pp$ collision data collected with the LHCb detector, approximately one third of which was collected during 2011 when the collision centre-of-mass energy was $\sqrt{s} = 7 \tev$ and the rest during 2012 with $\sqrt{s} = 8 \tev$.

The paper is organised as follows.
A brief description of the LHCb detector as well as reconstruction and simulation software is given in Sec.~\ref{sec:detector}.
The selection of signal candidates is described in Sec.~\ref{sec:selection},
and the branching fraction measurement is presented in Sec.~\ref{sec:BF}.
Studies of the backgrounds and the fit to the \B candidate invariant mass distribution are in Sec.~\ref{sec:mass-fit}, with studies of the signal efficiency and a definition of the square Dalitz plot (SDP) in Sec.~\ref{sec:efficiency}.
Systematic uncertainties on, and the results for, the branching fraction are discussed in Secs.~\ref{sec:BF-syst} and~\ref{sec:BF-results} respectively. 
A study of the angular moments of $\Bm\to\Dp\Km\pim$ decays is given in Sec.~\ref{sec:moments}, with results used to guide the Dalitz plot analysis that follows.
An overview of the Dalitz plot analysis formalism is given in Sec.~\ref{sec:dalitz-generalities}, and details of the implementation of the amplitude analysis are presented in Sec.~\ref{sec:dalitz}.
The evaluation of systematic uncertainties is described in Sec.~\ref{sec:systematics}.
The results and a summary are given in Sec.~\ref{sec:results}.

\section{LHCb detector}
\label{sec:detector}

The \lhcb detector~\cite{Alves:2008zz,LHCb-DP-2014-002} is a single-arm forward
spectrometer covering the \mbox{pseudorapidity} range $2<\eta <5$,
designed for the study of particles containing \bquark or \cquark
quarks. The detector includes a high-precision tracking system
consisting of a silicon-strip vertex detector~\cite{LHCb-DP-2014-001}
surrounding the $pp$ interaction region, a large-area silicon-strip detector
located upstream of a dipole magnet with a bending power of about
$4{\rm\,Tm}$, and three stations of silicon-strip detectors and straw
drift tubes~\cite{LHCb-DP-2013-003} placed downstream of the magnet.
The polarity of the dipole magnet is reversed periodically throughout data-taking.
The tracking system provides a measurement of momentum, \ptot, of charged particles with
a relative uncertainty that varies from 0.5\% at low momentum to 1.0\% at 200\gev.
The minimum distance of a track to a primary vertex, the impact parameter (IP), is measured with a resolution of $(15+29/\pt)\mum$,
where \pt is the component of the momentum transverse to the beam, in \gev.
Different types of charged hadrons are distinguished using information
from two ring-imaging Cherenkov detectors~\cite{LHCb-DP-2012-003}. 
Photon, electron and
hadron candidates are identified by a calorimeter system consisting of
scintillating-pad and preshower detectors, an electromagnetic
calorimeter and a hadronic calorimeter. Muons are identified by a
system composed of alternating layers of iron and multiwire
proportional chambers~\cite{LHCb-DP-2012-002}.

The trigger~\cite{LHCb-DP-2012-004} consists of a
hardware stage, based on information from the calorimeter and muon
systems, followed by a software stage, in which all tracks
with $\pt>500~(300)\mev$ are reconstructed for data collected in 2011 (2012).
The software trigger line used in the analysis reported in this paper requires
a two-, three- or four-track secondary vertex with significant displacement
from the primary $pp$ interaction vertices~(PVs). At least one charged particle
must have $\pt > 1.7\gev$ and be inconsistent with originating from the PV.
A multivariate algorithm~\cite{BBDT} is used for
the identification of secondary vertices consistent with the decay
of a \bquark hadron.

In the offline selection, the objects that fired the trigger are associated with reconstructed particles.  
Selection requirements can therefore be made not only on the trigger line that fired, but on whether the decision was due to the signal candidate, other particles produced in the $pp$ collision, or a combination of both.
Signal candidates are accepted offline if one of the final state particles created a cluster in the hadronic calorimeter with sufficient transverse energy to fire the hardware trigger. 
These candidates are referred to as ``triggered on signal'' or TOS.
Events that are triggered at the hardware level by another particle in the event, referred to as ``triggered independent of signal'' or TIS, are also retained. 
After all selection requirements are imposed, 57\,\% of events in the sample
were triggered by the decay products of the signal candidate (TOS), while the remainder were triggered only by another particle in the event (TIS-only).

Simulated events are used to characterise the detector response to signal and
certain types of background events.
In the simulation, $pp$ collisions are generated using
\pythia~\cite{Sjostrand:2006za,*Sjostrand:2007gs} with a specific \lhcb
configuration~\cite{LHCb-PROC-2010-056}.  Decays of hadronic particles
are described by \evtgen~\cite{Lange:2001uf}, in which final state
radiation is generated using \photos~\cite{Golonka:2005pn}. The
interaction of the generated particles with the detector and its
response are implemented using the \geant
toolkit~\cite{Allison:2006ve, *Agostinelli:2002hh} as described in
Ref.~\cite{LHCb-PROC-2011-006}.

\section{Selection requirements}
\label{sec:selection}

Most selection requirements are optimised using the $\Bm \to \Dp \pim \pim$ control channel. 
Loose initial selection requirements on the quality of the tracks combined to form the \B candidate, as well as on their $p$, $\pt$ and $\chi^2_{\rm{IP}}$, are applied to obtain a visible peak in the invariant mass distribution. 
The $\chi^2_{\rm{IP}}$ is the difference between the $\chi^2$ of the PV reconstruction with and without the considered particle.
Only candidates with invariant mass in the range $1770 < m(\Km\pip\pip) < 1968 \mev$ are retained.
Further requirements are imposed on the vertex quality ($\chi^2_{\rm{vtx}}$) and flight distance from the associated PV of the \B and \D candidates.
The \B candidate must also satisfy requirements on its invariant mass and on the cosine of the angle between the momentum vector and the line joining the PV under consideration to the $\B$ vertex ($\cos \theta_{\rm{dir}}$). 
The initial selection requirements are found to be about $90\,\%$ efficient on simulated signal decays.

Two neural networks~\cite{Feindt2006190} are used to further separate signal from background.
The first is designed to separate candidates that contain real $\Dp \to \Km \pip \pip$ decays from those that do not; the second separates $\Bm \to \Dp \pim \pim$ signal decays from background combinations.
Both networks are trained using the $\Dp \pim \pim$ control channel, where the {\it sPlot} technique~\cite{Pivk:2004ty} is used to statistically separate $\Bm \to \Dp \pim \pim$ signal decays from background combinations using the \D (\B) candidate mass as the discriminating variable for the first (second) network. 
The first network takes as input properties of the \D candidate and its daughter tracks, including information about kinematics, track and vertex quality.
The second uses a total of 27 input variables. 
They include the $\chi^2_{\rm{IP}}$ of the two ``bachelor'' pions (\ie\ pions that originate directly from the \B decay) and properties of the \D candidate including its $\chi^2_{\rm{IP}}$, $\chi^2_{\rm{vtx}}$, $\cos \theta_{\rm{dir}}$, the output of the \D neural network and the square of the flight distance divided by its uncertainty squared ($\chi^2_{\rm{flight}}$).
Variables associated to the \B candidate are also used, including $\pt$, $\chi^2_{\rm{IP}}$, $\chi^2_{\rm{vtx}}$, $\chi^2_{\rm{flight}}$ and $\cos \theta_{\rm{dir}}$.
The $\pt$ asymmetry and track multiplicity in a cone with half-angle of 1.5 units of the plane of pseudorapidity and azimuthal angle (measured in radians) around the \B candidate flight direction~\cite{LHCb-PAPER-2012-001}, which contain information about the isolation of the \B candidate from the rest of the event, are also used in the network.
The neural network input quantities depend only weakly on the kinematics of the $\B$ decay.
A requirement is imposed on the second neural network output that reduces the combinatorial background by an order of magnitude while retaining about $75\,\%$ of the signal.

The selection criteria for the $\Bm \to \Dp \Km \pim$ and $\Bm \to \Dp \pim \pim$ candidates are identical except for the particle identification (PID) requirement on the bachelor track that differs between the two modes.
All five final state particles for each decay mode have PID criteria applied to preferentially select either pions or kaons.
Tight requirements are placed on the higher momentum pion from the $\Dp$ decay and on the bachelor kaon in $\Bm \to \Dp \Km \pim$ to suppress backgrounds from $\Dsp \to \Km \Kp \pip$ and $\Bm \to \Dp \pim \pim$ decays, respectively.
The combined efficiency of the PID requirements on the five final state tracks is around $70\,\%$ for $\Bm \to \Dp \pim \pim$ decays and around $40\,\%$ for $\Bm \to \Dp \Km \pim$ decays.
The PID efficiency depends on the kinematics of the tracks, as described in detail in Sec.~\ref{sec:efficiency}, and is determined using samples of $D^{0} \to \Km \pip$ decays selected in data by exploiting the kinematics of the $D^{*+} \to D^{0}\pip$ decay chain to obtain clean samples without using the PID information.

To improve the \B candidate invariant mass resolution, track momenta are scaled~\cite{LHCb-PAPER-2012-048,LHCb-PAPER-2013-011} with calibration parameters determined by matching the measured peak of the $J/\psi \to \mu^{+} \mu^{-}$ decay to the known $J/\psi$ mass~\cite{PDG2014}.
Furthermore, a fit to the kinematics and topology of the decay chain~\cite{Hulsbergen:2005pu} is used to adjust the four-momenta of the tracks from the \D candidate so that their combined invariant mass matches the world average value for the $\Dp$ meson~\cite{PDG2014}. 
An additional \B mass constraint is applied in the calculation of the variables that are used in the Dalitz plot fit.

To remove potential background from misreconstructed $\Lc$ decays, candidates are rejected if the invariant mass of the \D candidate lies in the range $2280$--$2300\mev$ when the proton mass hypothesis is applied to the low momentum pion track. 
Possible backgrounds from $\Bm$ meson decays without an intermediate charm meson are suppressed by the requirement on the output value from the first neural network, and any surviving background of this type is removed by requiring that the \D candidate vertex is displaced by at least $1 \mm$ from the \B decay vertex. 
The efficiency of this requirement is about $85\,\%$.

Signal candidates are retained for further analysis if they have an invariant mass in the range $5100$--$5800 \mev$. 
After all selection requirements are applied, fewer than $1\,\%$ of events with one candidate also contain a second candidate.
Such multiple candidates are retained and treated in the same manner as other candidates; the associated systematic uncertainty is negligible.

\section{Branching fraction determination}
\label{sec:BF}

The ratio of branching fractions is calculated from the signal yields with event-by-event efficiency corrections applied as a function of square Dalitz plot position. 
The calculation is
\begin{equation}
\label{eq:BF-weighted}
\frac{\mathcal{B}(\Bm\to\Dp\Km\pim)}{\mathcal{B}(\Bm\to\Dp\pim\pim)} = 
\frac{N^{\rm corr}(\Bm\to\Dp\Km\pim)}{N^{\rm corr}(\Bm\to\Dp\pim\pim)}\,,
\end{equation}
where $N^{\rm corr} = \sum_{i}W_i/\epsilon_i$ is the efficiency-corrected yield.
The index $i$ sums over all candidates in the data sample and $W_i$ is the signal weight for each candidate, which is determined from the fits described in Sec.~\ref{sec:mass-fit} and shown in Figs.~\ref{fig:dpimi-fit} and~\ref{fig:dkpi-fit}, using the {\it sPlot} technique~\cite{Pivk:2004ty}.
Each fit is performed simultaneously to decays in the TOS and TIS-only categories.
The efficiency of candidate $i$, $\epsilon_i$, is obtained separately for each trigger subsample as described in Sec.~\ref{sec:efficiency}.

\subsection{Determination of signal and background yields}
\label{sec:mass-fit}

The candidates that survive the selection requirements are comprised of signal decays and various categories of background.
Combinatorial background arises from random combinations of tracks (possibly including a real $\Dp\to\Km\pip\pip$ decay).
Partially reconstructed backgrounds originate from \bquark~hadron decays with additional particles that are not part of the reconstructed decay chain.
Misidentified decays also originate from \bquark~hadron decays, but where one of the final state particles has been incorrectly identified (\eg\ a pion as a kaon).
The signal (normalisation channel) and background yields are obtained from unbinned maximum likelihood fits to the $\Dp\Km\pim$ ($\Dp\pim\pim$) invariant mass distributions. 

Both the $\Bm \to \Dp\Km\pim$ and $\Bm \to \Dp\pim\pim$ signal shapes are modelled by the sum of two Crystal Ball (CB) functions~\cite{Skwarnicki:1986xj} with a common mean and tails on opposite sides, where the high-mass tail accounts for non-Gaussian reconstruction effects.
The ratio of widths of the CB shapes and the relative normalisation of the narrower CB shape are constrained within their uncertainties to the values found in fits to simulated signal samples. 
The tail parameters of the CB shapes are also fixed to those found in simulation. 

The combinatorial backgrounds in both $\Dp\Km\pim$ and $\Dp\pim\pim$ samples are modelled with linear functions; the slope of this function is allowed to differ between the two trigger subsamples.
The decay $\Bm\to\Dstarp\Km\pim$ is a partially reconstructed background for $\Dp\Km\pim$ candidates, where the $\Dstarp$ decays to either $\Dp\gamma$ or $\Dp\piz$ and the neutral particle is not reconstructed.
Similarly the decay $\Bm\to\Dstarp\pim\pim$ forms a partially reconstructed background to the $\Dp\pim\pim$ final state.
These are modelled with non-parametric shapes determined from simulated samples.
The shapes are characterised by a sharp edge around $100 \mev$ below the \B peak, where the exact position of the edge depends on properties of the decay including the $\Dstarp$ polarisation.
The fit quality improves when the shape is allowed to be offset by a small shift that is determined from the data.

Most potential sources of misidentified backgrounds have broad $\B$ candidate invariant mass distributions, and hence are absorbed in the combinatorial background component in the fit. 
The decays $\Bm\to D^{(*)+} \pim\pim$ and $\Bm\to\Dsp\Km\pim$, however, give distinctive shapes in the mass distribution of $\Dp\Km\pim$ candidates. 
For $\Dp\pim\pim$ candidates the only significant misidentified background contribution is from $\Bm\to D^{(*)+} \Km\pim$ decays.
The misidentified background shapes are also modelled with non-parametric shapes determined from simulated samples.

\begin{figure}[!tb]
\centering
\includegraphics[scale=0.38]{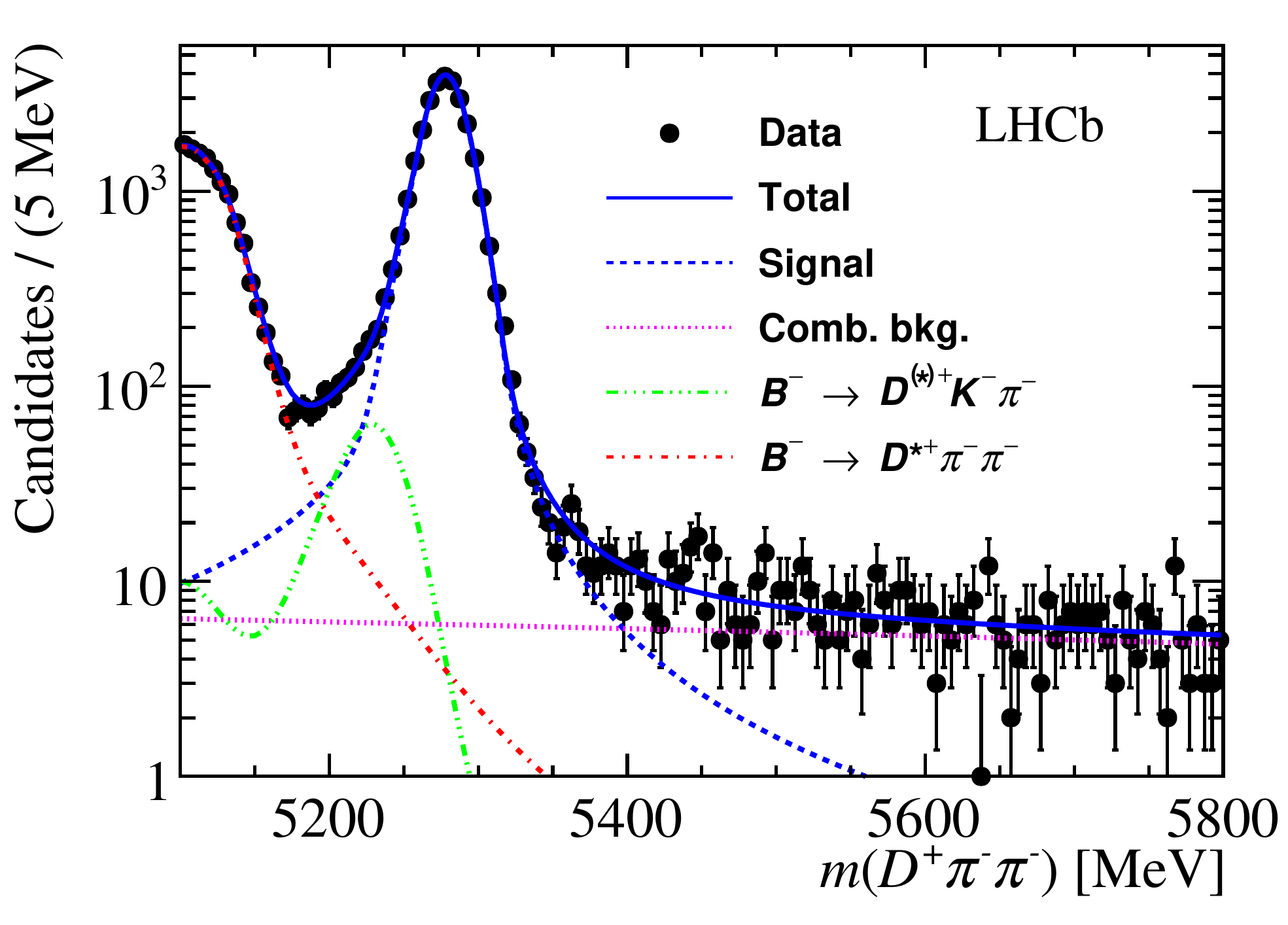}
\includegraphics[scale=0.38]{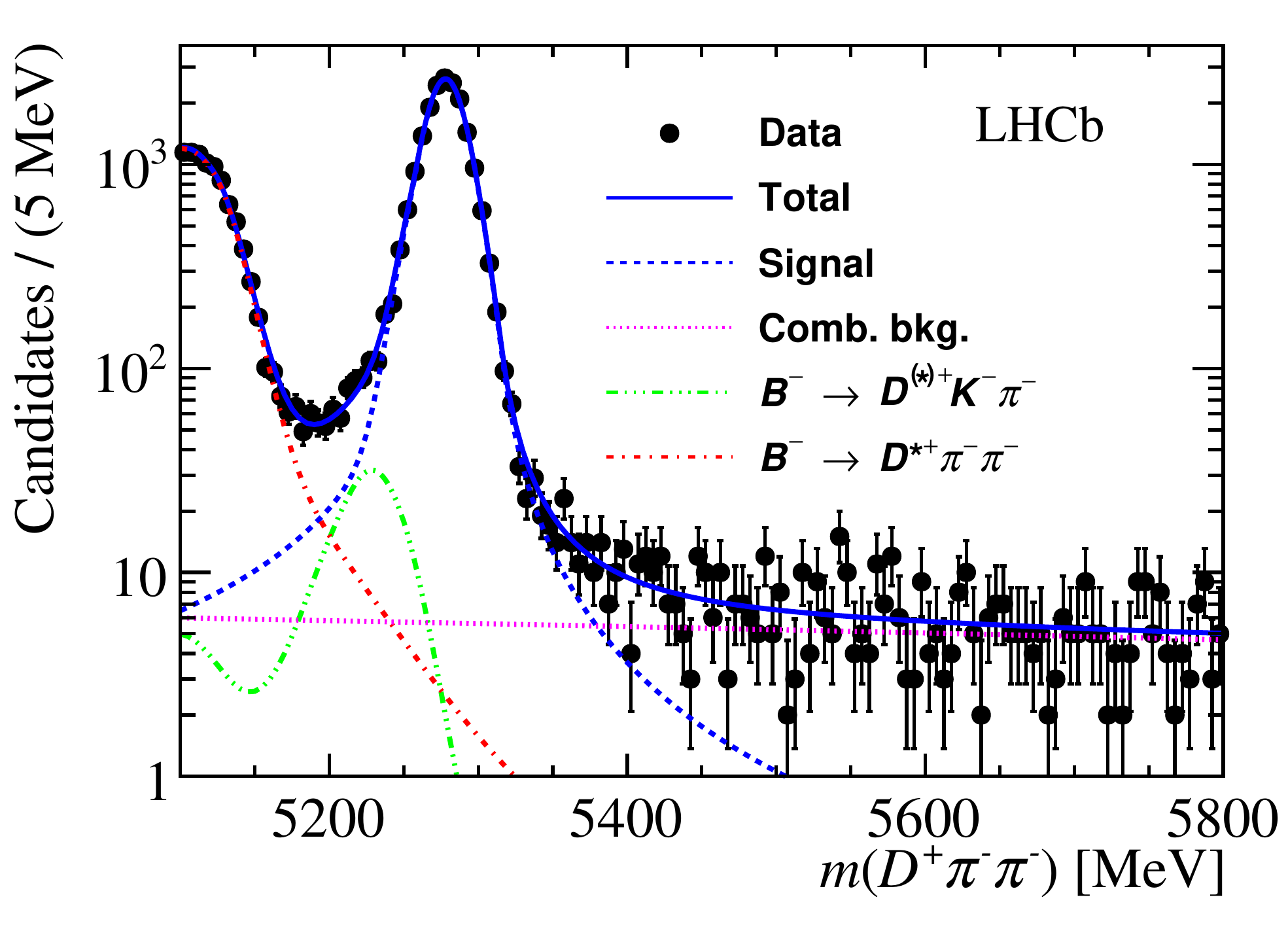}
\caption{\small 
  Results of the fit to the $\Bm \to \Dp\pim\pim$ candidate invariant mass distribution for the (left) TOS and (right) TIS-only subsamples.
  Data points are shown in black, the full fitted model as solid blue lines and the components as shown in the legend.}
\label{fig:dpimi-fit}
\end{figure}

The simulated samples used to obtain signal and background shapes are generated with flat distributions in the phase space of their SDPs.
For $\Bm\to \Dp \pim\pim$ and $\Bm\to \Dstarp \pim\pim$ decays, accurate models of the distributions across the SDP are known~\cite{Abe:2003zm,Aubert:2009wg}, so the simulated samples are reweighted using the $\Bm\to \Dp \pim \pim$ data sample; this affects the shape of the misidentified background component in the fit to the $\Dp\Km\pip$ sample.
Additionally, the $\Dp$ and $\Dstarp$ portions of this background are combined according to their known branching fractions. 
All of the shapes, except for that of the combinatorial background, are common between the two trigger subsamples in each fit, but the signal and background yields in the subsamples are independent. 
In total there are 15 free parameters in the fit to the $\Dp\pim\pim$ sample: 
yields in each subsample for signal, combinatorial, $\Bm\to D^{(*)+} \Km\pim$ and $\Bm\to\Dstarp\pim\pim$ backgrounds;
the combinatorial slope in each subsample; 
the double CB peak position, the width of the narrower CB, the ratio of CB widths and the fraction of entries in the narrower CB shape; 
and the shift parameter of the partially reconstructed background.
The result of the $\Dp\pim\pim$ fit is shown in Fig.~\ref{fig:dpimi-fit} for both trigger subsamples and gives a combined signal yield of approximately 49\,000 decays.
Component yields are given in Table~\ref{tab:DpipiFit_yields}.

\begin{table}[!tb] 
\centering
  \caption{\small
    Yields of the various components in the fit to $\Bm\to\Dp\pim\pim$ candidate invariant mass distribution.}
  \centering 
  \begin{tabular}{lcc} 
    \hline 
 Component & TOS & TIS-only \\
    \hline 
 $N(\Bm\to\Dp\pim\pim)$     & $29\,190 \pm 204$ & $19\,416 \pm 159$ \\
 $N(\Bm \to D^{(*)+}\Km\pim)$   & $\phantom{29\,}807 \pm 123$ & $\phantom{19\,}401 \pm 84\phantom{9}$ \\
 $N(\Bm \to \Dstarp\pim\pim)$ 	& $12\,120 \pm 115$ & $\phantom{1\,}8551 \pm 96\phantom{9}$ \\	      
 $N(\rm{comb. \ bkgd.})$  & $\phantom{29\,}784 \pm 54\phantom{4}$ & $\phantom{9\,4}746 \pm 47\phantom{9}$ \\
\hline 
  \end{tabular} 
\label{tab:DpipiFit_yields}
\end{table} 

\begin{figure}[!tb]
\centering
\includegraphics[scale=0.38]{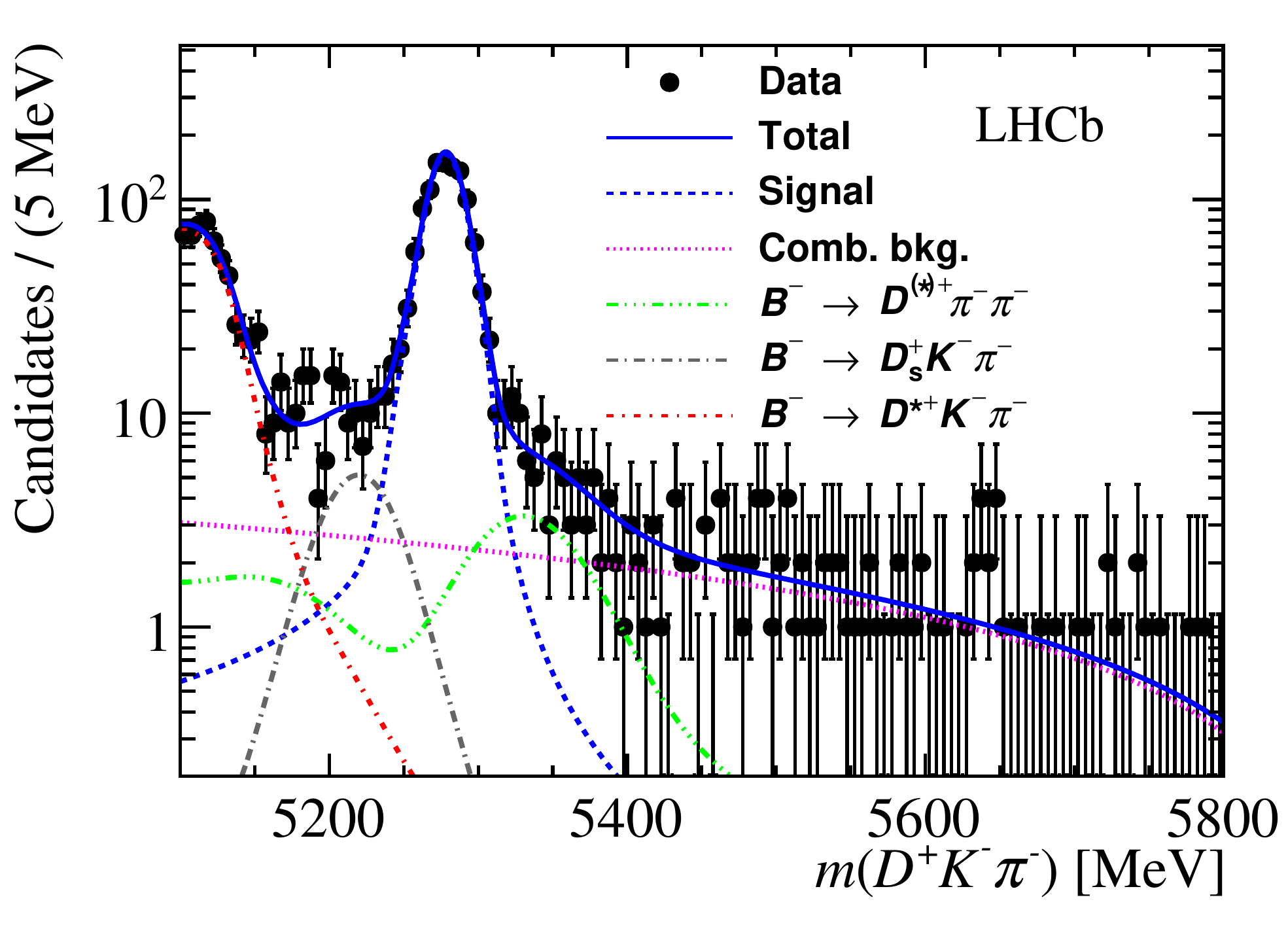}
\includegraphics[scale=0.38]{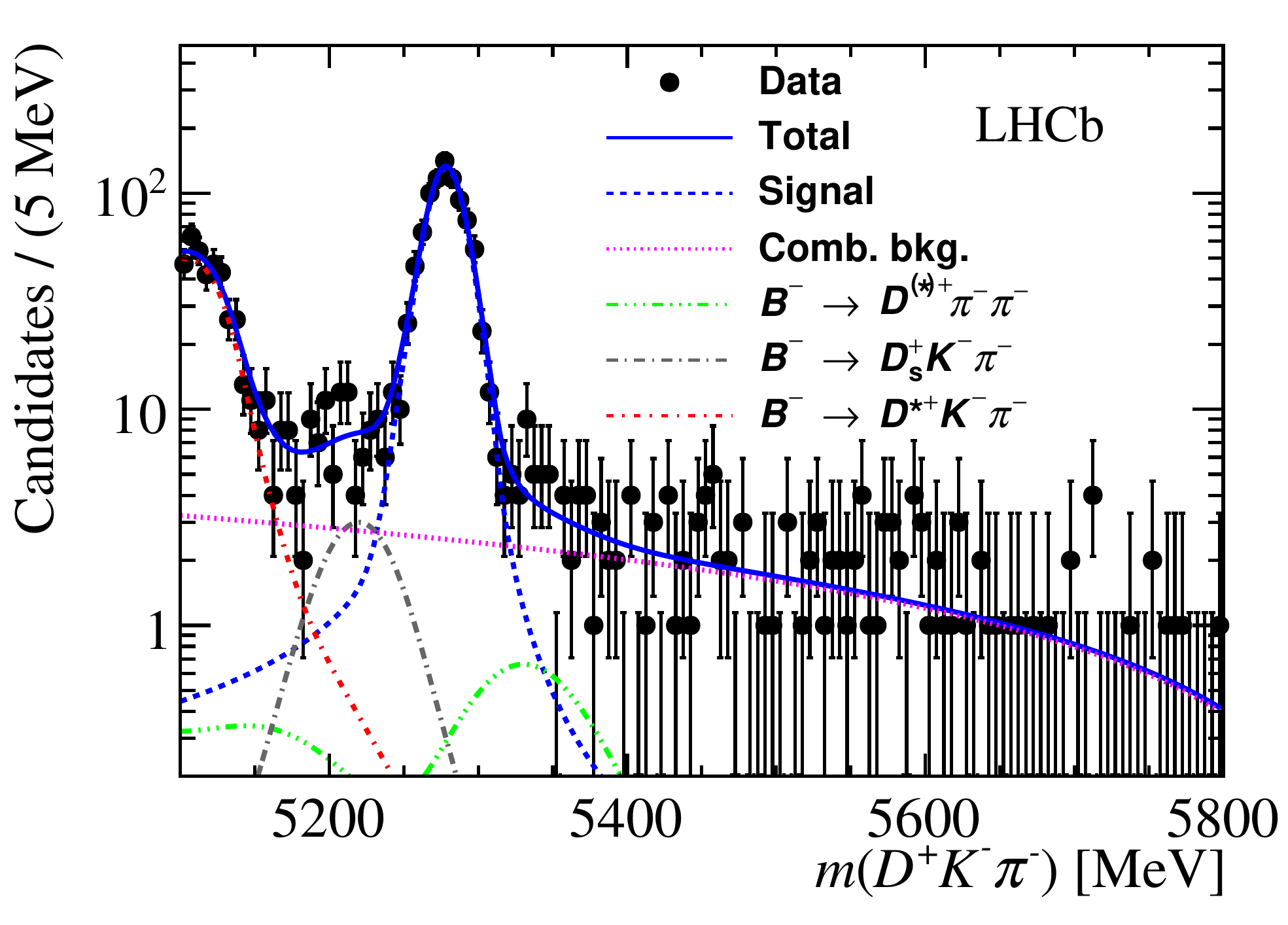}
\caption{\small 
  Results of the fit to the $\Bm \to \Dp\Km\pim$ candidate invariant mass distribution for the (left) TOS and (right) TIS-only subsamples.
  Data points are shown in black, the full fitted model as solid blue lines and the components as shown in the legend.}
\label{fig:dkpi-fit}
\end{figure}

There are a total of 17 free parameters in the fit to the $\Dp\Km\pim$ sample: 
yields in each subsample for signal, combinatorial, $\Bm\to\Dstarp\Km\pim$, $\Bm\to\Dsp\Km\pim$ and $\Bm\to D^{(*)+} \pim\pim$ backgrounds;
the combinatorial slope in each subsample;
the same signal shape parameters as for the $\Dp\pim\pim$ fit;
and the shift parameter of the partially reconstructed background.
Figure~\ref{fig:dkpi-fit} shows the result of the $\Dp\Km\pim$ fit for the two trigger subsamples that yield a total of approximately 2000 $\Bm\to\Dp\Km\pim$ decays.
The yields of all fit components are shown in Table~\ref{tab:DKpiFit_yields}.
The statistical signal significance, estimated in the conventional way from the change in negative log-likelihood from the fit when the signal component is removed, is in excess of $60$ standard deviations ($\sigma$).

\begin{table}[!tb] 
\centering
  \caption{\small
    Yields of the various components in the fit to $\Bm\to\Dp\Km\pim$ candidate invariant mass distribution.}
  \centering 
  \begin{tabular}{lcc} 
    \hline 
 Component & TOS & TIS-only \\
    \hline 
$N(\Bm\to\Dp\Km\pim)$ 	        & $1112 \pm 37$ & $891 \pm 32$ \\
$N(\Bm \to D^{(*)+}\pim\pim)$ 	& $\phantom{1}114 \pm 34$ & $\phantom{8}23 \pm 27$ \\
$N(\Bm \to \Dsp\Km\pim)$ 	& $\phantom{11}69 \pm 17$ & $\phantom{8}40 \pm 15$ \\
$N(\Bm \to \Dstarp\Km\pim)$ 	& $\phantom{1}518 \pm 26$ & $361 \pm 21$ \\
$N(\rm{comb. \ bkgd.})$	        & $\phantom{1}238 \pm 38$ & $253 \pm 36$ \\
\hline 
  \end{tabular} 
\label{tab:DKpiFit_yields}
\end{table} 

\subsection{Signal efficiency}
\label{sec:efficiency}

Since both $\Bm \to \Dp \Km \pim$ and $\Bm \to \Dp \pim \pim$ decays have
non-trivial DP distributions, it is necessary to understand the variation of
the efficiency across the phase space.
Since, moreover, the efficiency variation tends to be strongest close to the kinematic boundaries of the conventional Dalitz plot, it is convenient to model these effects in terms of the SDP defined by variables \mpr\ and \thpr\ which are valid in the range 0 to~1 and are given for the $\Dp \Km \pim$ case by 
\begin{equation}
\label{eq:sqdp-vars}
\mpr \equiv \frac{1}{\pi} \arccos\left(2\frac{m(\Dp\pim) - m^{\rm min}_{\Dp\pim}}{m^{\rm max}_{\Dp\pim} - m^{\rm min}_{\Dp\pim}} - 1 \right)
\hspace{10mm}{\rm and}\hspace{10mm}
\thpr \equiv \frac{1}{\pi}\theta(\Dp\pim)\,,
\end{equation}
where $m^{\rm max}_{\Dp\pim} = m_{\Bm} - m_{\Km}$ and $m^{\rm min}_{\Dp\pim} =
m_{\Dp} + m_{\pim}$ are the kinematic boundaries of $m(\Dp\pim)$ allowed in the $\Bm \to \Dp\Km\pim$ decay and $\theta(\Dp\pim)$ is the helicity angle of the $\Dp\pim$ system (the angle between the $\Km$ and the $\Dp$ meson momenta in the $\Dp\pim$ rest frame). 
For the $\Dp \pim \pim$ case, $\mpr$ and $\thpr$ are defined in terms of the $\pim\pim$ mass and helicity angle, respectively, since with this choice only the region of the SDP with $\thpr(\pim\pim) < 0.5$ is populated due to the symmetry of the two pions in the final state.

Efficiency variation across the SDP is caused by the detector acceptance and by trigger, selection and PID requirements.
The efficiency variation is evaluated for both $\Dp\Km\pim$ and $\Dp\pim\pim$ final states with simulated samples generated uniformly over the SDP.
Data-driven corrections are applied to correct for known differences between data and simulation in the tracking, trigger and PID efficiencies, using identical methods to those described in Ref.~\cite{LHCb-PAPER-2014-036}.
The efficiency functions are fitted with two-dimensional cubic splines to smooth out statistical fluctuations due to limited sample size.

\begin{figure}[!tb]
 \centering
 \includegraphics[scale=0.38]{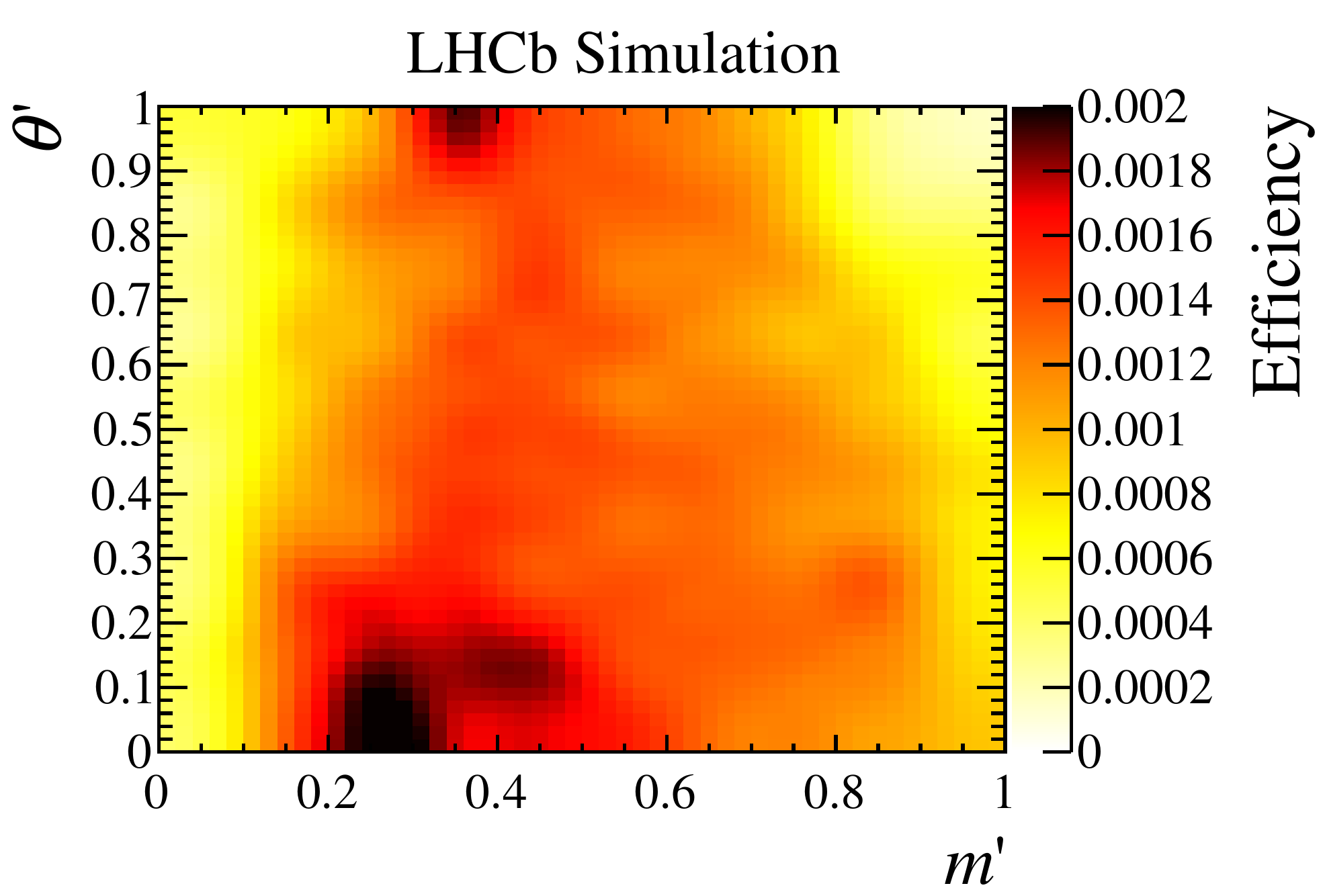}
 \includegraphics[scale=0.38]{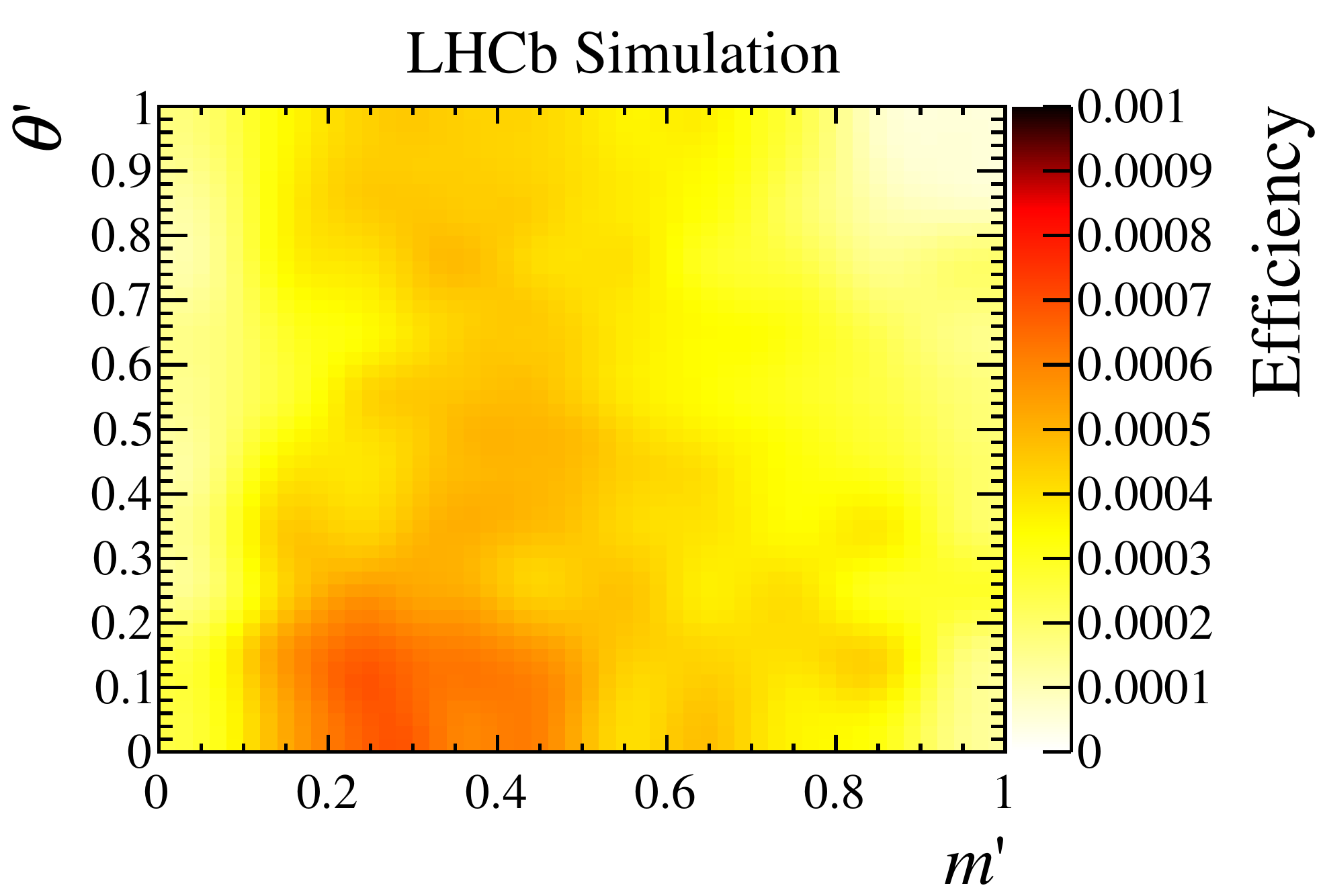}
 \caption{\small 
   Signal efficiency across the SDP for (left) TOS and (right) TIS-only $\Bm \to \Dp \Km \pim$ decays.
   The relative uncertainty at each point is typically $5\,\%$.
}
 \label{fig:eff}
\end{figure}

The efficiency is studied separately for the TOS and TIS-only categories.
The efficiency maps for each trigger subsample are shown for $\Bm \to \Dp \Km \pim$ decays in Fig.~\ref{fig:eff}. 
Regions of relatively high efficiency are seen where all decay products have comparable momentum in the \B rest frame; the efficiency drops sharply in regions with a low momentum bachelor track due to geometrical effects.
The efficiency maps are used to calculate the ratio of branching fractions and also as inputs to the $\Dp\Km\pim$ Dalitz plot fit. 

\subsection{Systematic uncertainties}
\label{sec:BF-syst}
Table~\ref{tab:bf-syst} summarises the systematic uncertainties on the
measurement of the ratio of branching fractions. 
Selection effects cancel in the ratio of branching fractions, except for inefficiency due to the $\Lc$ veto.
The invariant mass fits are repeated both with a wider veto ($2270$--$2310\mev$) and with no veto, and changes in the yields are used to assign a relative systematic uncertainty of $0.2\,\%$.

To estimate the uncertainty arising from the choice of invariant mass fit model, the $\Dp\Km\pim$ mass fit is varied by replacing the signal shape with the sum of two bifurcated Gaussian functions, removing the smoothing of the non-parametric functions, using exponential and second-order polynomial functions to describe the combinatorial background, varying fixed parameters within their uncertainties and varying the binning of histograms used to reweight the simulated background samples. 
For the $\Dp\pim\pim$ fit the same variations are made.
The relative changes in the yields are summed in quadrature to give a relative systematic uncertainty on the ratio of branching fractions of $2.0\,\%$.

The systematic uncertainty due to PID is estimated by accounting for three sources:
the intrinsic uncertainty of the calibration ($1.0\,\%$);
possible differences in the kinematics of tracks in simulated samples, used to reweight the calibration data samples, to those in the data ($1.7\,\%$);
the granularity of the binning in the reweighting procedure ($0.7\,\%$).
Combining these in quadrature, the total relative systematic uncertainty from PID is $2.1\,\%$.

The bins of the efficiency maps are varied within uncertainties to make 100 new efficiency maps, for both $\Dp\Km\pim$ and $\Dp\pim\pim$ modes. 
The efficiency-corrected yields are evaluated for each new map and their distributions are fitted with Gaussian functions.
The widths of these are used to assign a relative systematic uncertainty on the ratio of branching fractions of $0.8\,\%$. 

\begin{table}[!tb]
\centering
\caption{\small
  Relative systematic uncertainties on the measurement of the ratio of branching fractions for $\Bm \to \Dp\Km\pim$ and $\Bm \to \Dp\pim\pim$ decays.
}
\label{tab:bf-syst}
\begin{tabular}{lc}
\hline
Source & Uncertainty (\%) \\
\hline
$\Lc$ veto & 0.2 \\
Fit model & 2.0 \\
Particle identification & 2.1 \\
Efficiency modelling & 0.8 \\
\hline
Total & 3.0 \\
\hline
\end{tabular}
\end{table}

A number of additional cross-checks are performed to test the branching fraction result. The neural network and PID requirements are both tightened and loosened. The data sample is divided by dipole magnet polarity and year of data taking. The branching fraction is also calculated separately for TOS and TIS-only events. All cross-checks give consistent results.

\subsection{Results}
\label{sec:BF-results}
The ratio of branching fractions is found to be
\begin{equation}
\nonumber \frac{\mathcal{B}(\Bm\to\Dp\Km\pim)}{\mathcal{B}(\Bm\to\Dp\pim\pim)} = 0.0720 \pm 0.0019 \pm 0.0021\,, 
\end{equation}
where the first uncertainty is statistical and the second systematic. 
The statistical uncertainty includes contributions from the event weighting used in Eq.~(\ref{eq:BF-weighted}) and from the shape parameters that are allowed to vary in the fit~\cite{LHCb-PAPER-2012-018}.
The world average value of $\mathcal{B}(\Bm\to\Dp\pim\pim) = (1.07 \pm 0.05) \times 10^{-3}$~\cite{PDG2014} assumes that $\Bp\Bm$ and $\Bz\Bzb$ are produced equally in the decay of the $\Upsilon(4S)$ resonance.
Using $\Gamma(\Upsilon(4S)\to\Bp\Bm)/\Gamma(\Upsilon(4S)\to\Bz\Bzb) = 1.055 \pm 0.025$~\cite{PDG2014} gives a corrected value of $\mathcal{B}(\Bm\to\Dp\pim\pim) = (1.01 \pm 0.05) \times 10^{-3}$. 
This allows the branching fraction of $\Bm\to\Dp\Km\pim$ decays to be determined as 
\begin{equation}
\nonumber \mathcal{B}(\Bm\to\Dp\Km\pim) = (7.31 \pm 0.19 \pm 0.22 \pm 0.39) \times 10^{-5}\,,
\end{equation}
where the third uncertainty is from $\mathcal{B}(\Bm\to\Dp\pim\pim)$. 
This measurement represents the first observation of the $\Bm\to\Dp\Km\pim$ decay. 

\section{Study of angular moments}
\label{sec:moments}

To investigate which amplitudes should be included in the DP analysis of $\Bm
\to \Dp\Km\pim$ decays, a study of its angular moments is performed.
Such an analysis is particularly useful for $\Bm\to\Dp\Km\pim$ decays because resonant contributions are only expected to appear in the $\Dp\pim$ combination, and therefore the distributions should be free of effects from reflections that make them more difficult to interpret.

\begin{figure}[!htp]
\centering
 \includegraphics[scale=0.34]{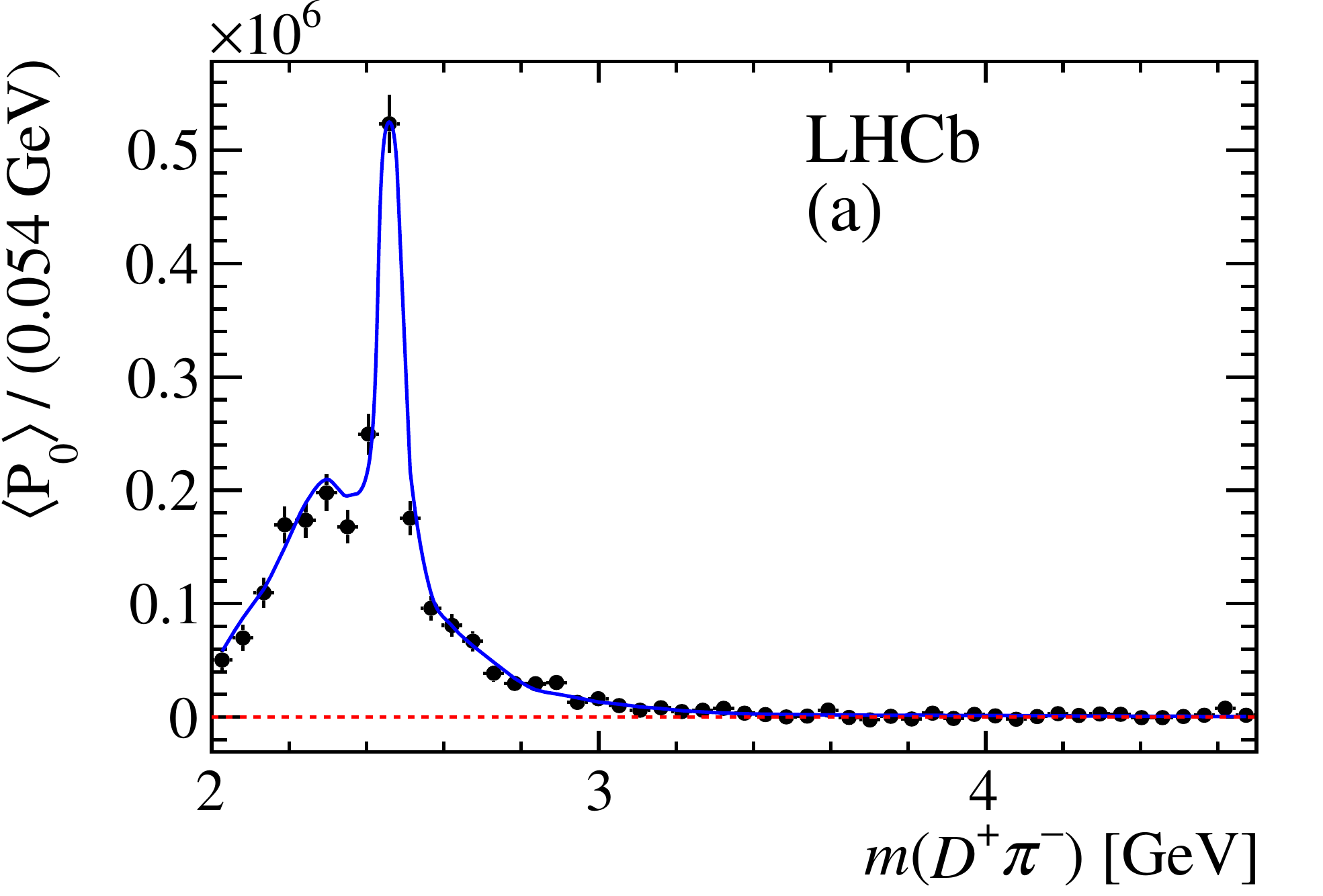}
 \includegraphics[scale=0.34]{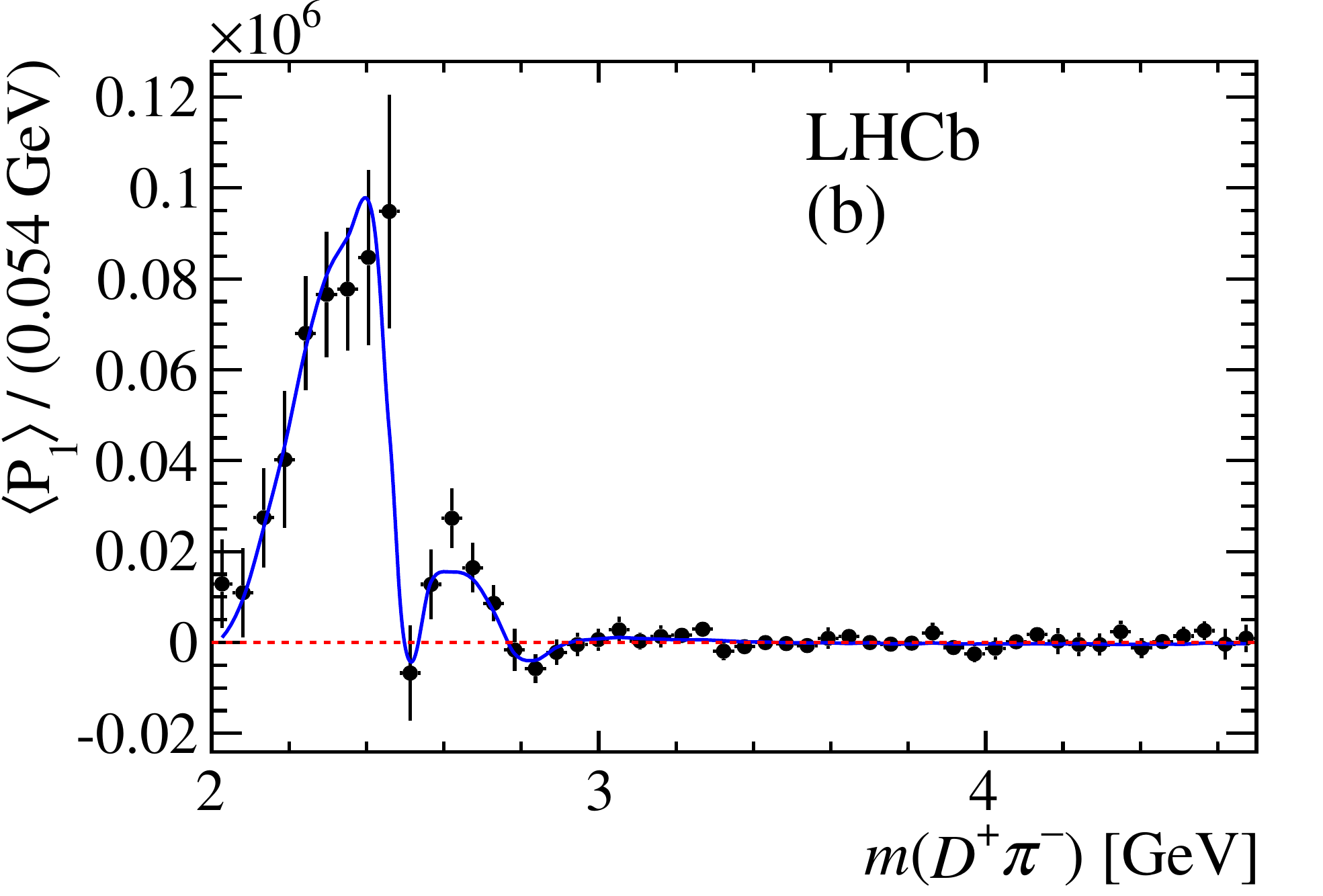}
 \includegraphics[scale=0.34]{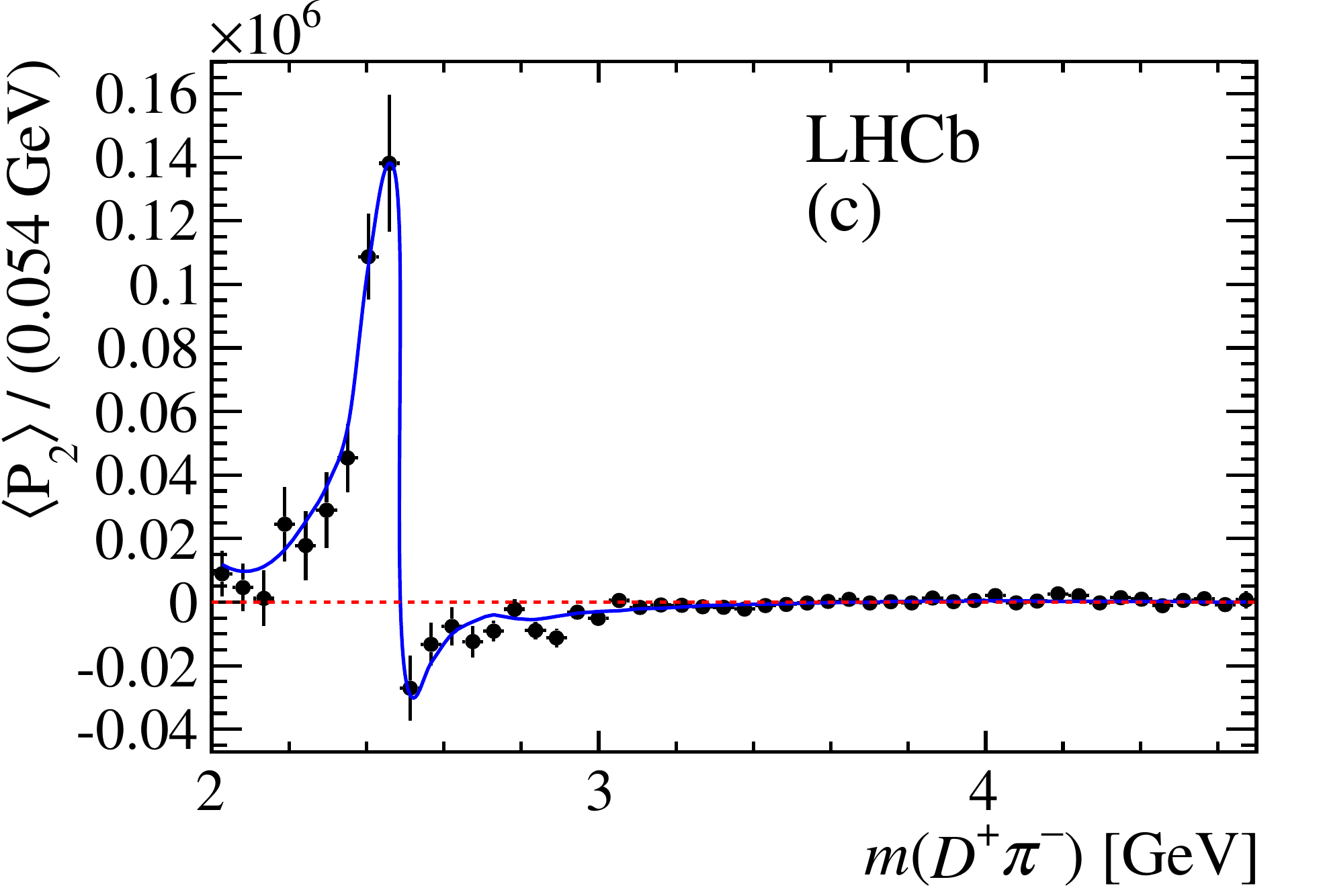}
 \includegraphics[scale=0.34]{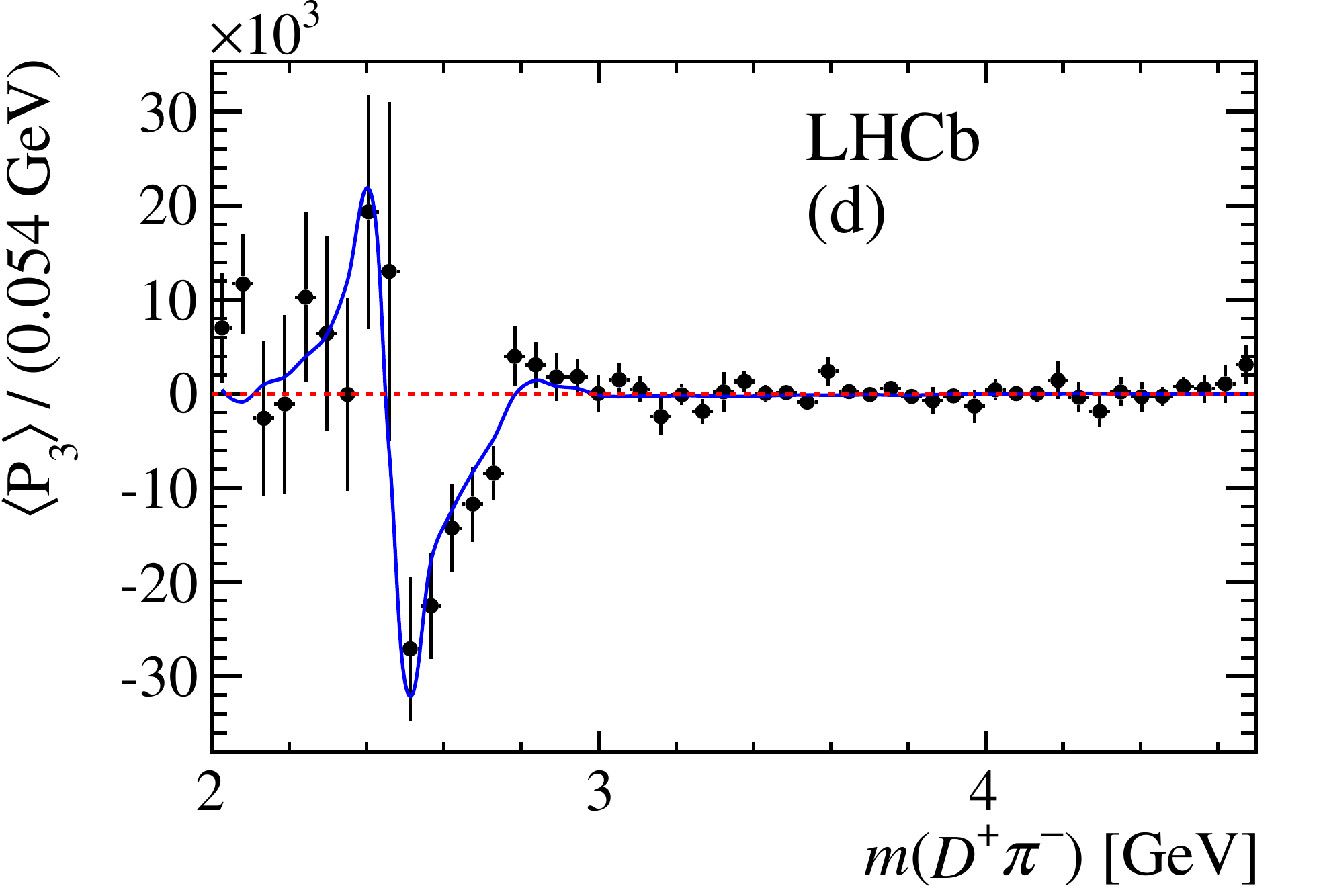}
 \includegraphics[scale=0.34]{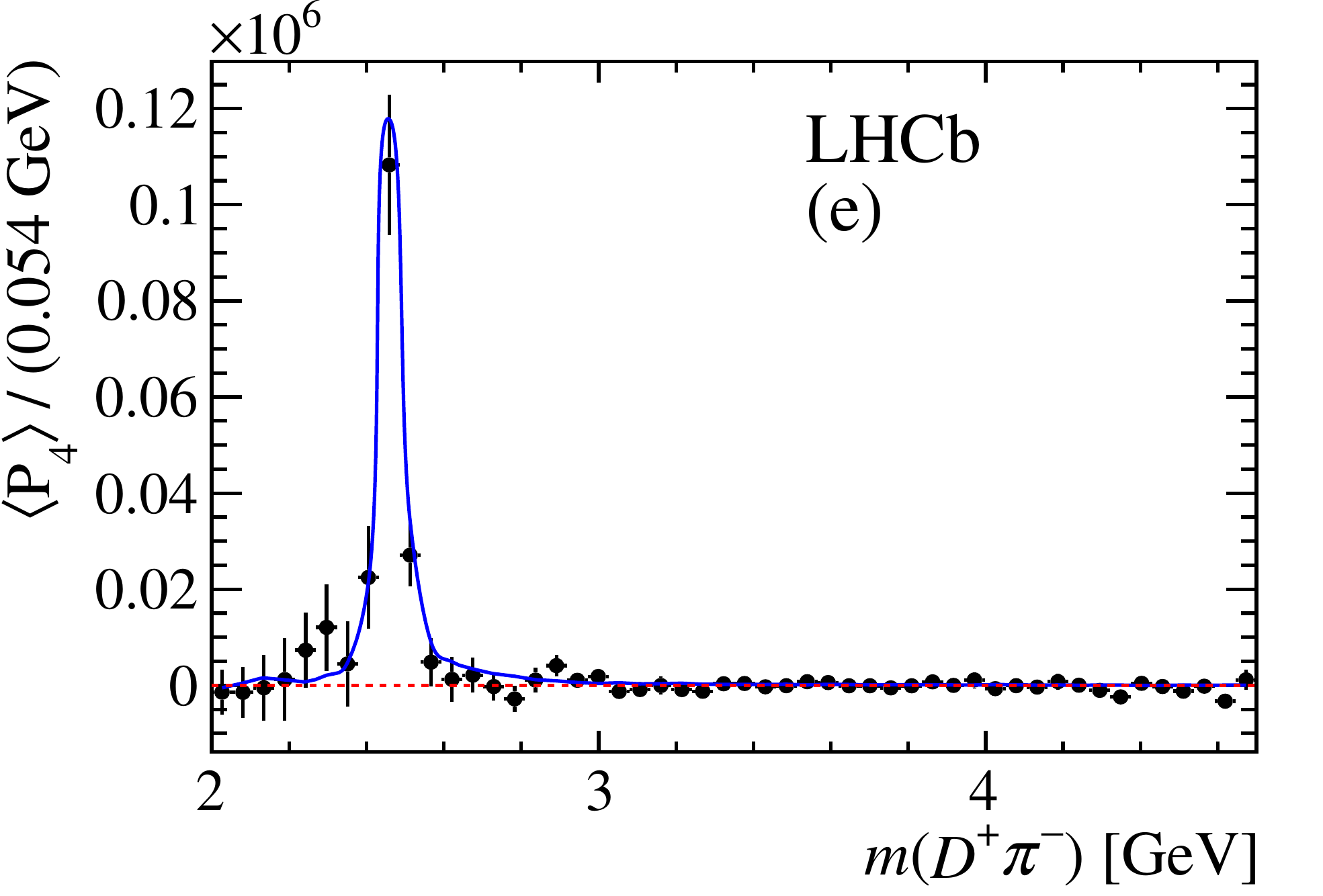}
 \includegraphics[scale=0.34]{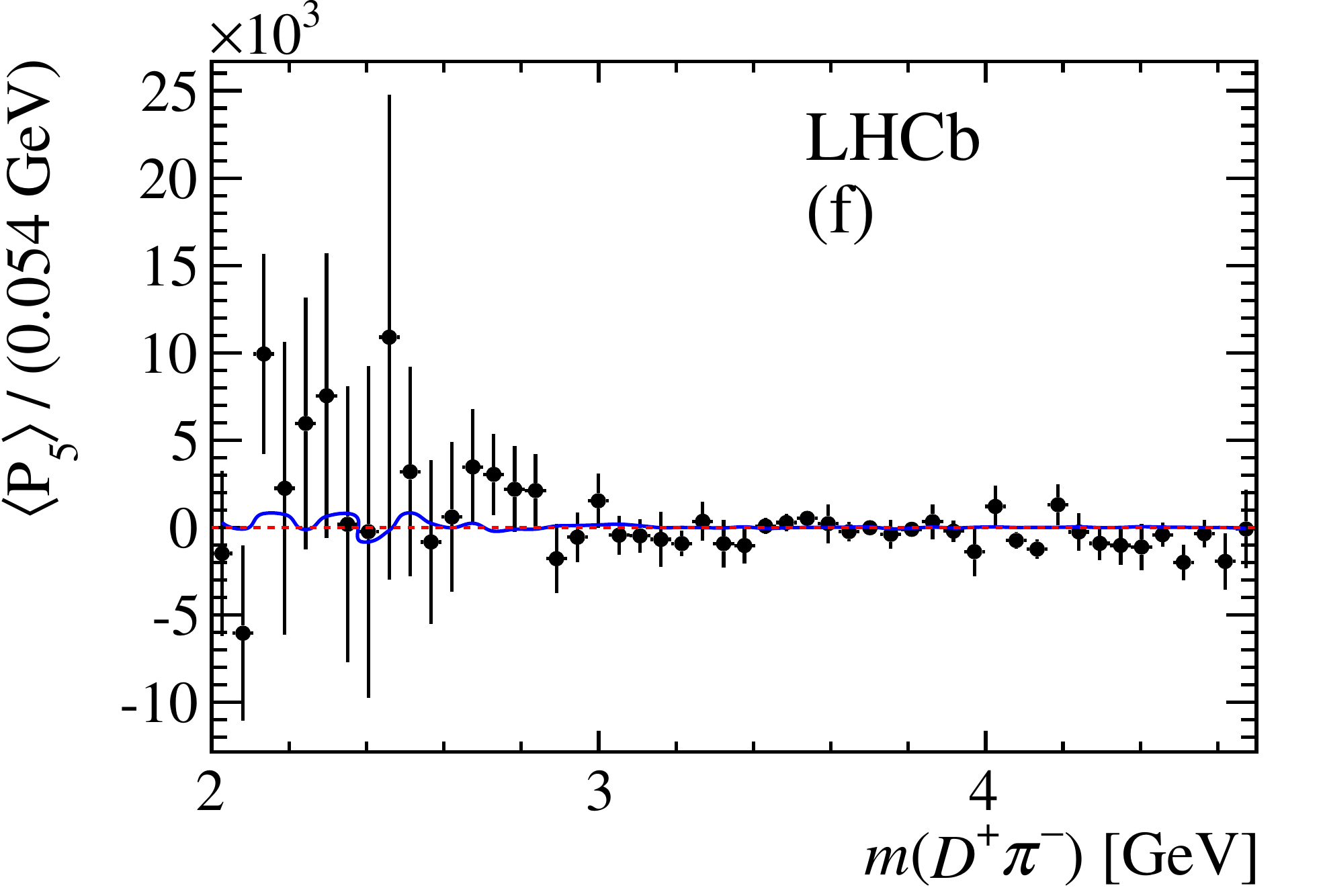}
 \includegraphics[scale=0.34]{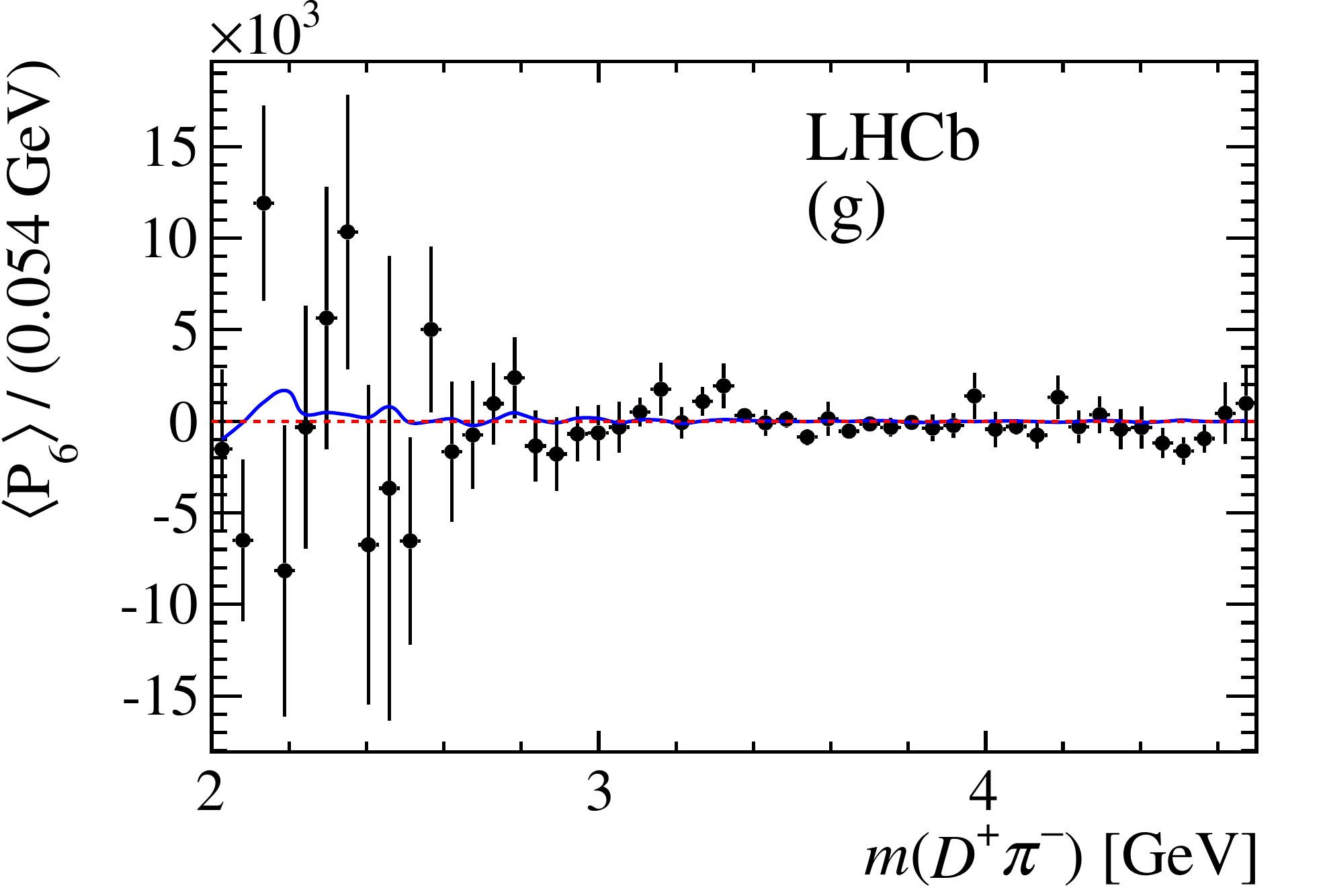}
\caption{\small
The first seven Legendre-polynomial weighted moments for background-subtracted and efficiency-corrected $\Bm \to \Dp\Km\pim$ data (black points) as a function of $m(\Dp\pim)$ in the range $2.0$--$3.0\gev$. 
Candidates from both TOS and TIS-only subsamples are included.
The blue line shows the result of the DP fit described in Sec.~\ref{sec:dalitz}.}
\label{fig:moments2}
\end{figure}

The analysis is performed by calculating moments from the Legendre polynomials $P_L$ of order up to $2 J_{\rm max}$, where $J_{\rm max}$ is the maximum spin of the resonances considered.
Each candidate is weighted according to its value of $P_L\left(\cos\theta(\Dp\pim)\right)$ with an efficiency correction applied, and background contributions subtracted.
The results for $J_{\rm max}=3$ are shown in Fig.~\ref{fig:moments2} for the $\Dp\pim$ invariant mass range $2.0$--$3.0\gev$.
The distributions of $\left\langle P_5 \right\rangle$ and $\left\langle P_6 \right\rangle$ are compatible with being flat, which implies that there are no significant spin~3 contributions. 
Considering only contributions up to spin~2, the following expressions are used to interpret Fig.~\ref{fig:moments2}
\begin{align}
 \label{eq:p0}
 \left\langle P_{0}\right\rangle \propto\, &
 \left|h_0\right|{}^2+\left|h_1\right|{}^2+\left|h_2\right|{}^2\,, \\
 \label{eq:p1}
 \left\langle P_1\right\rangle \propto\, &
 \frac{2}{\sqrt{3}} \left|h_0\right| \left|h_1\right| \cos \left(\delta _0-\delta
 _1\right)+\frac{4}{\sqrt{15}} \left|h_1\right| \left|h_2\right| \cos \left(\delta
 _1-\delta _2\right)\,, \\
 \label{eq:p2}
 \left\langle P_{2}\right\rangle \propto\, &
 \frac{2}{\sqrt{5}}\left|h_0\right| \left|h_2\right| \cos \left(\delta
 _0-\delta _2\right)+
 \frac{2}{5} \left|h_1\right|{}^2 + \frac{2}{7} \left|h_2\right|{}^2\,, \\
 \label{eq:p3}
 \left\langle P_{3}\right\rangle \propto\, &
 \frac{6}{7} \sqrt{\frac{3}{5}} \left|h_1\right| \left|h_2\right| \cos \left(\delta
 _1-\delta _2\right)\,, \\
 \label{eq:p4}
 \left\langle P_{4}\right\rangle \propto\, &
 \frac{2}{7}\left|h_2\right|{}^2\,, 
\end{align}
where S-, P- and D-wave contributions are denoted by amplitudes $h_j e^{i \delta_j}$ ($j=0,1,2$ respectively). 
The $D^*_2(2460)^0$ resonance is clearly seen in the $\left\langle P_4 \right\rangle$ distribution of Fig.~\ref{fig:moments2}(e).
The distribution of $\left\langle P_3 \right\rangle$ shows interference between spin~1 and 2 contributions, indicating the presence of a broad, possibly nonresonant, spin~1 contribution at low $m(\Dp\pim)$. 
The difference in shape between $\left\langle P_1 \right\rangle$ and $\left\langle P_3 \right\rangle$ shows interference between spin~1 and 0 indicating that a broad spin~0 component is similarly needed.

\section{Dalitz plot analysis formalism}
\label{sec:dalitz-generalities}

A Dalitz plot~\cite{Dalitz:1953cp} is a representation of the phase-space for a three-body decay in terms of two of the three possible two-body invariant mass squared combinations.
In $\Bm \to \Dp\Km\pim$ decays, resonances are expected in the $m^2(\Dp\pim)$ combination, therefore this and $m^2(\Dp\Km)$ are chosen to define the DP axes. 
For a fixed $\Bm$ mass, all other relevant kinematic quantities can be calculated from these two invariant mass squared combinations.

The complex decay amplitude is described using the isobar approach~\cite{Fleming:1964zz,Morgan:1968zza,Herndon:1973yn}, where the total amplitude is calculated as a coherent sum of amplitudes from resonant and nonresonant intermediate processes.
The total amplitude is then given by
\begin{equation}\label{eqn:amp}
  {\cal A}\left(m^2(\Dp\pim), m^2(\Dp\Km)\right) = \sum_{j=1}^{N} c_j F_j\left(m^2(\Dp\pim), m^2(\Dp\Km)\right) \,,
\end{equation}
where $c_j$ are complex coefficients giving the relative contribution of each intermediate process. 
The $F_j\left(m^2(\Dp\pim),m^2(\Dp\Km)\right)$ terms contain the resonance dynamics, which are 
composed of several terms and 
are normalised such that the integral of the squared magnitude over the DP is unity for each term.
For a $\Dp\pim$ resonance
\begin{equation}
  \label{eq:ResDynEqn}
  F\left(m^2(\Dp\pim), m^2(\Dp\Km)\right) = 
  R\left(m(\Dp\pim)\right) \times X(|\vec{p}\,|\,r_{\rm BW}) \times X(|\vec{q}\,|\,r_{\rm BW}) 
  \times T(\vec{p},\vec{q}\,) \, ,
\end{equation}
where the functions $R$, $X$ and $T$ are described below, and $\vec{p}$ and $\vec{q}$ are the bachelor particle momentum and the momentum of one of the resonance daughters, respectively, both evaluated in the $\Dp\pim$ rest frame.

The $X(z)$ terms, where $z=|\vec{q}\,|\,r_{\rm BW}$ or $|\vec{p}\,|\,r_{\rm BW}$, are Blatt--Weisskopf barrier factors~\cite{blatt-weisskopf} with barrier radius $r_{\rm BW}$, and are given by
\begin{equation}\begin{array}{rcl}
L = 0 \ : \ X(z) & = & 1\,, \\
L = 1 \ : \ X(z) & = & \sqrt{\frac{1 + z_0^2}{1 + z^2}}\,, \\
L = 2 \ : \ X(z) & = & \sqrt{\frac{z_0^4 + 3z_0^2 + 9}{z^4 + 3z^2 + 9}}\,,\\
L = 3 \ : \ X(z) & = & \sqrt{\frac{z_0^6 + 6z_0^4 + 45z_0^2 + 225}{z^6 + 6z^4 + 45z^2 + 225}}\,,
\end{array}\label{eq:BWFormFactors}\end{equation}
where $z_0$ is the value of $z$ when the invariant mass is equal to the pole mass of the resonance and $L$ is the spin of the resonance. For a $\Dp\pim$ resonance, since the \Bm meson has zero spin, $L$ is also the orbital angular momentum between the resonance and the kaon.
The barrier radius, $r_{\rm BW}$, is taken to be $4.0\gev^{-1}\approx 0.8\fm$~\cite{Aubert:2005ce,LHCb-PAPER-2014-036} for all resonances.

The terms $T(\vec{p},\vec{q})$ describe the angular probability distribution and are
given in the Zemach tensor formalism~\cite{Zemach:1963bc,Zemach:1968zz} by
\begin{equation}\begin{array}{rcl}
L = 0 \ : \ T(\vec{p},\vec{q}) & = & 1\,,\\
L = 1 \ : \ T(\vec{p},\vec{q}) & = & -\,2\,\vec{p}\cdot\vec{q}\,,\\
L = 2 \ : \ T(\vec{p},\vec{q}) & = & \frac{4}{3} \left[3(\vec{p}\cdot\vec{q}\,)^2 - (|\vec{p}\,||\vec{q}\,|)^2\right]\,,\\
L = 3 \ : \ T(\vec{p},\vec{q}) & = & -\,\frac{24}{15} \left[5(\vec{p}\cdot\vec{q}\,)^3 - 3(\vec{p}\cdot\vec{q}\,)(|\vec{p}\,||\vec{q}\,|)^2\right]\,,
\end{array}\label{eq:ZTFactors}\end{equation}
which are proportional to the Legendre polynomials, $P_L(x)$, where $x$ is the cosine of the angle between $\vec{p}$ and $\vec{q}$ (referred to as the helicity angle). 

The function $R\left(m(\Dp\pim)\right)$ of Eq.~(\ref{eq:ResDynEqn}) is the mass lineshape.
The resonant contributions considered in the DP model are described by the relativistic Breit--Wigner (RBW) function
\begin{equation}
\label{eq:RelBWEqn}
R(m) = \frac{1}{(m_0^2 - m^2) - i\, m_0 \Gamma(m)} \,,
\end{equation}
where the mass-dependent decay width is
\begin{equation}
\label{eq:GammaEqn}
\Gamma(m) = \Gamma_0 \left(\frac{q}{q_0}\right)^{2L+1}
\left(\frac{m_0}{m}\right) X^2(q\,r_{\rm BW}) \,,
\end{equation}
where $q_0$ is the value of $q = |\vec{q}\,|$ for $m = m_0$.
Virtual contributions, from resonances
with pole masses outside the kinematically accessible region of the phase space, can also be modelled by this shape with one modification:
the pole mass $m_0$ is replaced with $m_0^{\rm{eff}}$, a mass in the kinematically allowed region, in the calculation of the parameter $q_{0}$. 
This effective mass is defined by the {\it ad hoc} formula~\cite{LHCb-PAPER-2014-036}
\begin{equation}\label{eqn:effmass}
  m_0^{\rm{eff}}(m_0) = m^{\rm{min}} + (m^{\rm{max}} - m^{\rm{min}}) \left( 1 + \tanh\left( \frac{m_0 - \frac{m^{\rm{min}}+m^{\rm{max}}}{2}}{m^{\rm{max}}-m^{\rm{min}}} \right) \right)\, ,
\end{equation}
where $m^{\rm{max}}$ and $m^{\rm{min}}$ are the upper and lower limits of the kinematically allowed range, respectively.
For virtual contributions, only the tail of the RBW function enters the Dalitz plot.  

Given the large available phase-space in the \B decay, it is possible to have nonresonant amplitudes (\ie\ contributions that are not from any known resonance, including virtual states) that vary across the Dalitz plot.
A model that has been found to describe well nonresonant contributions in several \B decay DP analyses is an exponential form factor (EFF)~\cite{Garmash:2004wa}, 
\begin{equation}
  R(m) = e^{-\alpha m^2} \, ,
  \label{eq:nonres}
\end{equation}
where $m$ is a two-body (in this case $D\pi$) invariant mass and $\alpha$ is a shape parameter that must be determined from the data.

Neglecting reconstruction effects, the DP probability density function would be
\begin{equation}
\label{eq:SigDPLike}
{\cal{P}}_{\rm phys}\left(m^2(\Dp\pim), m^2(\Dp\Km)\right) =
\frac
{|{\cal A}\left(m^2(\Dp\pim), m^2(\Dp\Km)\right)|^2}
{\int\!\!\int_{\rm DP}~{|{\cal A}|^2}~dm^2(\Dp\pim)\,dm^2(\Dp\Km)} \, ,
\end{equation}
where the dependence of ${\cal A}$ on the DP position has been suppressed in the denominator for brevity.
The complex coefficients, given by $c_j$ in Eq.~(\ref{eqn:amp}), are the primary results of most Dalitz plot analyses.
However, these depend on the choice of normalisation, phase convention and amplitude formalism in each analysis.
Fit fractions and interference fit fractions are also reported as these provide a convention-independent method to allow meaningful comparisons of results. 
The fit fraction is defined as the integral of the amplitude for a single component squared divided by that of the coherent matrix element squared for the complete Dalitz plot, 
\begin{equation}
{\it FF}_j =
\frac
{\int\!\!\int_{\rm DP}\left|c_j F_j\left(m^2(\Dp\pim), m^2(\Dp\Km)\right)\right|^2~dm^2(\Dp\pim)\,dm^2(\Dp\Km)}
{\int\!\!\int_{\rm DP}\left|{\cal A}\right|^2~dm^2(\Dp\pim)\,dm^2(\Dp\Km)} \, .
\label{eq:fitfraction}
\end{equation}
The fit fractions do not necessarily sum to unity due to the potential presence of net constructive or destructive interference, described by interference fit fractions defined for $i<j$ only by 
\begin{equation}
  {\it FF}_{ij} =
  \frac
  {\int\!\!\int_{\rm DP} 2 \, \Real\left[c_ic_j^* F_iF_j^*\right]~dm^2(\Dp\pim)\,dm^2(\Dp\Km)}
  {\int\!\!\int_{\rm DP}\left|{\cal A}\right|^2~dm^2(\Dp\pim)\,dm^2(\Dp\Km)} \, ,
  \label{eq:intfitfraction}
\end{equation}
where the dependence of $F_i^{(*)}$ and ${\cal A}$ on the DP position has been omitted.

\section{Dalitz plot fit}
\label{sec:dalitz}

The {\sc Laura++}~\cite{Laura++} package is used to perform the Dalitz plot fit, with the two trigger subsamples fitted simultaneously using the {\it J}{\sc fit} method~\cite{Ben-Haim:2014afa}. 
The two subsamples have separate signal and background yields, efficiency maps and background SDP distributions, but all parameters of the signal model are common.
The likelihood function that is used is
\begin{equation}
 {\cal L} =
 \prod_i^{N_c}
 \Bigg[
 \sum_k N_k {\cal P}_k\left(m^2_i(\Dp\pim),m^2_i(\Dp\Km)\right)
 \Bigg] \,,
\end{equation}
where the index $i$ runs over $N_c$ candidates, while $k$ distinguishes the signal and background components with $N_k$ the yield in each component.
The probability density function for signal events, ${\cal P}_{\rm sig}$, is given by Eq.~(\ref{eq:SigDPLike}) where 
the $|{\cal A}\left(m^2(\Dp\pim), m^2(\Dp\Km)\right)|^2$ terms are multiplied by the efficiency function described in 
Sec.~\ref{sec:efficiency}.
The mass resolution is approximately $2.4\mev$, which is much lower than the width of the narrowest contribution to the Dalitz plot ($\sim 50 \mev$); therefore, this has negligible effect on the likelihood and is not considered further.

\begin{figure}[!tb]
\centering
\includegraphics[scale=0.396]{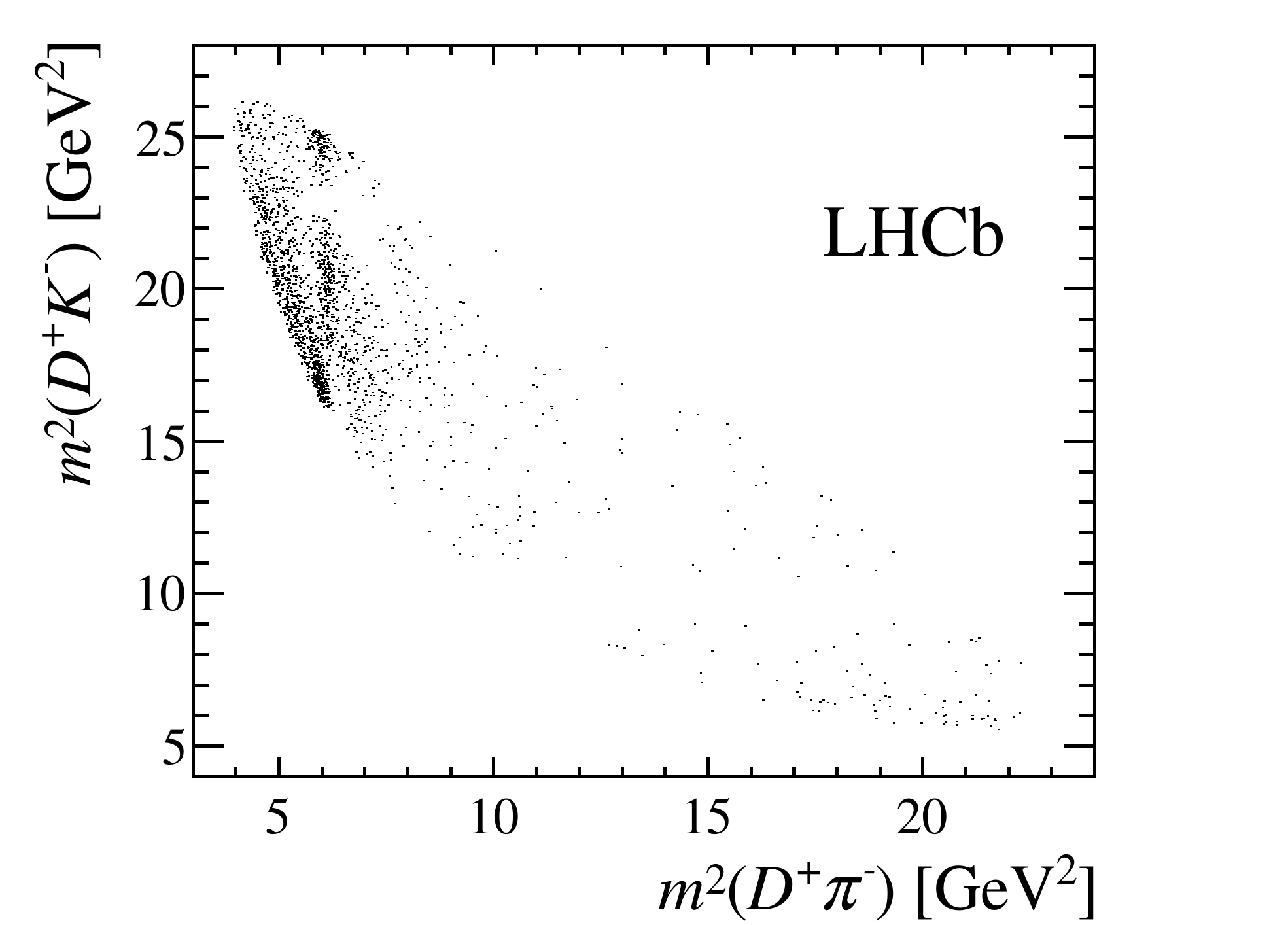}
\includegraphics[scale=0.396]{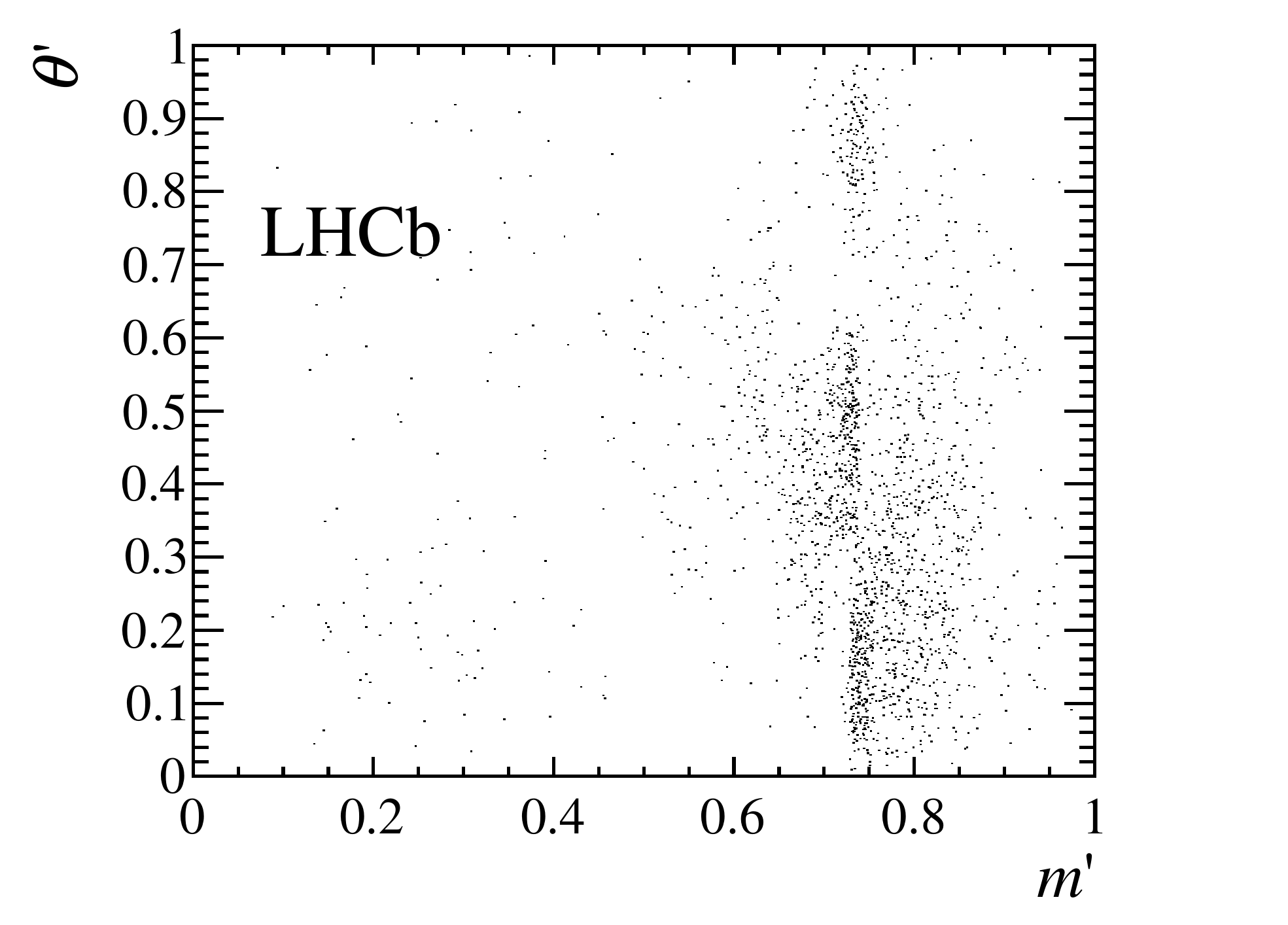}
\caption{\small Distribution of $\Bm\to\Dp\Km\pim$ candidates in the signal
  region over (left) the DP and (right) the SDP.
Candidates from both TOS and TIS-only subsamples are included.}
\label{fig:canddp}
\end{figure}

The signal and background yields that enter the Dalitz plot fit are taken from the mass fit described in Sec.~\ref{sec:mass-fit}.
Only candidates in the signal region, defined as $\pm 2.5\sigma$ around the \B signal peak, where $\sigma$ is the width of the peak, are used in the Dalitz plot fit. 
Within this region, in the TOS subsample the result of the \B candidate invariant mass fit corresponds to yields of $1060 \pm 35$, $37 \pm 6$, $26 \pm 8$ and $16 \pm 4$ in the signal, combinatorial background, $D^{(*)+}\pim\pim$ and $\Dsp\Km\pim$  components, respectively. 
The equivalent yields in the TIS-only subsample are $849 \pm 30$, $39 \pm 6$, $5 \pm 5$ and $9 \pm 3$ candidates. 
The contribution from $\Dstarp\Km\pim$ decays is negligible in the signal window.
The distributions of the candidates in the signal region over the DP and SDP are shown in Fig.~\ref{fig:canddp}.

\begin{figure}
\centering
\includegraphics[scale=0.37]{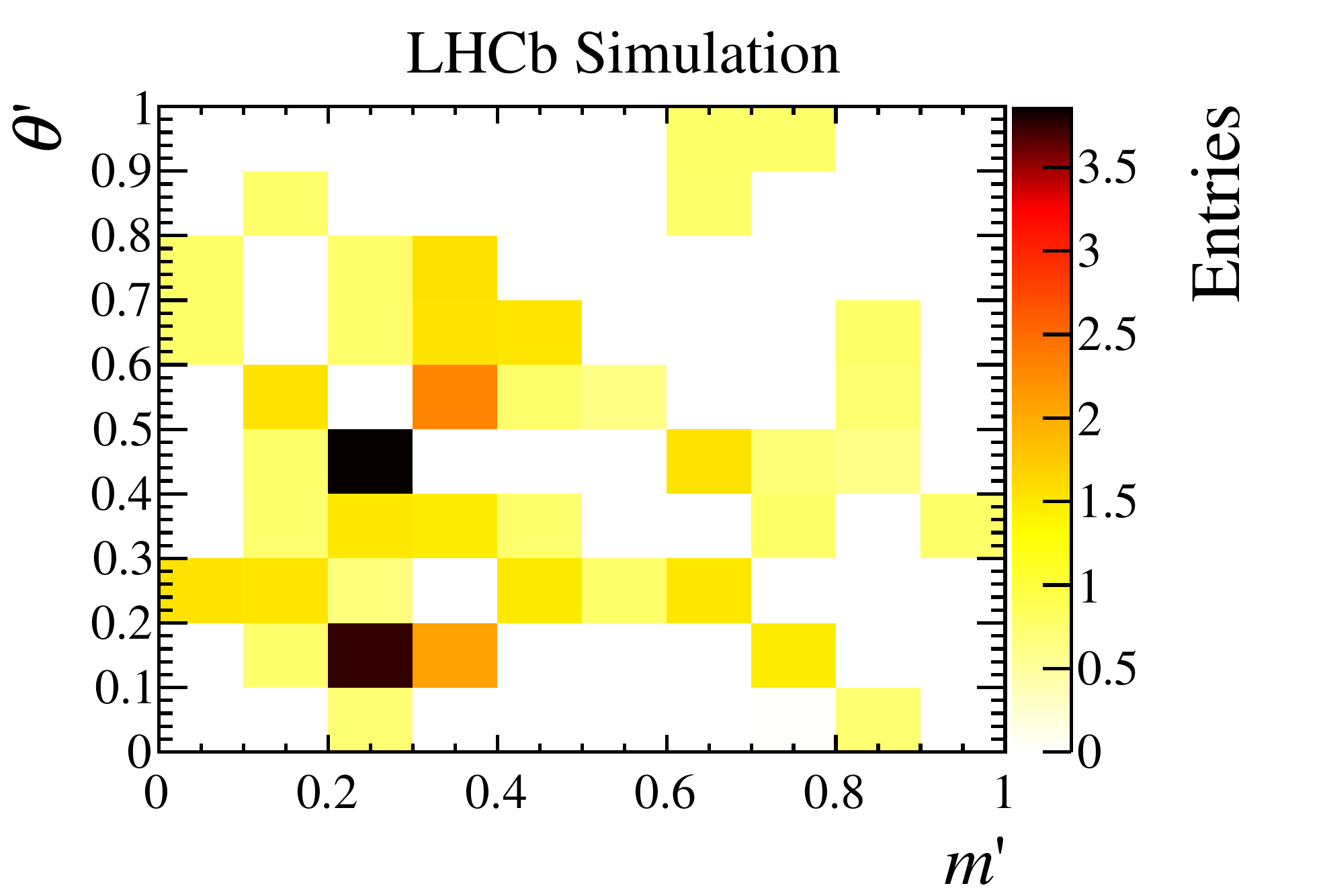}
\includegraphics[scale=0.37]{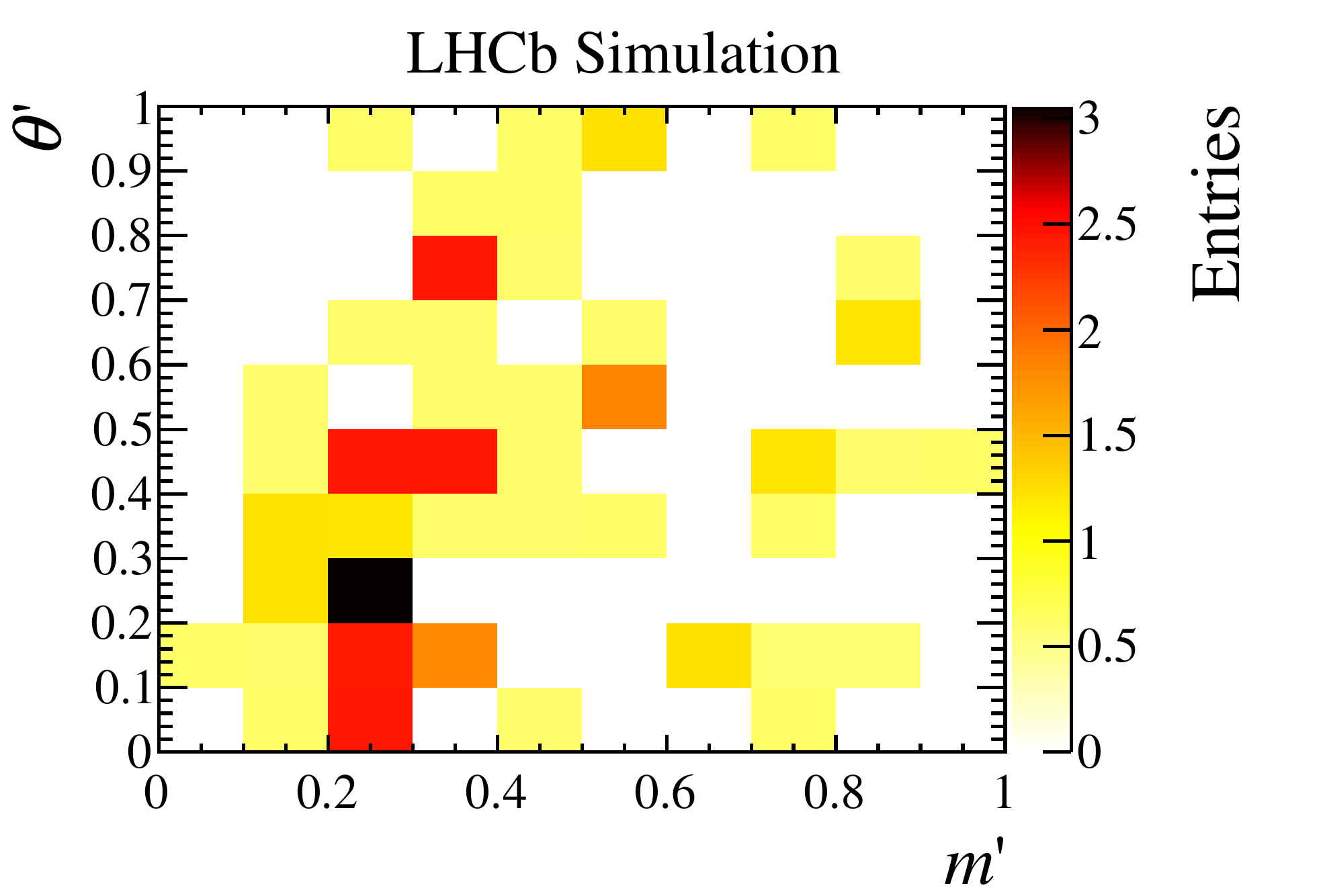}
\includegraphics[scale=0.37]{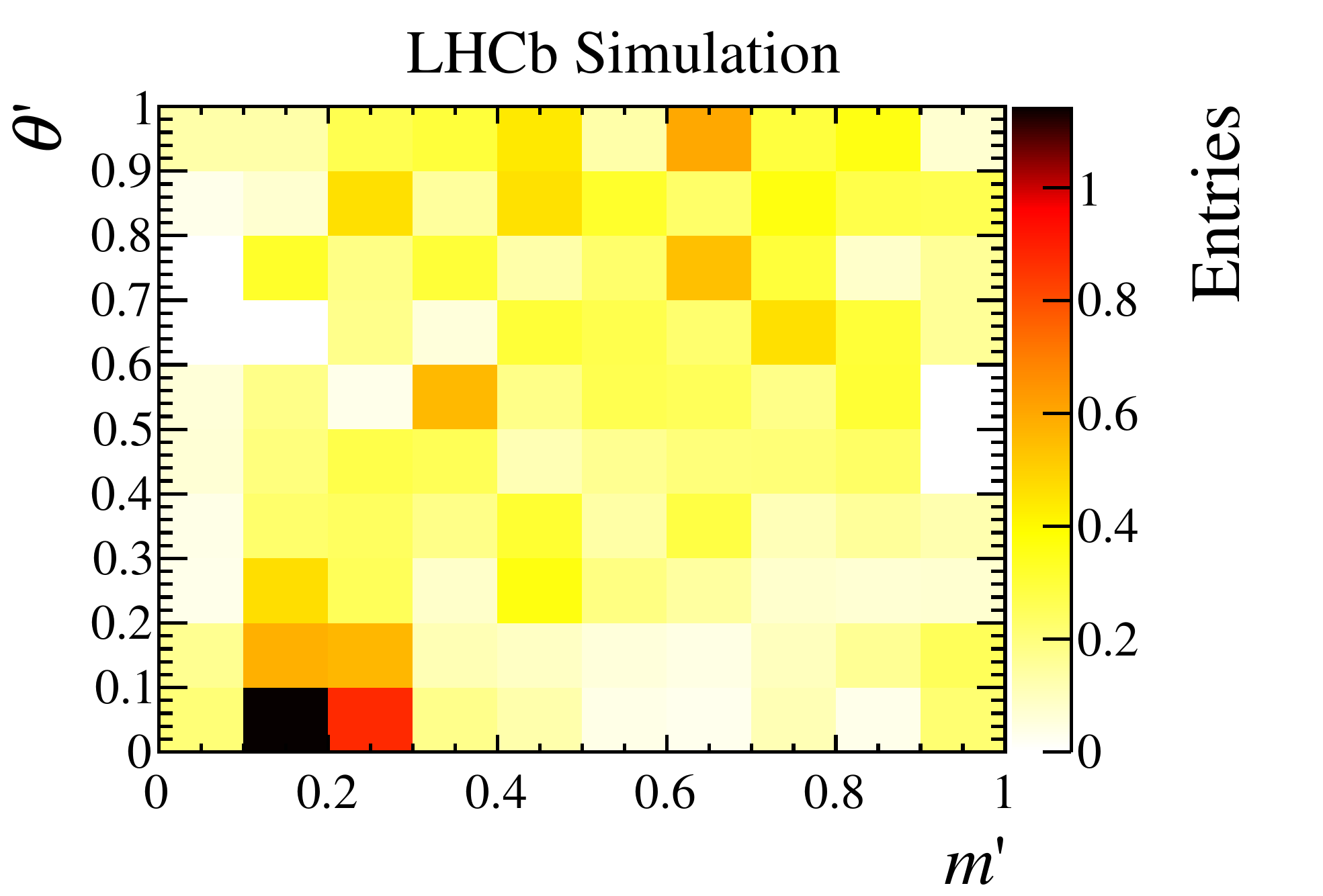}
\includegraphics[scale=0.37]{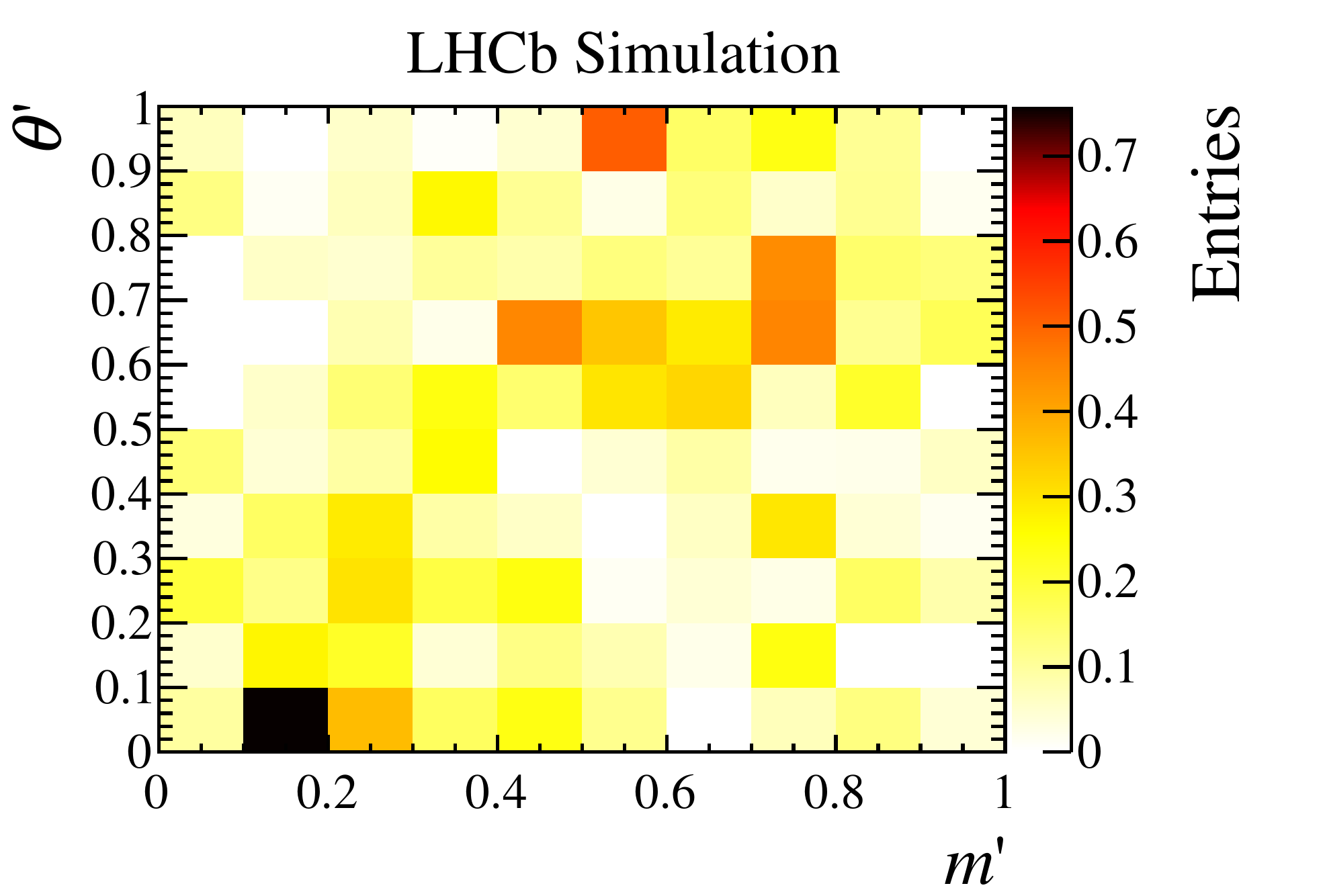}
\includegraphics[scale=0.37]{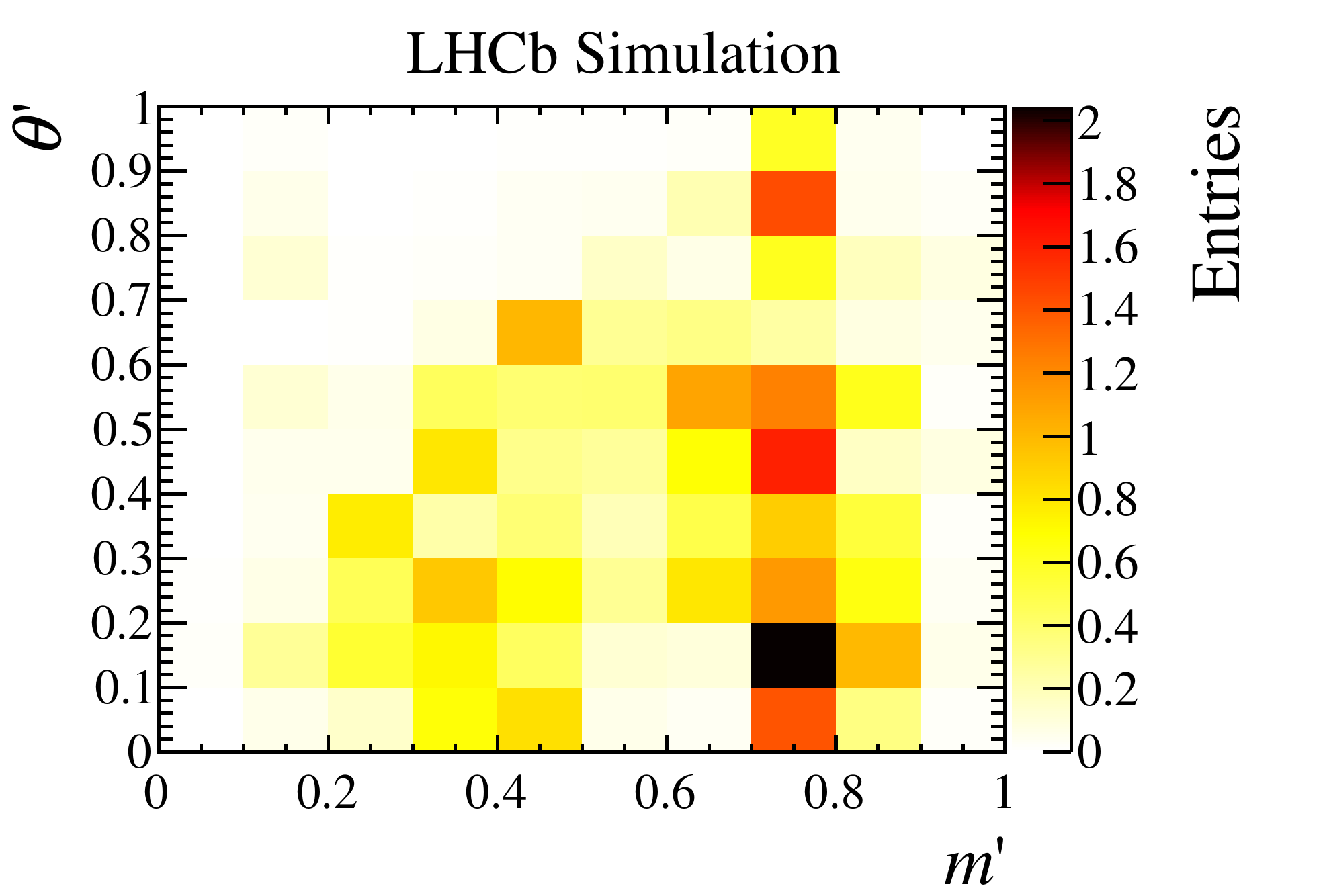}
\includegraphics[scale=0.37]{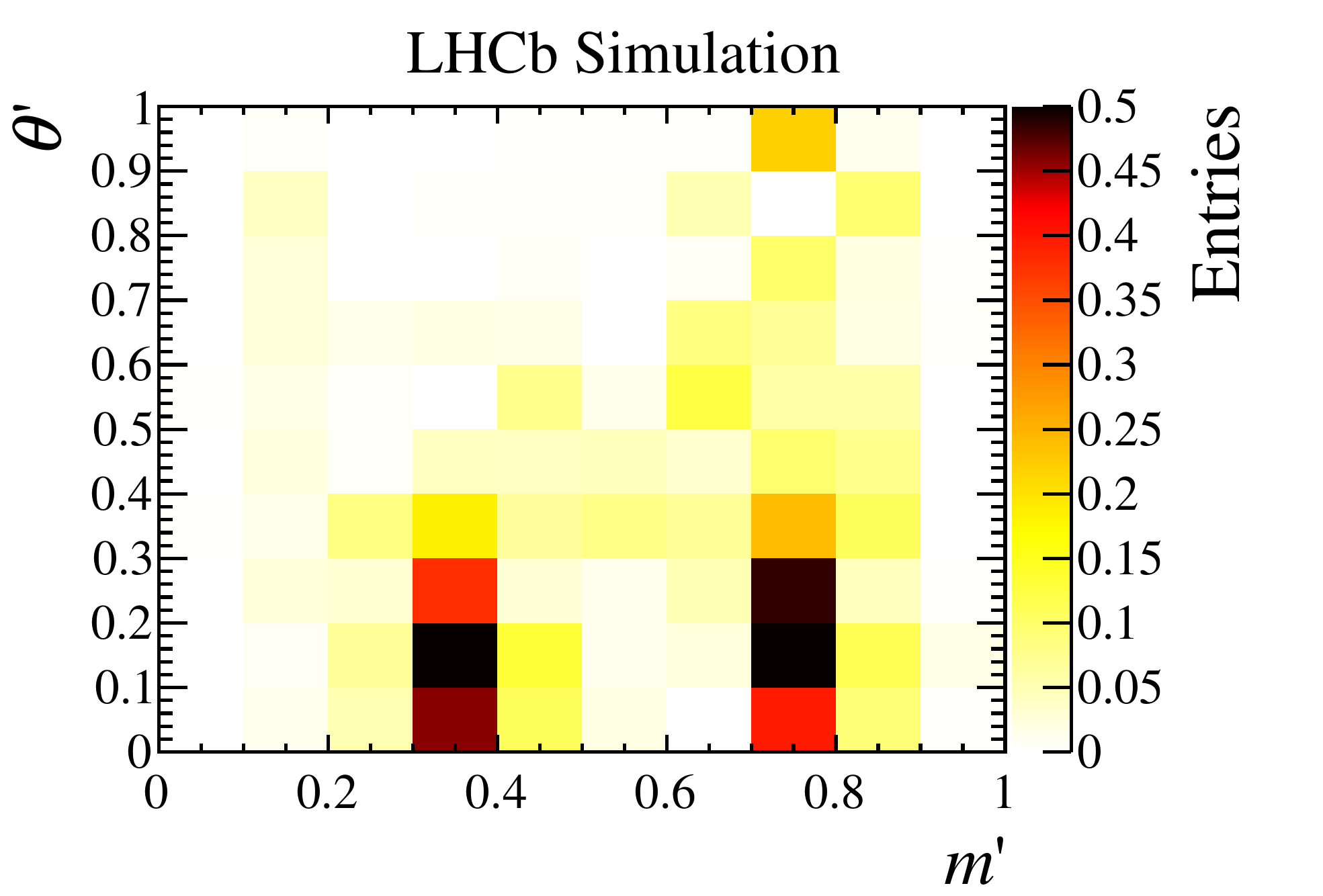}
\caption{\small Square Dalitz plot distributions used in the Dalitz plot fit for (top) combinatorial background, (middle) $\Bm \to D^{(*)+}\pim\pim$ decays and (bottom) $\Bm \to \Dsp\Km\pim$ decays. 
  Candidates the TOS (TIS-only) subsamples are shown in the left (right) column.}
\label{fig:bkgs}
\end{figure}

The SDP distributions of the $D^{(*)+}\pim\pim$ and $\Dsp\Km\pim$ background sources are obtained from simulated samples using the same procedures as described for their invariant mass distributions in Sec.~\ref{sec:mass-fit}. The distribution of combinatorial background events is modelled by considering $\Dp\Km\pim$ candidates in the sideband high-mass range $5500$--$5800\mev$, with contributions from $D^{(*)+}\pim\pim$ in this region subtracted.
The dependence of the SDP distribution on \B candidate mass was investigated and found to be negligible.
The SDP distributions of these backgrounds are shown in Fig.~\ref{fig:bkgs}.
These histograms are used to model the background contributions in the Dalitz plot fit.

Using the results of the moments analysis of Sec.~\ref{sec:moments} as a guide, the nominal Dalitz plot fit model for $\Bm\to\Dp\Km\pim$ decays is determined by considering several resonant, nonresonant and virtual amplitudes. 
Those that do not contribute significantly and that do not aid the stability of the fit are removed. 
Only natural spin-parity intermediate states are considered, as unnatural spin-parity states do not decay to two pseudoscalars.
The resulting signal model, referred to below as the nominal DP model, consists of the seven amplitudes shown in Table~\ref{tab:resonances}: three resonances, two virtual resonances and two nonresonant terms. 
Parts of the model are known to be approximations.
In particular both S- and P-waves in the $D\pi$ system are modelled with overlapping broad structures.
The nominal model gives a better description of the data than any of the alternative models considered; alternative models are used to assign systematic uncertainties as discussed in Sec.~\ref{sec:systematics}. 

\begin{table}[!tb]
\centering
\caption{\small
  Signal contributions to the fit model, where parameters and uncertainties are taken from Ref.~\cite{PDG2014}. 
  States labelled with subscript $v$ are virtual contributions.
}
\label{tab:resonances}
\begin{tabular}{lcccc}
\hline
Resonance & Spin & DP axis & Model & Parameters \\
\hline \\ [-2.5ex] 
$D^{*}_{0}(2400)^{0}$ &0& $m^{2}(D\pi)$ & RBW & $m = 2318 \pm 29\mev$, $\Gamma = 267 \pm 40\mev$ \\
$D^{*}_{2}(2460)^{0}$ &2& $m^{2}(D\pi)$ & RBW & \multirow{2}{*}{Determined from data (see Table~\ref{tab:masswidth})} \\ 
$D^{*}_{J}(2760)^{0}$ &1& $m^{2}(D\pi)$ & RBW \\
\hline
Nonresonant &0& $m^{2}(D\pi)$ & EFF & \multirow{2}{*}{Determined from data (see text)} \\
Nonresonant &1& $m^{2}(D\pi)$ & EFF \\
\hline \\ [-2.5ex] 
$D^{*}_{v}(2007)^{0}$ &1& $m^{2}(D\pi)$ & RBW & $m = 2006.98 \pm 0.15 \mev$, $\Gamma = 2.1 \mev$ \\
$B^{*0}_{v}$ &1& $m^{2}(DK)$ & RBW & $m = 5325.2 \pm 0.4 \mev$, $\Gamma = 0.0 \mev$  \\
\hline
\end{tabular}
\end{table}

The free parameters in the fit are the $c_j$ terms introduced in Eq.~(\ref{eqn:amp}), with the real and imaginary parts of these complex coefficients determined for each amplitude in the fit model. 
The $D^*_2(2460)^0$ component, as the reference amplitude, is the exception with real and imaginary parts fixed to 1 and 0, respectively. 
Fit fractions and interference fit fractions are derived from these free parameters, as are the magnitudes and phases of the complex coefficients. 
Statistical uncertainties for the derived parameters are calculated using large samples of simulated pseudoexperiments to ensure that non-trivial correlations are accounted for.
Several other parameters are also determined from the fit as described below.

In Dalitz plot fits it is common for the minimisation procedure to find local minima of the likelihood function.
To find the global minimum, the fit is performed many times using randomised starting values for the complex coefficients.
In addition to the global minimum of the likelihood, corresponding to the results reported below, several additional minima are found.
Two of these have negative log-likelihood (NLL) values close to that of global minimum.
The main differences between secondary minima and the global minimum are the interference patterns in the $D\pi$ S- and P-waves, as shown in App.~\ref{app:minima}.

The shape parameters, defined in Eq.~(\ref{eq:nonres}), for the nonresonant components are determined from the fit to data to be $0.36 \pm 0.03 \gev^{-2}$ and $0.36 \pm 0.04 \gev^{-2}$ for the S-wave and P-wave, respectively, where the uncertainties are statistical only. 
The mass and width of the $D^{*}_{2}(2460)^{0}$ resonance are determined from the fit to improve the fit quality.
Since the mass and width of the $D^{*}_{J}(2760)^{0}$ state have not been precisely determined by previous experiments, these parameters are also allowed to vary in the fit.
The masses and widths of the $D^{*}_{2}(2460)^{0}$ and $D^{*}_{J}(2760)^{0}$ are reported in Table~\ref{tab:masswidth}.

The spin of the $D^{*}_{J}(2760)^{0}$ state has not been determined previously.
Fits are performed with all values up to 3, and spin~1 is found to be preferred with changes relative to the spin~0, 2 and 3 hypotheses of $2\Delta{\rm NLL} = 37.3, 49.5$ and $48.2$ units, respectively. 
For comparison, the value of $2\Delta{\rm NLL}$ obtained from a fit with the $D^{*}_{1}(2760)^{0}$ state excluded is 75.0 units.
The alternative models discussed in Sec.~\ref{sec:systematics} give very similar values and therefore do not affect the conclusion that the $D^{*}_{J}(2760)^{0}$ state has spin~1.

\begin{table}[!tb]
\centering
\caption{\small Masses and widths determined in the fit to data, with statistical uncertainties only.}
\label{tab:masswidth}
\begin{tabular}{lcc}
\hline
Resonance & Mass $(\mevnsp)$ & Width $(\mevnsp)$ \\
\hline \\ [-2.5ex] 
$D^{*}_{2}(2460)^{0}$ & $ 2464.0 \pm 1.4\phantom{1} $ & $ 43.8 \pm 2.9 $\\ 
$D^{*}_{J}(2760)^{0}$ & $ 2781 \pm 18 $ & $ 177 \pm 32 $\\ 
\hline
\end{tabular}
\end{table}

The values of the complex coefficients and fit fractions returned by the fit are shown in Table~\ref{tab:ffstat}.
Results for the interference fit fractions are given in App.~\ref{app:iffstat}. The total fit fraction exceeds unity 
mostly due to interference between the $D^{*}_{0}(2400)^{0}$ and S-wave nonresonant contributions. 

\begin{table}[!tb]
\centering
\caption{\small Complex coefficients and fit fractions determined from the
  Dalitz plot fit. Uncertainties are statistical only.
}
\label{tab:ffstat}
\resizebox{\textwidth}{!}{
\begin{tabular}{lccccc}
\hline
& & \multicolumn{4}{c}{Isobar model coefficients} \\
Resonance & Fit fraction (\%) & Real part & Imaginary part & Magnitude & Phase \\
\hline \\ [-2.5ex] 
$D^{*}_{0}(2400)^{0}$ & $\phantom{1}8.3 \pm 2.6$ &  $-0.04 \pm 0.07$ &           $-0.51 \pm 0.07$ &           $0.51 \pm 0.09$ & $-1.65 \pm 0.16$\\
$D^{*}_{2}(2460)^{0}$ & $31.8 \pm 1.5$ & $\phantom{-}1.00$ & $\phantom{-}0.00$ & $1.00$ & $\phantom{-}0.00$\\
$D^{*}_{1}(2760)^{0}$ & $\phantom{1}4.9 \pm 1.2$ &  $-0.32 \pm 0.06$ &           $-0.23 \pm 0.07$ &           $0.39 \pm 0.05$ & $-2.53 \pm 0.24$\\
\hline
S-wave nonresonant &  $38.0 \pm 7.4$ & $\phantom{-}0.93 \pm 0.09$ & $-0.58 \pm 0.08$ &           $1.09 \pm 0.09$ & $-0.56 \pm 0.09$\\
P-wave nonresonant &  $23.8 \pm 5.6$ & $-0.43 \pm 0.09$ &           $\phantom{-}0.75 \pm 0.09$ & $0.87 \pm 0.09$ & $\phantom{-}2.09 \pm 0.15$\\
\hline \\ [-2.5ex] 
$D^{*}_{v}(2007)^{0}$ & $\phantom{1}7.6 \pm 2.3$ &  $\phantom{-}0.16 \pm 0.08$ & $\phantom{-}0.46 \pm 0.09$ & $0.49 \pm 0.07$ & $\phantom{-}1.24 \pm 0.17$\\
$B^{*}_{v}$ &           $\phantom{1}3.6 \pm 1.9$ &  $-0.07 \pm 0.08$ &           $\phantom{-}0.33 \pm 0.07$ & $0.34 \pm 0.06$ & $\phantom{-}1.78 \pm 0.23$\\
\hline
Total fit fraction & $118.1$ \\
\hline
\end{tabular}
}
\end{table}

\begin{figure}[!tb]
\centering
\includegraphics[scale=0.5]{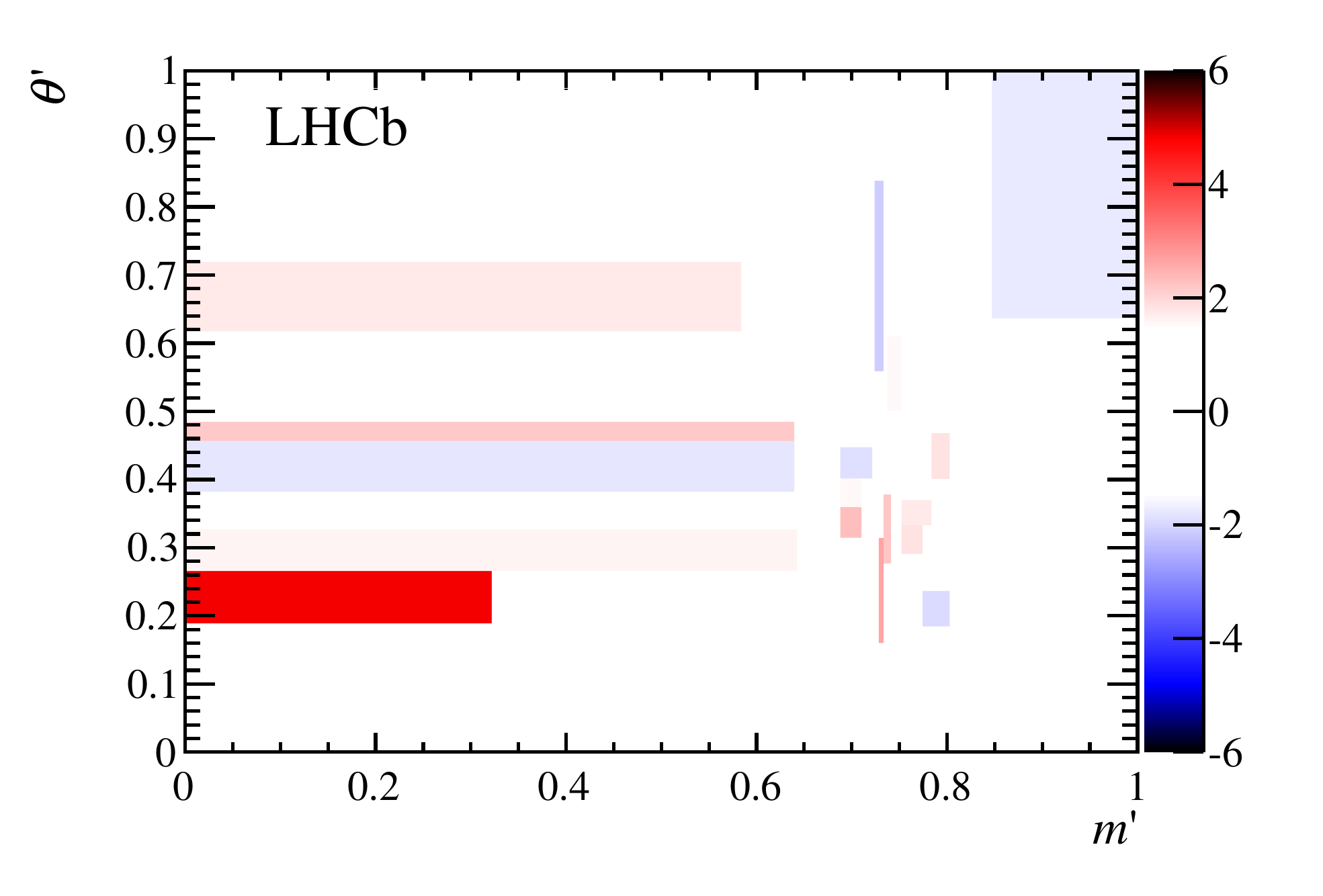}
\caption{\small Differences between the data SDP distribution and the fit model across the SDP, in terms of the per-bin pull.}
\label{fig:sdppull}
\end{figure}

The consistency of the fit model and the data is evaluated in several ways.
Numerous one-dimensional projections (including several shown below and those shown in Sec.~\ref{sec:moments}) show good agreement.
A two-dimensional $\chi^2$ value is determined by comparing the data and the fit model in $100$ equally populated bins across the SDP. 
The pull, \ie\ the difference between the data and fit model divided by the uncertainty, is shown with this SDP binning in Fig.~\ref{fig:sdppull}.
The $\chi^2$ value obtained is found to be within the bulk of the distribution expected from simulated pseudoexperiments.
Other unbinned fit quality tests~\cite{Williams:2010vh} also show acceptable agreement between the data and the fit model.

\begin{figure}[!tb]
\centering
\includegraphics[scale=0.395]{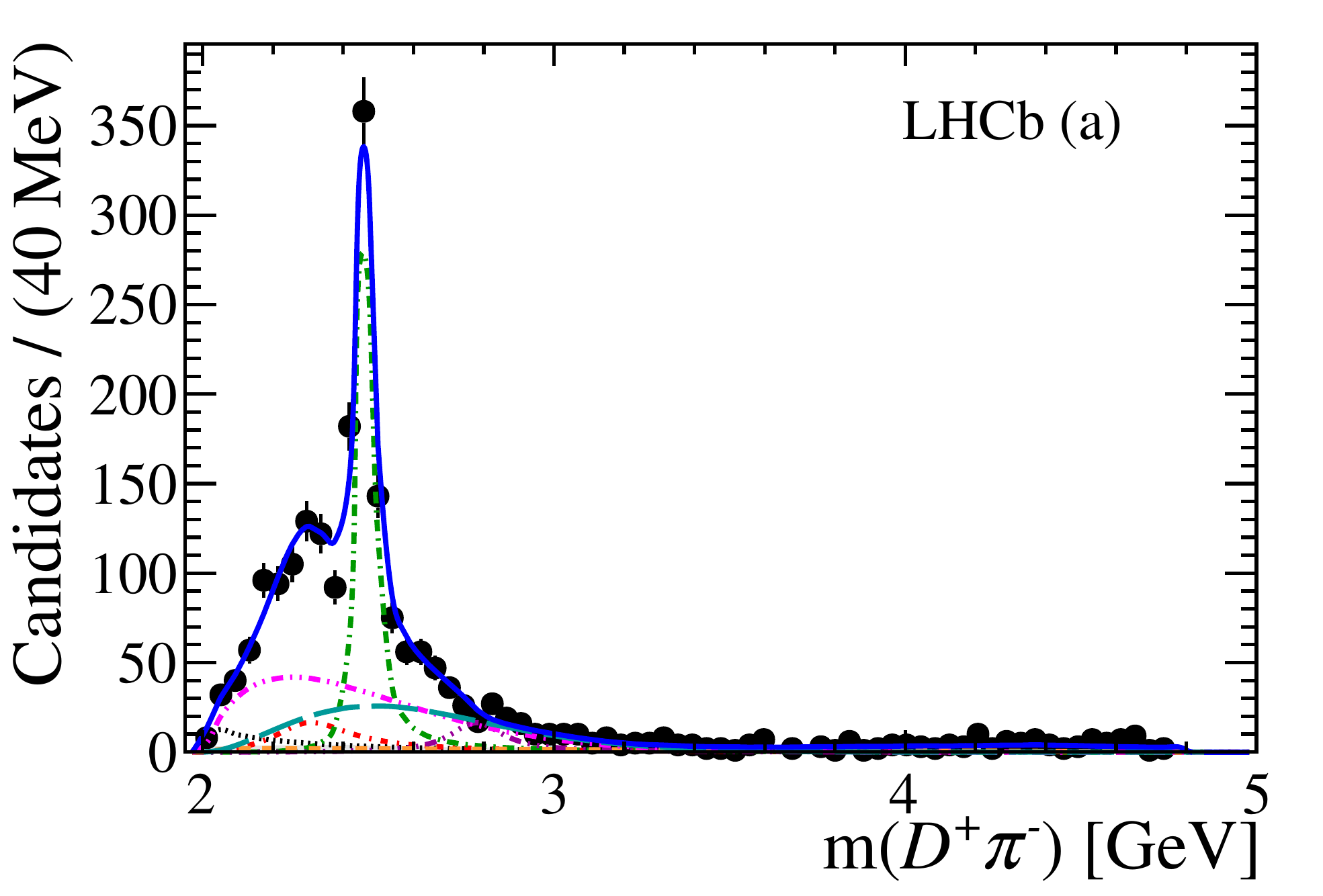}
\includegraphics[scale=0.395]{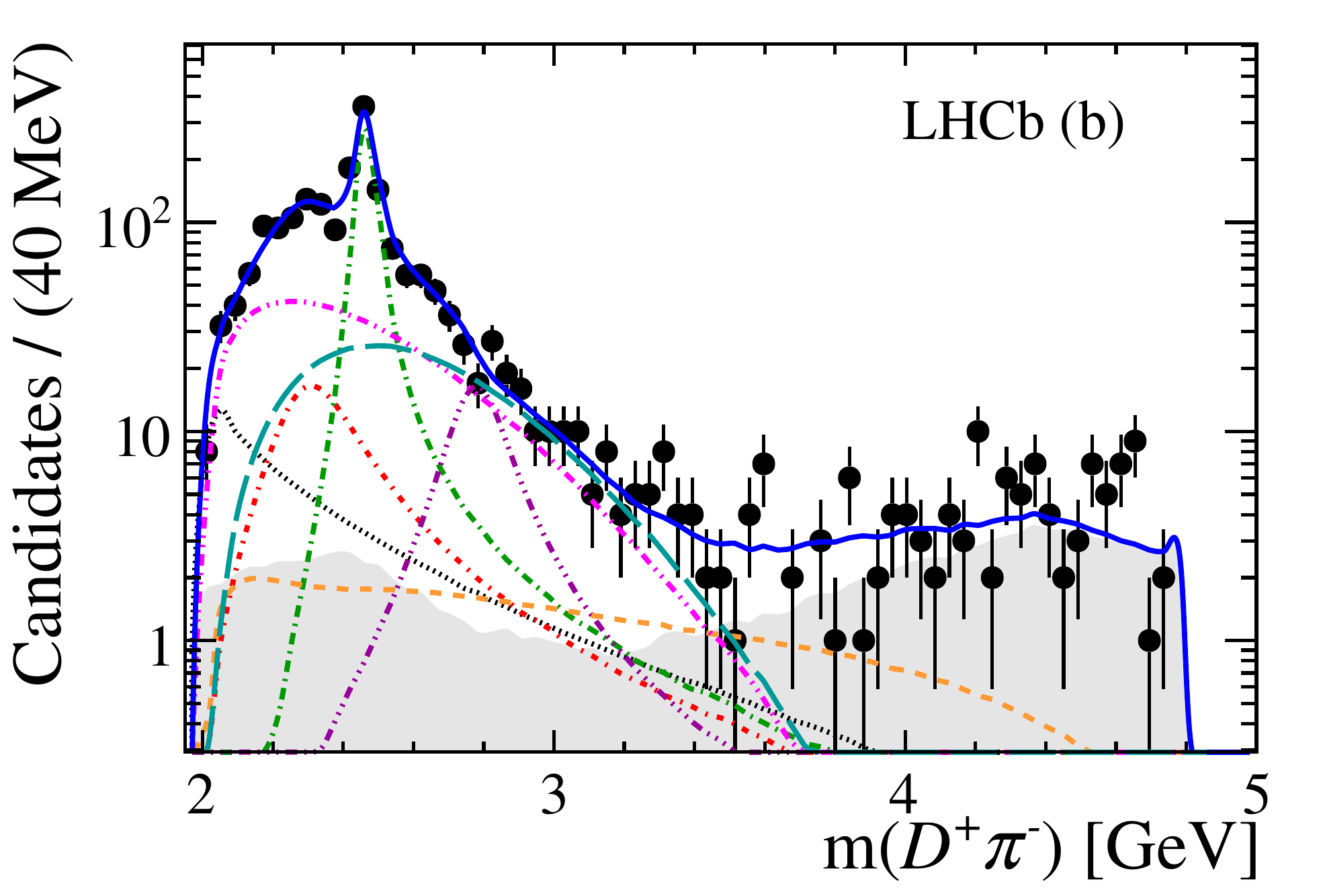}
\includegraphics[scale=0.395]{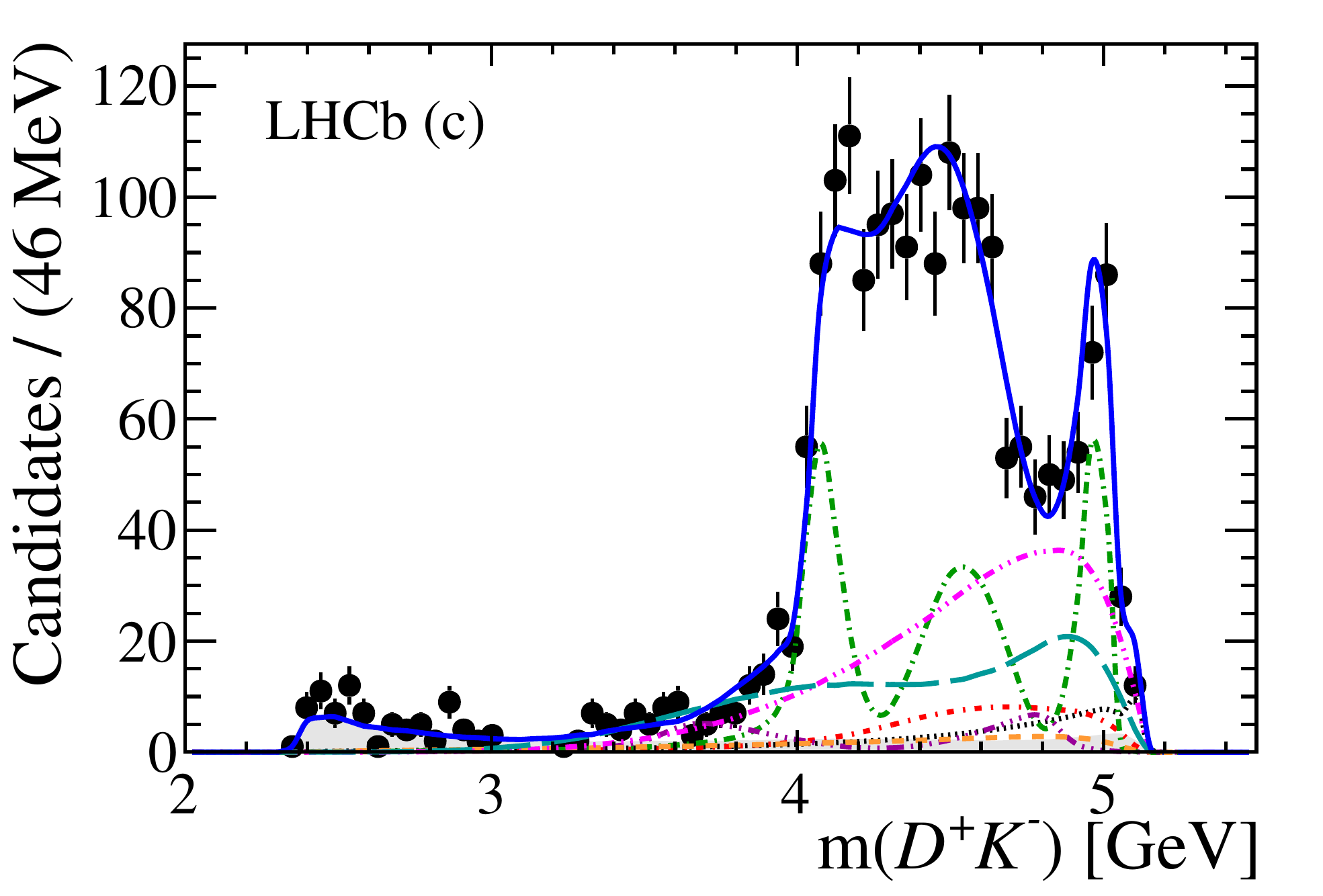}
\includegraphics[scale=0.395]{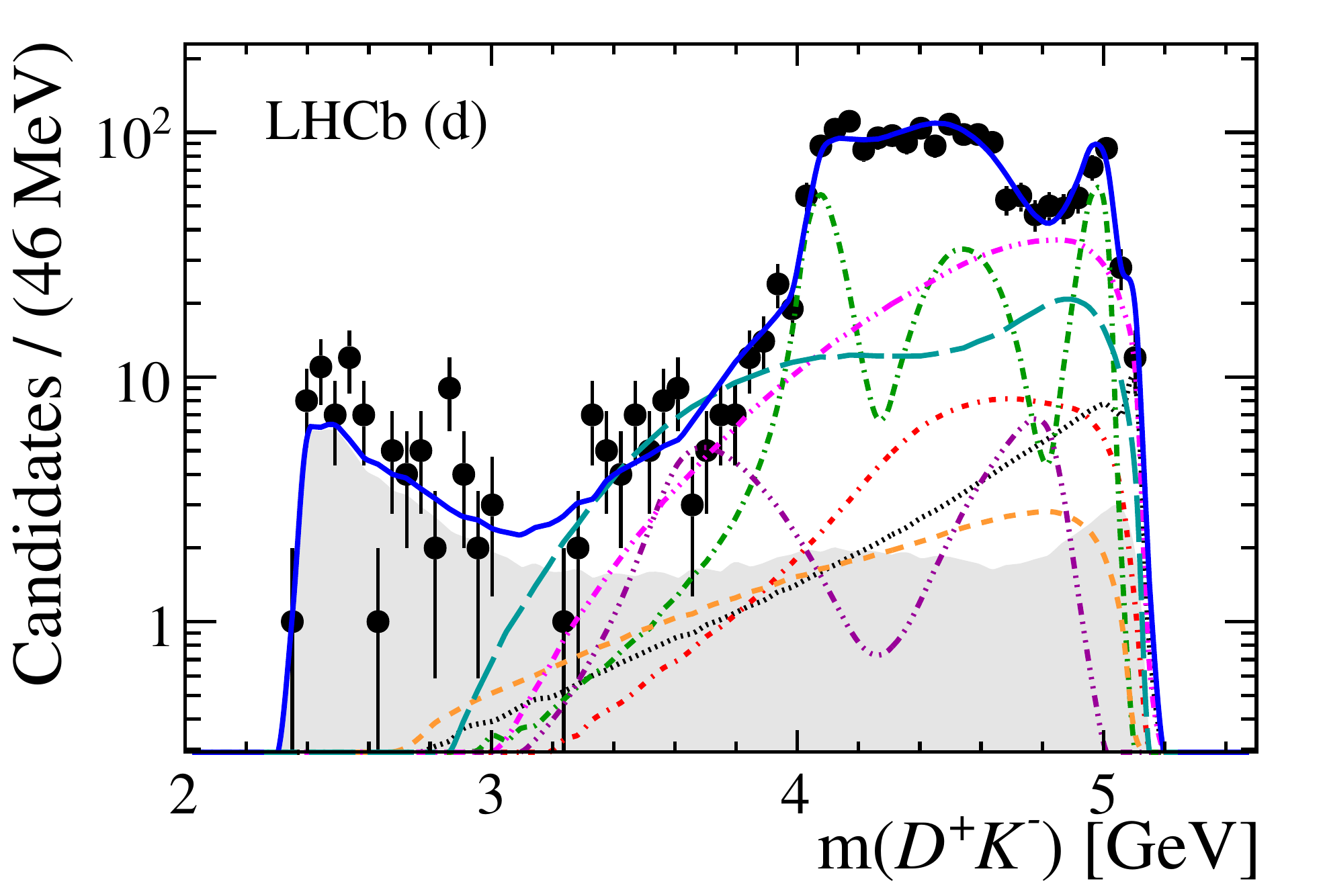}
\includegraphics[scale=0.395]{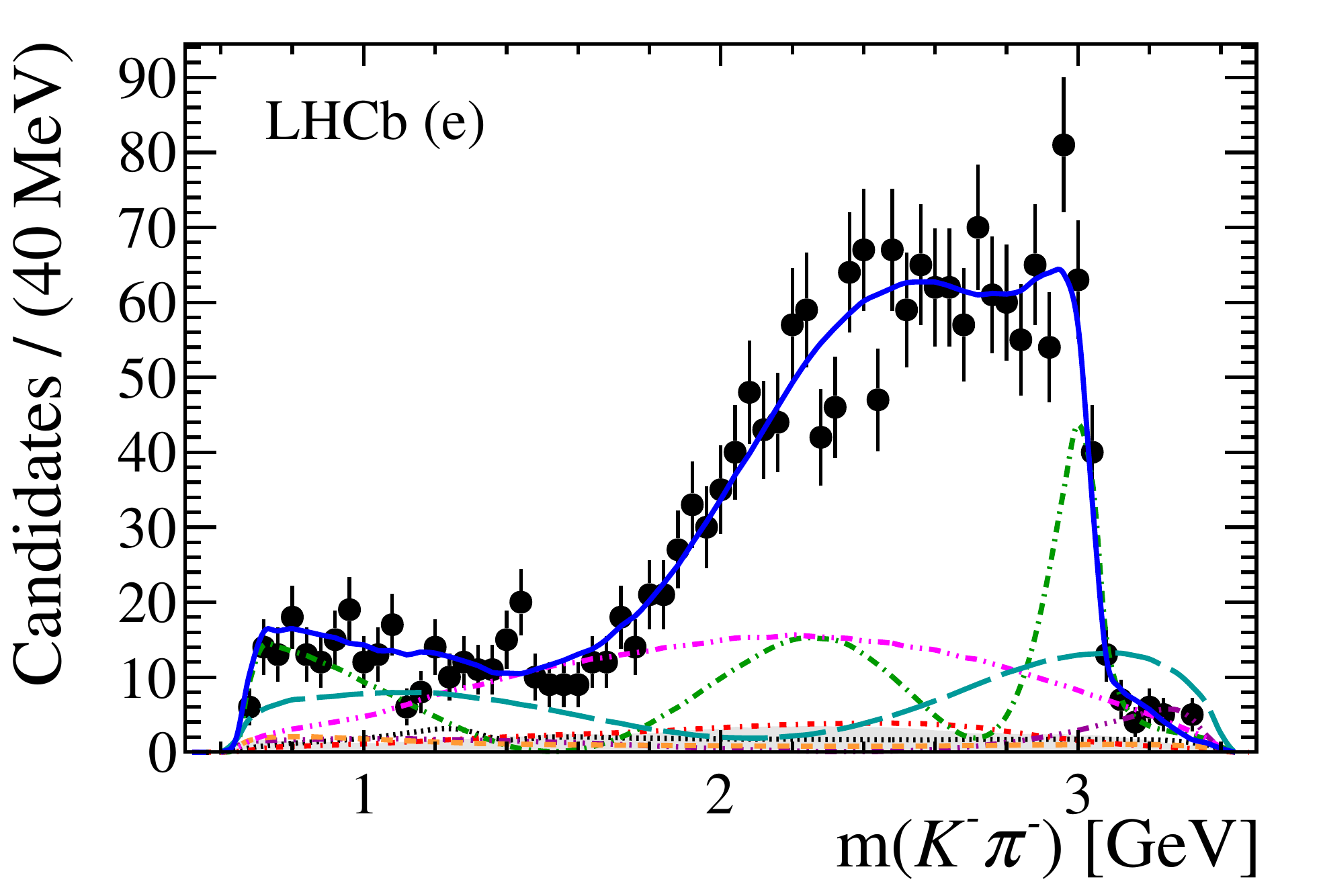}
\includegraphics[scale=0.395]{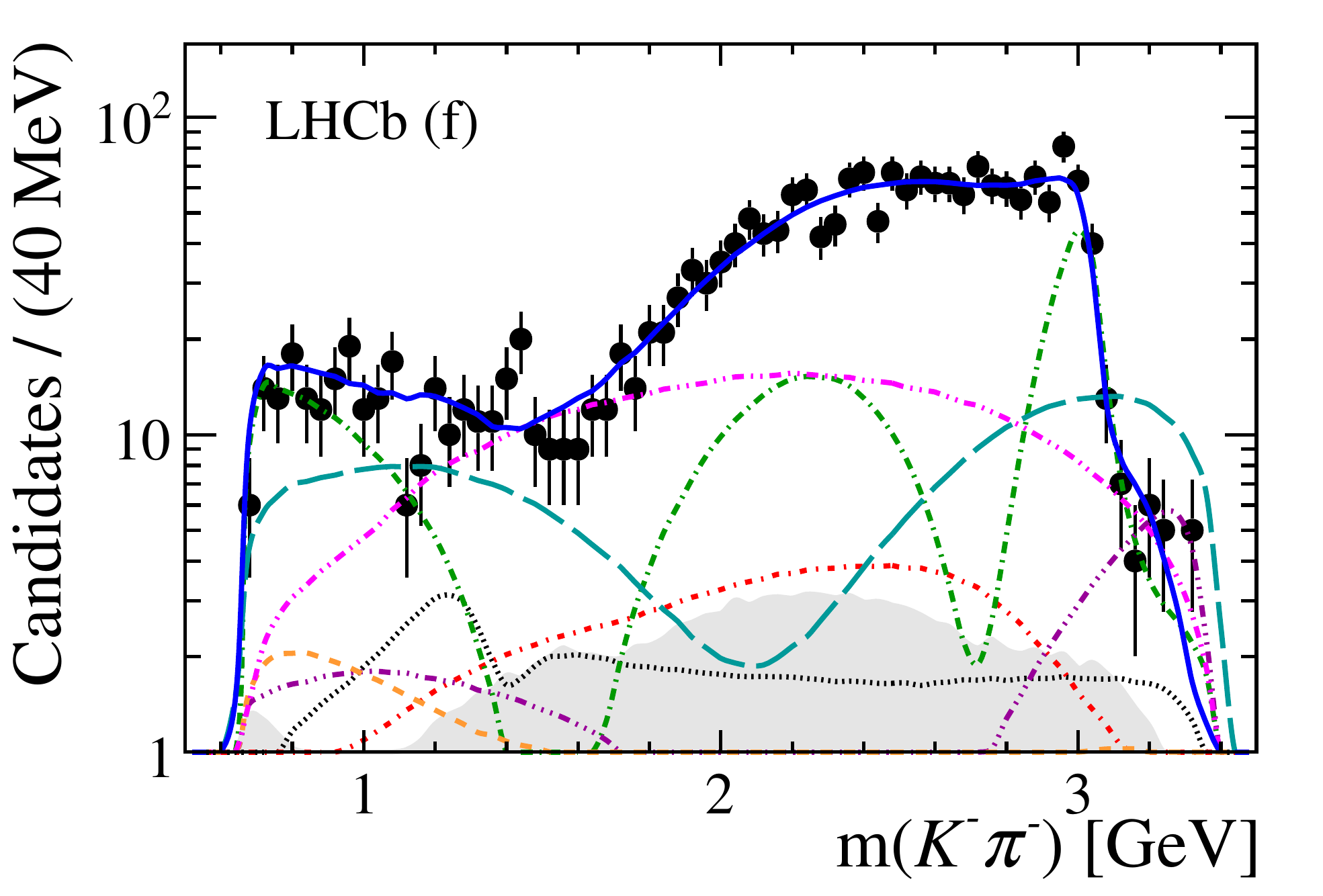}
\includegraphics[scale=0.395]{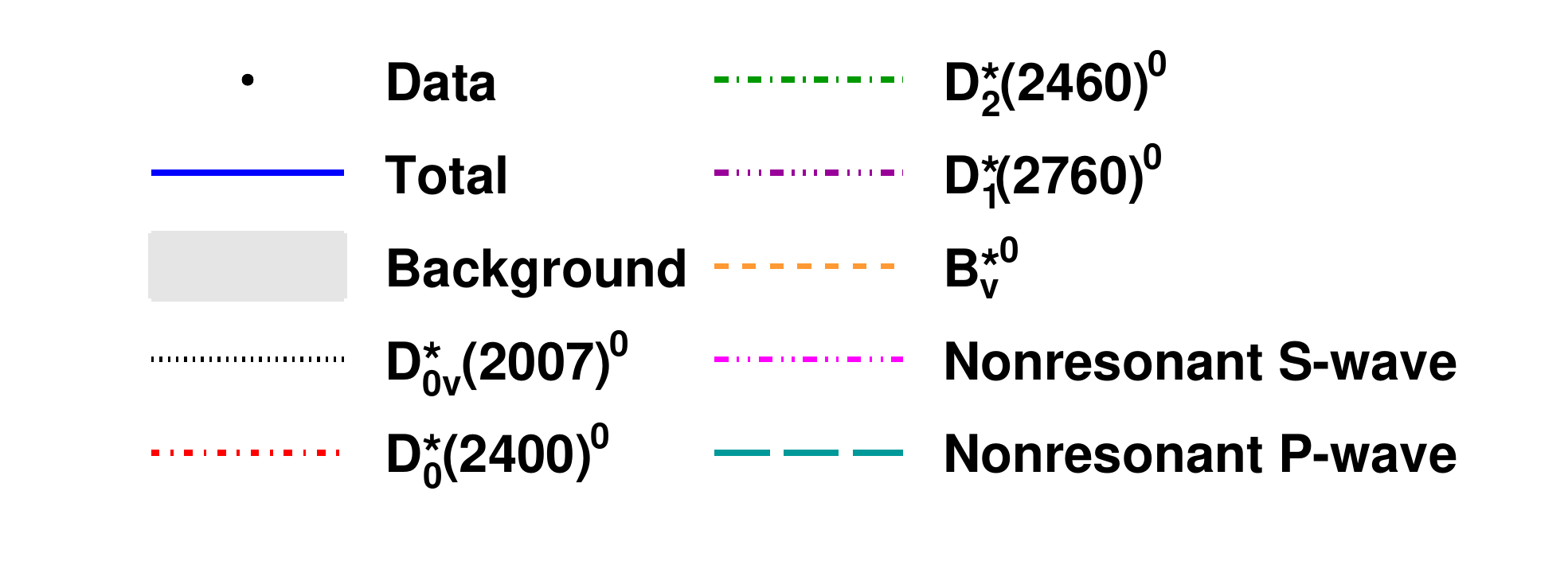}
\caption{\small Projections of the data and amplitude fit onto (a)~$m(D\pi)$, (c)~$m(DK)$ and (e)~$m(K\pi)$, with the same 
projections shown in (b), (d) and (f) with a logarithmic $y$-axis scale. Components are described in the legend.}
\label{fig:fitproj}
\end{figure}

\begin{figure}[!tb]
\centering
\includegraphics[scale=0.395]{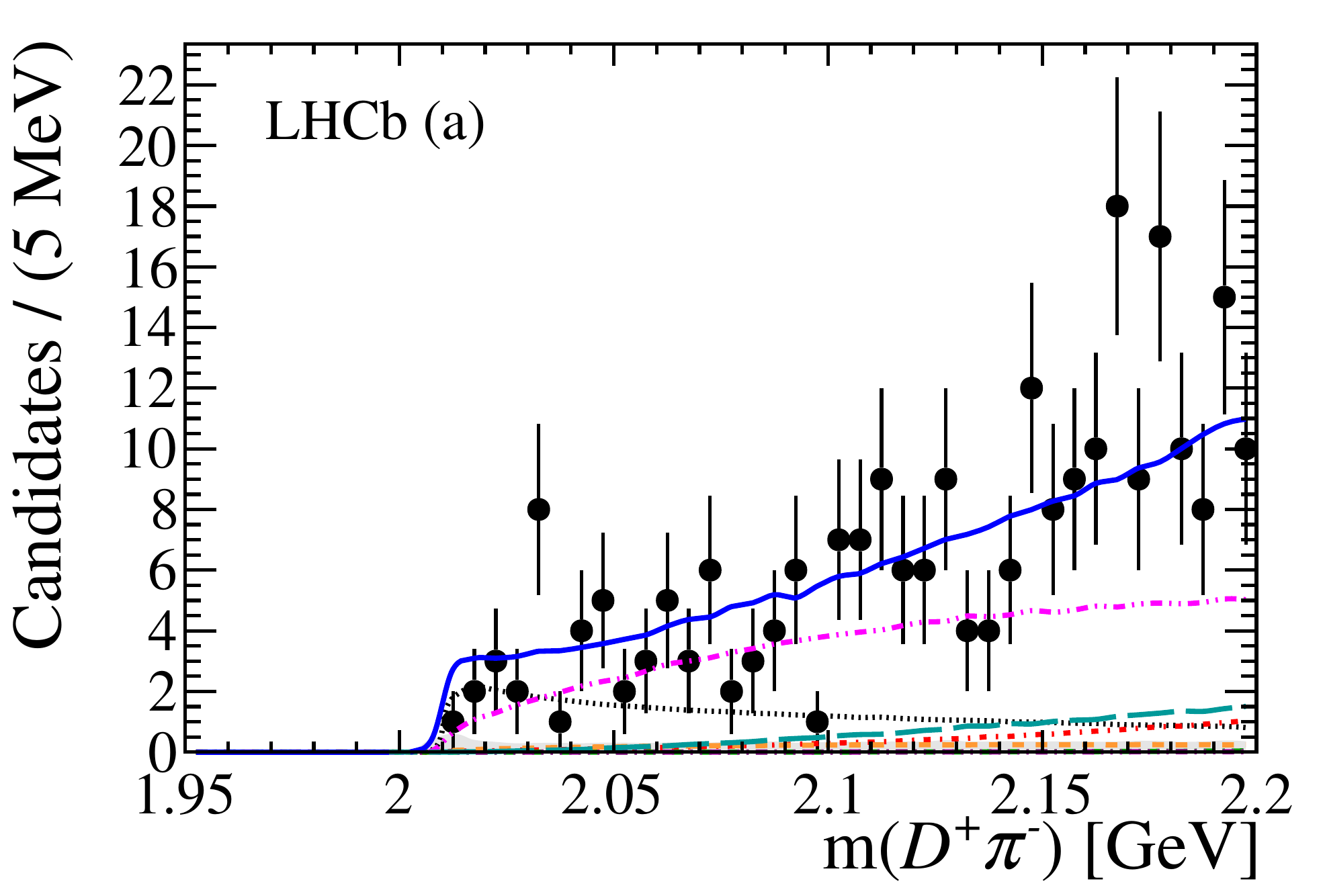}
\includegraphics[scale=0.395]{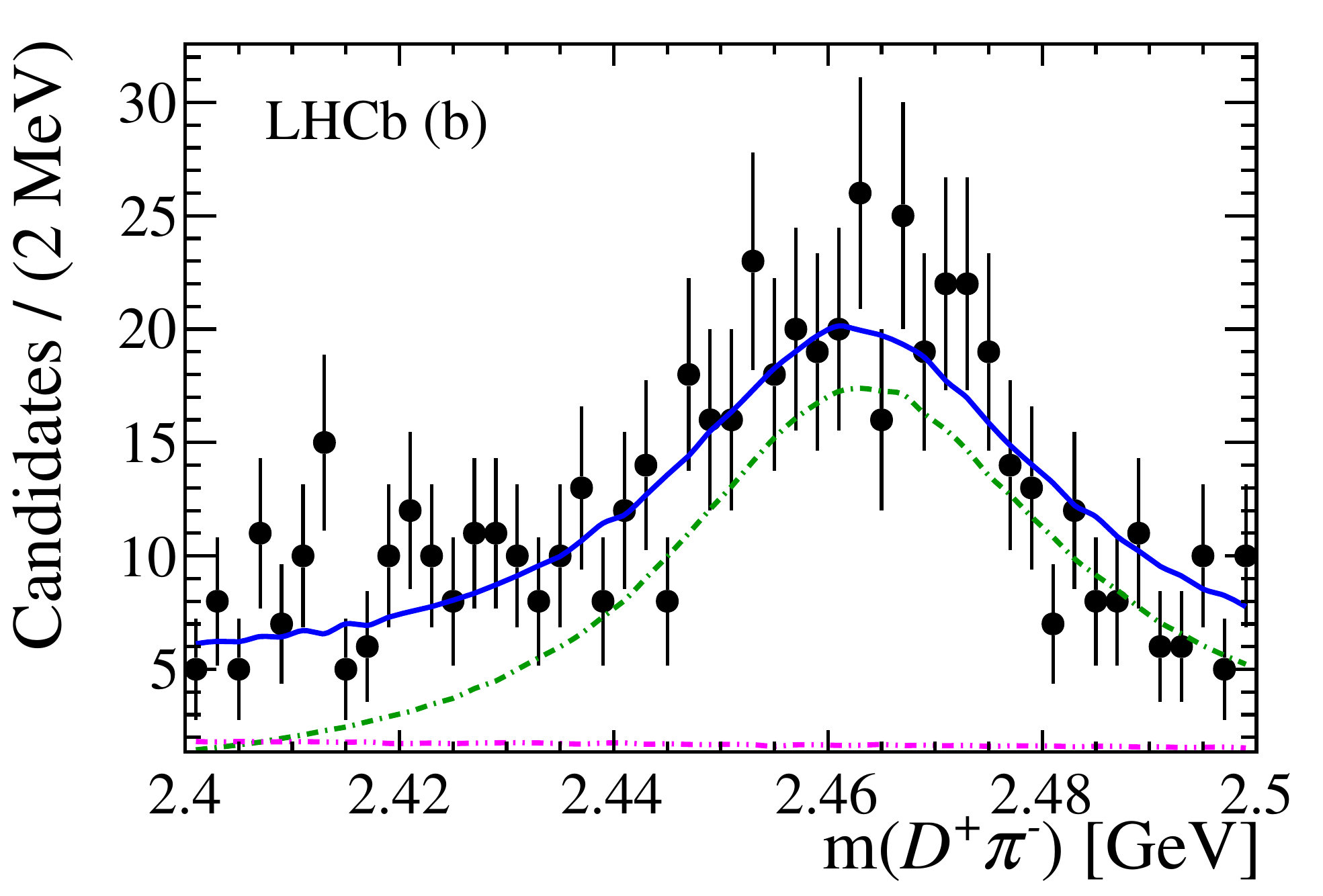}
\includegraphics[scale=0.395]{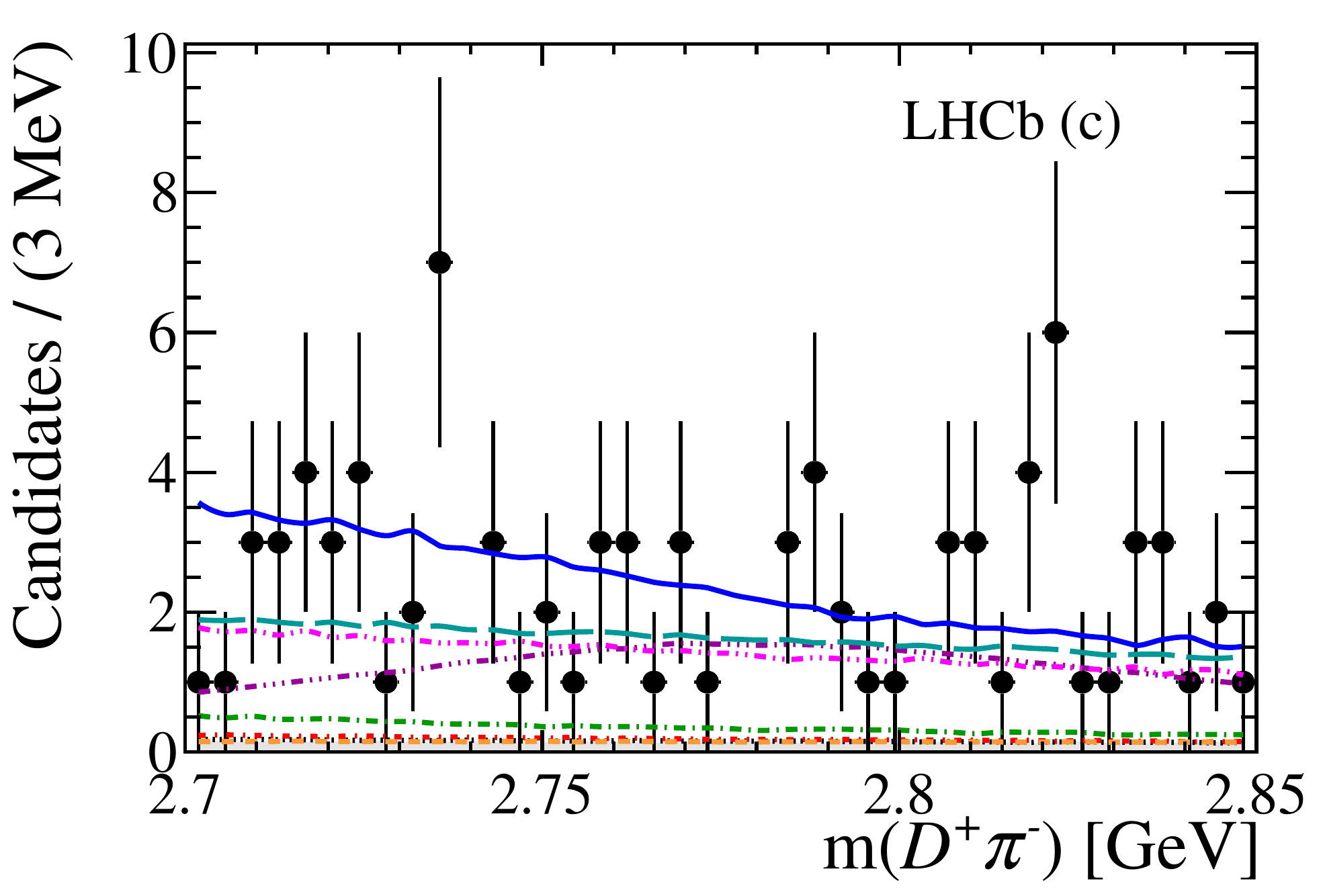}
\caption{\small Projections of the data and amplitude fit onto $m(D\pi)$ in (a)~the threshold region, (b)~the $D^*_2(2460)^0$ region and (c)~the $D^*_1(2760)^0$ region. Components are as shown in Fig.~\ref{fig:fitproj}.}
\label{fig:fitprojzoom}
\end{figure}

\begin{figure}[!tb]
\centering
\includegraphics[scale=0.395]{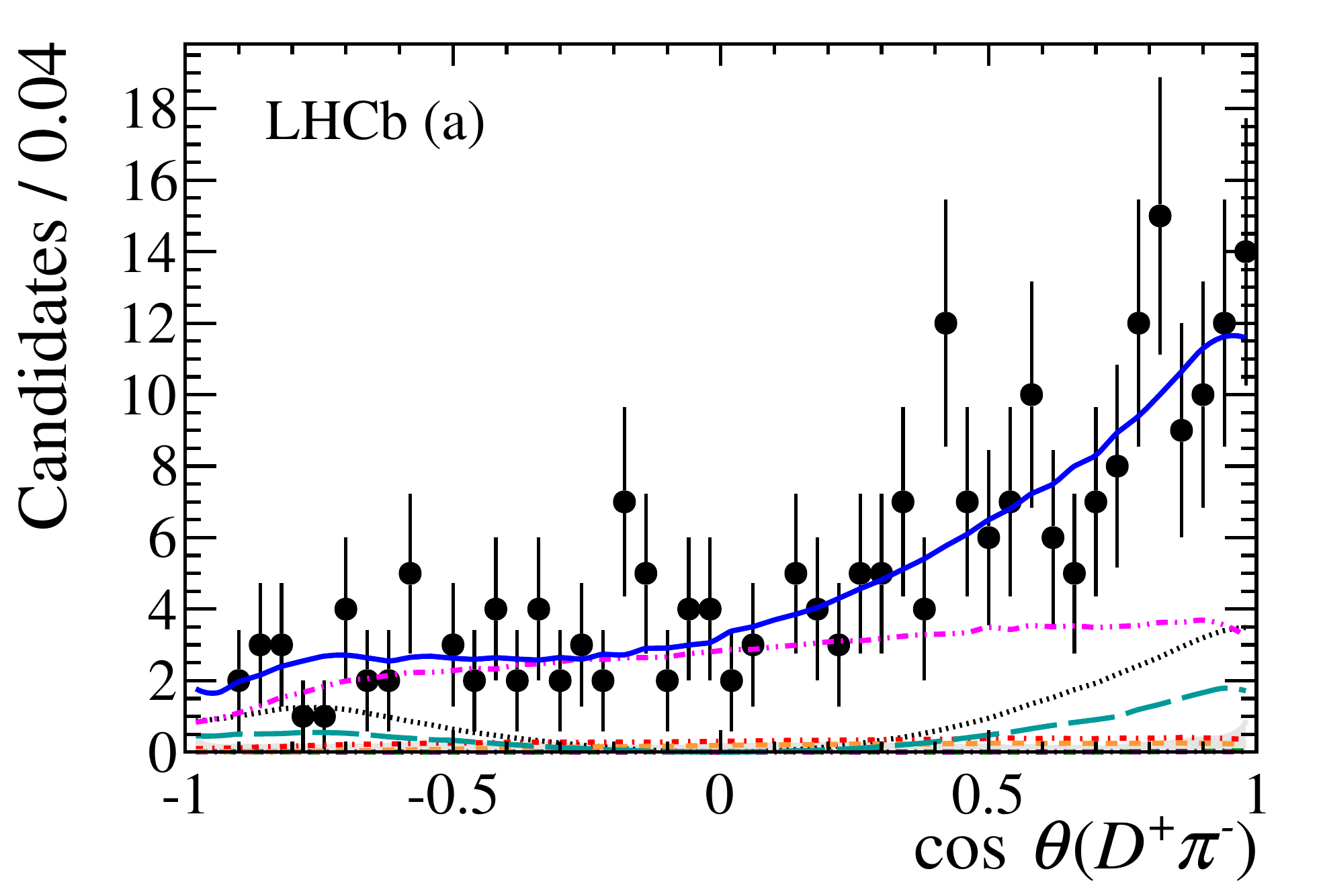}
\includegraphics[scale=0.395]{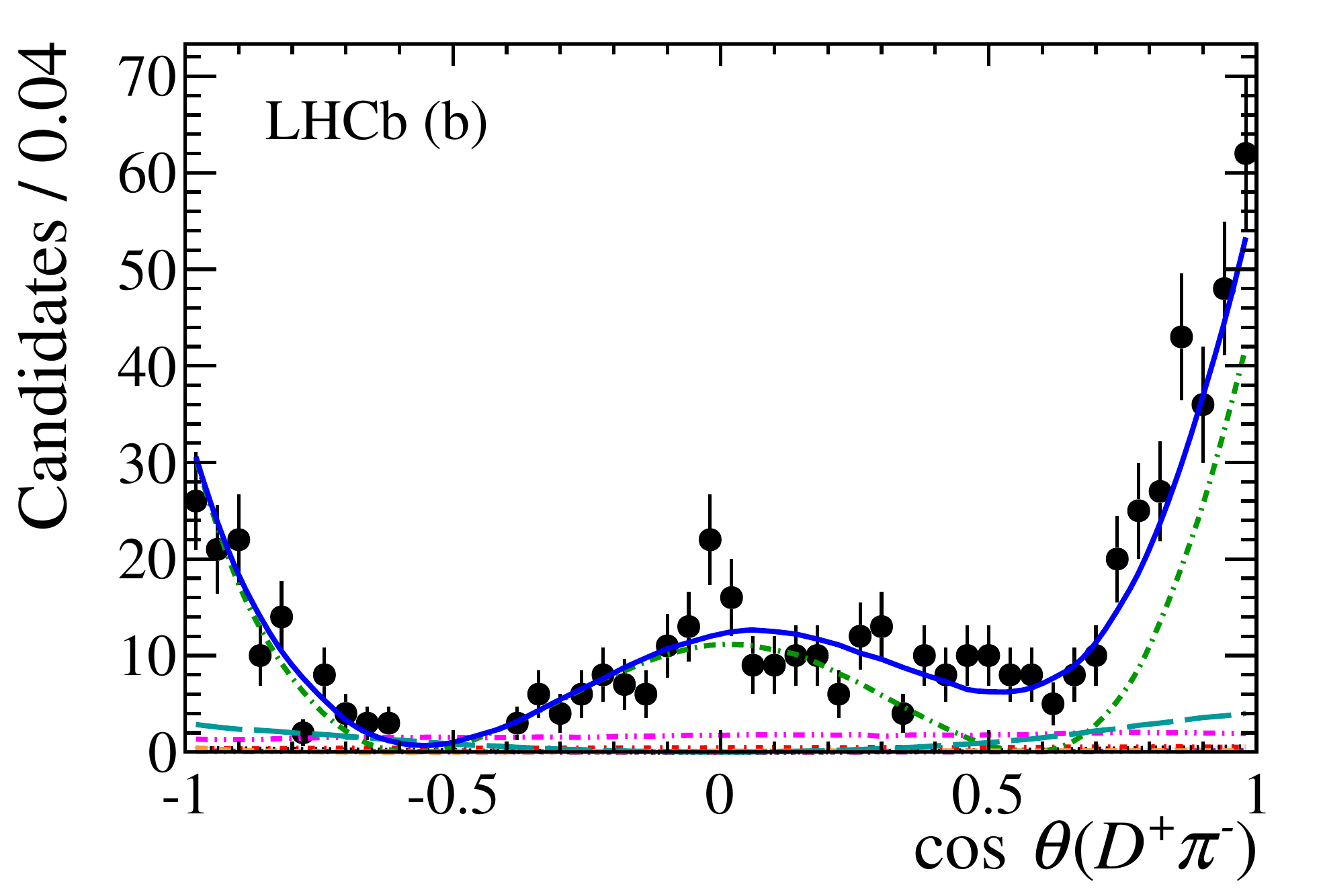}
\includegraphics[scale=0.395]{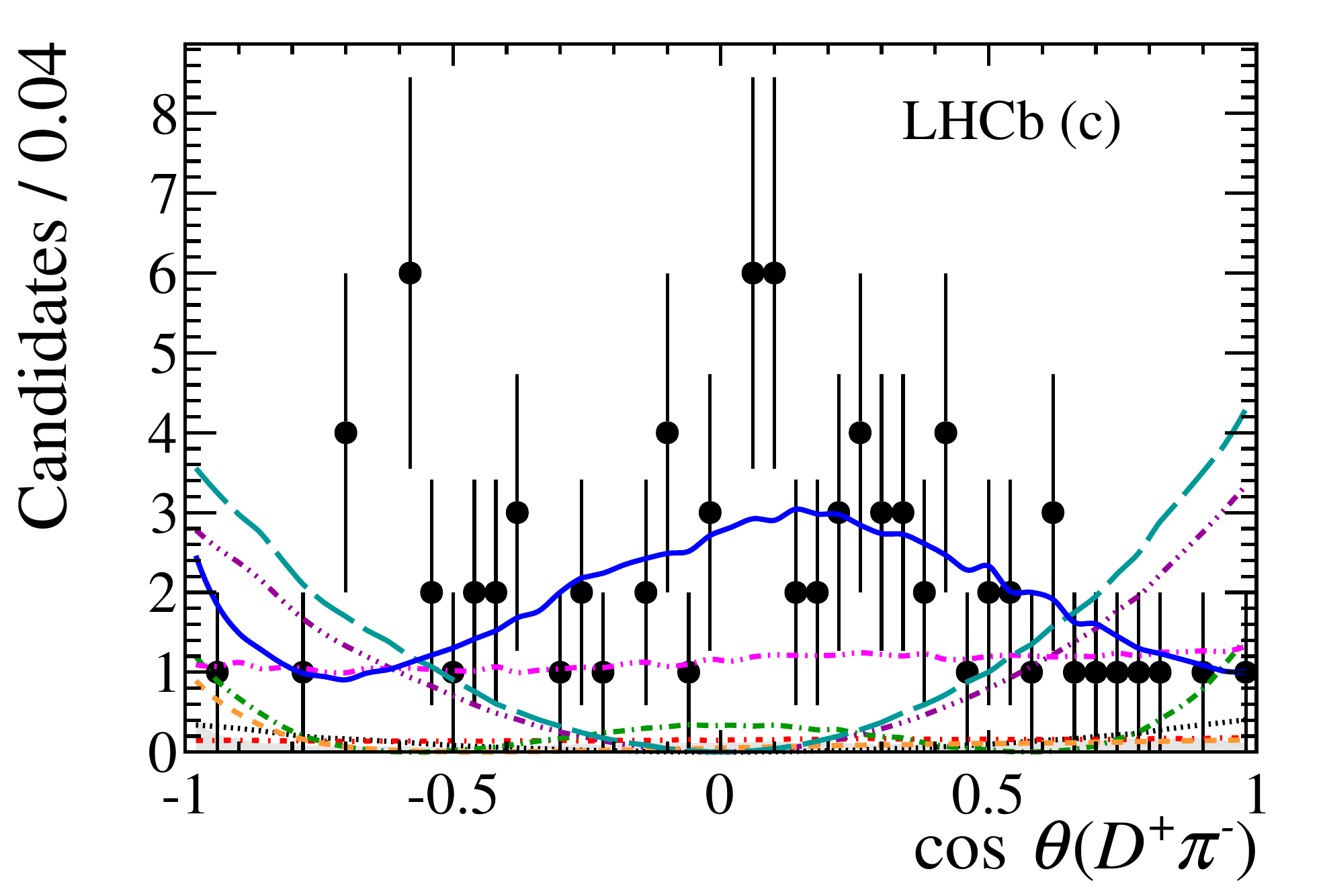}
\caption{\small Projections of the data and amplitude fit onto the cosine of the helicity angle for the $D\pi$ system in (a)~the threshold region, (b)~the $D^*_2(2460)^0$ region and (c)~the $D^*_1(2760)^0$ region. Components are as shown in Fig.~\ref{fig:fitproj}.}
\label{fig:fitprojcos}
\end{figure}

Figure~\ref{fig:fitproj} shows projections of the nominal fit model and the data onto $m(D\pi)$, $m(DK)$ and $m(K\pi)$. 
Zooms are provided around the resonant structures on $m(D\pi)$ in Fig.~\ref{fig:fitprojzoom}. 
Projections of the cosine of the helicity angle of the $D\pi$ system are shown in Fig.~\ref{fig:fitprojcos}.
Good agreement is seen between the data and the fit model.

\section{Systematic uncertainties}
\label{sec:systematics}

Sources of systematic uncertainty are divided into two categories: experimental and model uncertainties. 
The sources of experimental systematic uncertainty are: the signal and background yields in the signal region; the SDP distributions of the background components; the efficiency variation across the SDP; possible fit bias. 
The considered model uncertainties are: the fixed parameters in the amplitude model; the addition or removal of marginal amplitudes; the choice of models for the nonresonant contributions. 
The systematic uncertainties from each source are combined in quadrature.

The signal and background yields in the signal region are determined from the fit to the $B$ candidate invariant mass distribution, as described in Sec.~\ref{sec:mass-fit}. 
The uncertainty on each yield (including systematic uncertainty evaluated as in Sec.~\ref{sec:BF-syst}) is calculated, and the yields varied accordingly in the DP fit. 
The deviations from the nominal DP fit result are assigned as systematic uncertainties.

The effect of imperfect knowledge of the background distributions over the SDP is tested by varying the histograms used to model the shapes within their statistical uncertainties. 
For $D^{(*)+}\pim\pim$ decays the ratio of the $\Dstarp$ and $\Dp$ contributions is varied. 
Where applicable, the reweighting of the SDP distribution of the simulated samples is removed.

The uncertainty related to the knowledge of the variation of efficiency across the SDP is determined by varying the efficiency histograms before the spline fit is performed. 
The central bin in each cell of $3 \times 3$ bins is varied by its statistical uncertainty and the surrounding bins in the cell are varied by interpolation. 
This procedure accounts for possible correlations between the bins, since a systematic effect on a given bin is likely also to affect neighbouring bins.
The effects on the DP fit results are assigned as systematic uncertainties. 
An additional systematic uncertainty is assigned by varying the binning scheme of the control sample used to determine the PID efficiencies.

Systematic uncertainties related to possible intrinsic fit bias are investigated using an ensemble of pseudoexperiments. 
Differences between the input and fitted values from the ensemble for the fit parameters are found to be small. 
Systematic uncertainties are assigned as the sum in quadrature of the difference between the input and output values and the uncertainty on the mean of the output value determined from a fit to the ensemble.

Systematic uncertainties due to fixed parameters in the fit model are determined by varying the parameters within their 
uncertainties and repeating the fit. The fixed parameters considered are the mass and width of the $D^*_0(2400)^0$ resonance
and the \mbox{Blatt--Weisskopf} barrier radius, $r_{\rm BW}$. The mass and width are varied by the uncertainties shown in 
Table~\ref{tab:resonances} and the barrier radius is varied between 3 and 5\,$\gev^{-1}$~\cite{LHCb-PAPER-2014-036}.
For each fit parameter, the difference compared to the nominal fit model is assigned as a systematic uncertainty for each source.

The marginal $B^{*0}_{v}$ component is removed from the model and the changes in the other parameters are assigned as the systematic uncertainties.
Dalitz plot analysis of $\Bsb \to \Dz \Kp\pim$ revealed that a structure at $m(\Dz\Kp) \sim 2.86 \gev$ has both spin~1 and spin~3 components~\cite{LHCb-PAPER-2014-035,LHCb-PAPER-2014-036}.
Although there is no evidence for a spin~3 resonance in this analysis, the excess at $m(\Dp\pim) \sim 2.76 \gev$ could have a similar composition.
A putative $D^*_3(2760)$ resonance is added to the fit model, and the effect on the other parameters is used to assign systematic uncertainties. 

The EFF lineshapes used to model the nonresonant S- and P-wave contributions are replaced by a power-law model and the change in the fit parameters used as a systematic uncertainty. 
The dependence of the results on the effective pole mass description of Eq.~(\ref{eqn:effmass}) that is used for the virtual resonance contributions is found by using a fixed width in Eq.~(\ref{eq:RelBWEqn}), removing the dependency on $m^{\rm eff}_0$.

The total experimental and model systematic uncertainties for fit fractions and complex coefficients are summarised in Tables~\ref{tab:exptsyst} and~\ref{tab:modsyst}, respectively.
The contributions for the fit fractions, masses and widths are broken down in Tables~\ref{tab:exptsystbreak} and~\ref{tab:modsystbreak}.
Similar tables summarising the systematic uncertainties on the interference fit fractions are given in App.~\ref{app:iffstat}.
The largest source of experimental systematic uncertainty on the fit fractions is due to the efficiency variation. 
For the model uncertainty on the fit fractions, the addition and removal of marginal components and variation of fixed parameters dominate.
In general, the model uncertainties are larger than the experimental systematic uncertainties for the fit fractions and the masses and widths.

\begin{table}[!tb]
\centering
\caption{\small Experimental systematic uncertainties on the fit fractions and complex amplitudes.}
\label{tab:exptsyst}
\resizebox{\textwidth}{!}{ 
\begin{tabular}{lccccc} 
\hline 
& & \multicolumn{4}{c}{Isobar model coefficients} \\ 
Resonance & Fit fraction (\%) & Real part & Imaginary part & Magnitude & Phase \\ 
\hline  \\ [-2.5ex] 
$D^{*}_{0}(2400)^{0}$ & 0.6 & 0.03 & 0.02 & 0.02 & 0.06\\ 
$D^{*}_{2}(2460)^{0}$ & 0.9 & -- & -- & -- & -- \\ 
$D^{*}_{1}(2760)^{0}$ & 0.4 & 0.03 & 0.03 & 0.01 & 0.08\\ 
\hline 
S-wave nonresonant & 1.5 & 0.03 & 0.03 & 0.02 & 0.04\\ 
P-wave nonresonant & 2.1 & 0.03 & 0.05 & 0.03 & 0.05\\ 
\hline  \\ [-2.5ex] 
$D^{*}_{v}(2007)^{0}$ & 1.3 & 0.03 & 0.04 & 0.04 & 0.07\\ 
$B^{*}_{v}$ & 0.9 & 0.22 & 0.02 & 0.03 & 0.11\\ 
\hline 
\end{tabular} 
}
\end{table}

\begin{table}[!tb]
\centering
\caption{\small Model uncertainties on the fit fractions and complex amplitudes.}
\label{tab:modsyst}
\resizebox{\textwidth}{!}{ 
\begin{tabular}{lccccc} 
\hline 
& & \multicolumn{4}{c}{Isobar model coefficients} \\ 
Resonance & Fit fraction (\%) & Real part & Imaginary part & Magnitude & Phase \\ 
\hline  \\ [-2.5ex] 
$D^{*}_{0}(2400)^{0}$ & 1.9 & 0.28 & 0.13 & 0.15 & 0.51\\ 
$D^{*}_{2}(2460)^{0}$ & 1.4 & -- & -- & -- & -- \\ 
$D^{*}_{1}(2760)^{0}$ & 0.9 & 0.03 & 0.03 & 0.03 & 0.08\\ 
\hline 
S-wave nonresonant & 10.8 & 0.17 & 0.15 & 0.20 & 0.11\\ 
P-wave nonresonant & 3.7 & 0.34 & 0.68 & 0.12 & 0.95\\ 
\hline  \\ [-2.5ex] 
$D^{*}_{v}(2007)^{0}$ & 1.5 & 0.56 & 0.77 & 0.05 & 0.60\\ 
$B^{*}_{v}$ & 1.6 & 0.09 & 0.08 & 0.07 & 0.27\\ 
\hline 
\end{tabular} 
}
\end{table}

\begin{table}[!tb]
\centering
\caption{\small Breakdown of experimental systematic uncertainties on the fit fractions (\%) and masses $(\mevnsp)$ and widths $(\mevnsp)$.}
\label{tab:exptsystbreak}
\begin{tabular}{lrccccc} 
\hline 
 & Nominal & S/B frac. & Eff. & Bkg. & Fit bias & Total \\ 
\hline  \\ [-2.5ex] 
$D^{*}_{0}(2400)^{0}$ & $8.3 \pm 2.6$ & 0.2 & 0.5 & 0.1 & 0.3 & 0.6 \\ 
$D^{*}_{2}(2460)^{0}$ & $31.8 \pm 1.5$ & 0.2 & 0.8 & 0.0 & 0.2 & 0.9 \\ 
$D^{*}_{1}(2760)^{0}$ & $4.9 \pm 1.2$ & 0.2 & 0.2 & 0.1 & 0.2 & 0.3 \\ 
\hline 
S-wave nonresonant & $38.0 \pm 7.4$ & 0.7 & 0.5 & 0.4 & 1.2 & 1.5 \\ 
P-wave nonresonant & $23.8 \pm 5.6$ & 1.0 & 1.6 & 0.7 & 0.5 & 2.1 \\ 
\hline  \\ [-2.5ex] 
$D^{*}_{v}(2007)^{0}$ & $7.6 \pm 2.3$ & 0.7 & 1.0 & 0.3 & 0.3 & 1.3 \\ 
$B^{*}_{v}$ & $3.6 \pm 1.9$ & 0.3 & 0.3 & 0.2 & 0.8 & 0.9 \\ 
\hline 
\hline \\ [-2.5ex] 
$m\left( D^{*}_{2}(2460)^{0}\right)$ & $2464.0 \pm 1.4$ & 0.1 & 0.1 & 0.0 & 0.2 & 0.2 \\ 
$\Gamma\left( D^{*}_{2}(2460)^{0}\right)$ & $43.8 \pm 2.9$ & 0.3 & 0.3 & 0.0 & 0.4 & 0.6 \\ 
\hline  \\ [-2.5ex] 
$m\left( D^{*}_{1}(2760)^{0}\right)$ & $2781 \pm 18\phantom{.}$ &1 & 4 & 0 & 2 & 6 \\ 
$\Gamma\left( D^{*}_{1}(2760)^{0}\right)$ & $177 \pm 32\phantom{.}$ & 3 & 1 & 2 & 5 & 7 \\ 
\hline 
\end{tabular} 
\end{table}

\begin{table}[!tb]
\centering
\caption{\small Breakdown of model uncertainties on the fit fractions (\%) and masses $(\mevnsp)$ and widths $(\mevnsp)$.}
\label{tab:modsystbreak}
\begin{tabular}{lrcccc} 
\hline 
 & Nominal & Add/rem & Alt. models & Fixed params & Total \\ 
\hline  \\ [-2.5ex] 
$D^{*}_{0}(2400)^{0}$ & $8.3 \pm 2.6$ & 2.0 & 0.1 & 0.2 & 2.0 \\ 
$D^{*}_{2}(2460)^{0}$ & $31.8 \pm 1.5$ & 1.3 & 0.2 & 0.4 & 1.4 \\ 
$D^{*}_{1}(2760)^{0}$ & $4.9 \pm 1.2$ & 0.8 & 0.1 & 0.3 & 0.9 \\ 
\hline 
S-wave nonresonant & $38.0 \pm 7.4$ & 4.8 & 4.5 & 5.4 & 10.8 \\ 
P-wave nonresonant & $23.8 \pm 5.6$ & 2.6 & 2.1 & 3.0 & 3.7 \\ 
\hline  \\ [-2.5ex] 
$D^{*}_{v}(2007)^{0}$ & $7.6 \pm 2.3$ & 0.6 & 0.1 & 1.4 & 1.5 \\ 
$B^{*}_{v}$ & $3.6 \pm 1.9$ & 0.7 & 1.0 & 1.1 & 1.6 \\ 
\hline 
\hline  \\ [-2.5ex] 
$m\left( D^{*}_{2}(2460)^{0}\right)$ & $2464.0 \pm 1.4$ & 0.5 & 0.1 & 0.1 & 0.5 \\ 
$\Gamma\left( D^{*}_{2}(2460)^{0}\right)$ & $43.8 \pm 2.9$ & 0.8 & 1.4 & 0.6 & 1.7 \\ 
\hline  \\ [-2.5ex] 
$m\left( D^{*}_{1}(2760)^{0}\right)$ & $2781 \pm 18\phantom{.}$ & 6 & 6 & 1 & 11 \\ 
$\Gamma\left( D^{*}_{1}(2760)^{0}\right)$ & $177 \pm 32\phantom{.}$ & 16 & 9 & 1 & 20 \\ 
\hline 
\end{tabular} 
\end{table}

Several cross-checks are performed to confirm the stability of the results. The data sample is divided into two parts depending 
on the charge of the \B candidate, the polarity of the magnet and the year of data taking. 
Selection effects are also checked by varying the requirement on the neural network output variable and the PID criteria applied to the bachelor kaon.
A fit is performed for each of the subsamples individually and each is seen to be consistent with the default fit results, although in some cases one of the secondary minima described in App.~\ref{app:minima} becomes the preferred solution. 
To cross-check the amplitude model, the fit is repeated many times with an extra resonance with fixed mass, width and spin included in the model. 
All possible mass and width values, and spin up to 3, were considered.
None of the additional resonances are found to contribute significantly.

\section{Results and summary}
\label{sec:results}

The results for the complex coefficients are reported in Tables~\ref{tab:cf-results2} and~\ref{tab:cf-results} in terms of real and imaginary parts and of magnitudes and phases, respectively. 
The results for the fit fractions are given in Table~\ref{tab:cfitfrac-results} and the results for the interference fit fractions are given in App.~\ref{app:iffstat}.
The fit fractions for resonant contributions are converted into quasi-two-body product branching fractions by multiplying by $\mathcal{B}(\Bm\to\Dp\Km\pim) = (7.31 \pm 0.19 \pm 0.22 \pm 0.39) \times 10^{-5}$, as determined in Sec.~\ref{sec:BF-results}.
These product branching fractions are shown in Table~\ref{tab:BFresults}; they cannot be converted into absolute branching fractions because the branching fractions for the resonance decays to $\Dp\pim$ are unknown.

\begin{table}[!tb]
\centering
\caption{\small Results for the complex amplitudes and their uncertainties. The three quoted errors are statistical, 
experimental systematic and model uncertainties, respectively.}
\label{tab:cf-results2}
\begin{tabular}{lcc} 
\hline 
& \multicolumn{2}{c}{Isobar model coefficients} \\ 
Resonance & Real part & Imaginary part \\ 
\hline  \\ [-2.5ex] 
$D^{*}_{0}(2400)^{0}$ & $-0.04 \pm 0.07 \pm 0.03 \pm 0.28$ & $-0.51 \pm 0.07 \pm 0.02 \pm 0.13$ \\
$D^{*}_{2}(2460)^{0}$ &  $1.00$ & $0.00$ \\ 
$D^{*}_{1}(2760)^{0}$ & $-0.32 \pm 0.06 \pm 0.03 \pm 0.03$ & $-0.23 \pm 0.07 \pm 0.03 \pm 0.03$ \\
\hline 
S-wave nonresonant & $\phantom{-}0.93 \pm 0.09 \pm 0.03 \pm 0.17$ & $-0.58 \pm 0.08 \pm 0.03 \pm 0.15$ \\
P-wave nonresonant & $-0.43 \pm 0.09 \pm 0.03 \pm 0.34$ & $\phantom{-}0.75 \pm 0.09 \pm 0.05 \pm 0.68$ \\
\hline \\ [-2.5ex] 
$D^{*}_{v}(2007)^{0}$ & $\phantom{-}0.16 \pm 0.08 \pm 0.03 \pm 0.56$ & $\phantom{-}0.46 \pm 0.09 \pm 0.04 \pm 0.77$ \\
$B^*_v$ & $-0.07 \pm 0.08 \pm 0.22 \pm 0.09$ & $\phantom{-}0.33 \pm 0.07 \pm 0.02 \pm 0.08$ \\
\hline 
\end{tabular} 
\end{table}

\begin{table}[!tb]
\centering
\caption{\small Results for the complex amplitudes and their uncertainties. The three quoted errors are statistical, 
experimental systematic and model uncertainties, respectively.}
\label{tab:cf-results}
\begin{tabular}{lcc} 
\hline 
& \multicolumn{2}{c}{Isobar model coefficients} \\ 
Resonance & Magnitude & Phase \\ 
\hline \\ [-2.5ex] 
$D^{*}_{0}(2400)^{0}$ & $0.51 \pm 0.09 \pm 0.02 \pm 0.15$ & $-1.65 \pm 0.16 \pm 0.06 \pm 0.50$ \\ 
$D^{*}_{2}(2460)^{0}$ & $1.00$ & $0.00$ \\ 
$D^{*}_{1}(2760)^{0}$ & $0.39 \pm 0.05 \pm 0.01 \pm 0.03$ & $-2.53 \pm 0.24 \pm 0.08 \pm 0.08$ \\ 
\hline 
S-wave nonresonant & $1.09 \pm 0.09 \pm 0.02 \pm 0.20$ & $-0.56 \pm 0.09 \pm 0.04 \pm 0.11$ \\ 
P-wave nonresonant & $0.87 \pm 0.09 \pm 0.03 \pm 0.11$ & $\phantom{-}2.09 \pm 0.15 \pm 0.05 \pm 0.95$ \\ 
\hline \\ [-2.5ex] 
$D^{*}_{v}(2007)^{0}$ & $0.49 \pm 0.07 \pm 0.04 \pm 0.05$ & $\phantom{-}1.24 \pm 0.17 \pm 0.07 \pm 0.60$ \\ 
$B^*_v$ & $0.34 \pm 0.06 \pm 0.03 \pm 0.07$ & $\phantom{-}1.78 \pm 0.23 \pm 0.11 \pm 0.27$ \\ 
\hline 
\end{tabular} 
\end{table}

\begin{table}[!tb]
\centering
\caption{\small Results for the fit fractions and their uncertainties (\%). The three quoted errors are statistical, experimental 
systematic and model uncertainties, respectively.}
\label{tab:cfitfrac-results}
\begin{tabular}{lc} 
\hline 
 Resonance & Fit fraction \\ 
 \hline \\ [-2.5ex] 
$D^{*}_{0}(2400)^{0}$ & $\phantom{2}8.3 \pm 2.6 \pm 0.6 \pm \phantom{2}1.9$ \\ 
$D^{*}_{2}(2460)^{0}$ & $31.8 \pm 1.5 \pm 0.9 \pm \phantom{2}1.4$ \\ 
$D^{*}_{1}(2760)^{0}$ & $\phantom{2}4.9 \pm 1.2 \pm 0.3 \pm \phantom{2}0.9$ \\ 
\hline 
S-wave nonresonant & $38.0 \pm 7.4 \pm 1.5 \pm 10.8$ \\ 
P-wave nonresonant & $23.8 \pm 5.6 \pm 2.1 \pm \phantom{2}3.7$ \\ 
\hline \\ [-2.5ex] 
$D^{*}_{v}(2007)^{0}$ & $\phantom{2}7.6 \pm 2.3 \pm 1.3 \pm \phantom{2}1.5$ \\ 
$B^*_v$ & $\phantom{2}3.6 \pm 1.9 \pm 0.9 \pm \phantom{2}1.6$ \\ 
\hline 
\end{tabular} 
\end{table}

\begin{table}[!tb]
\centering
\caption{\small Results for the product branching fractions ${\cal B}(\Bm \to R\Km) \times {\cal B}(R \to \Dp\pim)$ ($10^{-6}$). The four quoted errors are statistical, experimental systematic, model and inclusive branching fraction uncertainties, respectively.}
\label{tab:BFresults}
\begin{tabular}{lc} 
\hline 
 Resonance & Branching fraction \\ 
 \hline \\ [-2.5ex] 
$D^{*}_{0}(2400)^{0}$ & $\phantom{2}6.1 \pm 1.9 \pm 0.5 \pm 1.4 \pm 0.4$ \\ 
$D^{*}_{2}(2460)^{0}$ & $23.2 \pm 1.1 \pm 0.6 \pm 1.0 \pm 1.6$ \\ 
$D^{*}_{1}(2760)^{0}$ & $\phantom{2}3.6 \pm 0.9 \pm 0.3 \pm 0.7 \pm 0.2$ \\ 
\hline 
S-wave nonresonant & $27.8 \pm 5.4 \pm 1.1 \pm 7.9 \pm 1.9$ \\
P-wave nonresonant & $17.4 \pm 4.1 \pm 1.5 \pm 2.7 \pm 1.2$ \\
\hline \\ [-2.5ex] 
$D^{*}_{v}(2007)^{0}$ & $\phantom{2}5.6 \pm 1.7 \pm 1.0 \pm 1.1 \pm 0.4$ \\ 
$B^*_v$ & $\phantom{2}2.6 \pm 1.4 \pm 0.6 \pm 1.2 \pm 0.2$ \\ 
\hline 
\end{tabular} 
\end{table}

The masses and widths of the $D^{*}_{2}(2460)^{0}$ and $D^{*}_{1}(2760)^{0}$ are determined to be
\begin{eqnarray*}
m(D^{*}_{2}(2460)^{0})      & = & (2464.0 \pm 1.4 \pm 0.5 \pm 0.2) \mev \, ,\\
\Gamma(D^{*}_{2}(2460)^{0}) & = & \phantom{24}(43.8   \pm 2.9  \pm 1.7 \pm 0.6) \mev \, ,\\
m(D^{*}_{1}(2760)^{0})      & = & \phantom{.0}(2781 \pm \phantom{.}18 \pm \phantom{.}11 \pm \phantom{.0}6) \mev \, ,\\
\Gamma(D^{*}_{1}(2760)^{0}) & = & \phantom{.02}(177  \pm \phantom{.}32 \pm \phantom{.}20 \pm \phantom{.0}7) \mev \, ,
\end{eqnarray*}
where the three quoted errors are statistical, experimental systematic and model uncertainties, respectively.
The results for the $D^{*}_{2}(2460)^{0}$ are within $2\,\sigma$ of the world average values~\cite{PDG2014}.
The mass of the $D^{*}_{1}(2760)^{0}$ resonance is similarly consistent with previous measurements. 
The measured width of this state is larger than previous measurements by 2 to 3 times the uncertainties. 
Future studies based on much larger data samples will be required to better understand these states.

The measurement of $\mathcal{B}(\Bm\to\Dp\Km\pim)$ corresponds to the first observation of this decay mode. 
Therefore, the resonant contributions to the decay are also first observations.
The significance of the $\Bm \to D^{*}_{1}(2760)^{0}\Km$ observation is investigated by removing the corresponding resonance from the DP model.
A fit without the $D^{*}_{1}(2760)^{0}$ component increases the value of $2\Delta{\rm NLL}$ by $75.0$ units, corresponding to a high statistical significance. 
Only the systematic effects due to uncertainties in the DP model could in principle significantly change the conclusion regarding the need for this resonance.
However, in alternative DP models where a $D\pi$ resonance with spin~3 is added and where the $B^*_v$ contribution is removed, the shift in $2\Delta{\rm NLL}$ remains above 50 units.
The alternative models also do not significantly impact the level at which the $D^{*}_{1}(2760)^{0}$ state is preferred to be spin~1.
Therefore, these results represent the first observation of the $\Bm \to D^{*}_{1}(2760)^{0}\Km$ and the measurement of the spin of the $D^{*}_{1}(2760)^{0}$ resonance.

In summary, the $\Bm\to\Dp\Km\pim$ decay has been observed in a data sample corresponding to $3.0\invfb$ of $pp$ collision data recorded by the LHCb experiment.
An amplitude analysis of its Dalitz plot distribution has been performed, in which a model containing resonant contributions from the $D^{*}_{0}(2400)^{0}$, $D^{*}_{2}(2460)^{0}$ and $D^{*}_{1}(2760)^{0}$ states in addition to both S-wave and P-wave nonresonant amplitudes and components due to virtual $D^{*}_{v}(2007)^{0}$ and $B^{*0}_{v}$ resonances was found to give a good description of the data.
The $\Bm \to D^{*}_{2}(2460)^{0}\Km$ decay may in future be used to determine the angle $\gamma$ of the CKM unitarity triangle.
The results provide insight into the spectroscopy of charm mesons, and demonstrate that further progress may be obtained with Dalitz plot analyses of larger data samples.

\section*{Acknowledgements}

\noindent We express our gratitude to our colleagues in the CERN
accelerator departments for the excellent performance of the LHC. We
thank the technical and administrative staff at the LHCb
institutes. We acknowledge support from CERN and from the national
agencies: CAPES, CNPq, FAPERJ and FINEP (Brazil); NSFC (China);
CNRS/IN2P3 (France); BMBF, DFG, HGF and MPG (Germany); INFN (Italy); 
FOM and NWO (The Netherlands); MNiSW and NCN (Poland); MEN/IFA (Romania); 
MinES and FANO (Russia); MinECo (Spain); SNSF and SER (Switzerland); 
NASU (Ukraine); STFC (United Kingdom); NSF (USA).
The Tier1 computing centres are supported by IN2P3 (France), KIT and BMBF 
(Germany), INFN (Italy), NWO and SURF (The Netherlands), PIC (Spain), GridPP 
(United Kingdom).
We are indebted to the communities behind the multiple open 
source software packages on which we depend. We are also thankful for the 
computing resources and the access to software R\&D tools provided by Yandex LLC (Russia).
Individual groups or members have received support from 
EPLANET, Marie Sk\l{}odowska-Curie Actions and ERC (European Union), 
Conseil g\'{e}n\'{e}ral de Haute-Savoie, Labex ENIGMASS and OCEVU, 
R\'{e}gion Auvergne (France), RFBR (Russia), XuntaGal and GENCAT (Spain), Royal Society and Royal
Commission for the Exhibition of 1851 (United Kingdom).

\addcontentsline{toc}{section}{References}
\setboolean{inbibliography}{true}

\begin{mcitethebibliography}{10}
\mciteSetBstSublistMode{n}
\mciteSetBstMaxWidthForm{subitem}{\alph{mcitesubitemcount})}
\mciteSetBstSublistLabelBeginEnd{\mcitemaxwidthsubitemform\space}
{\relax}{\relax}

\bibitem{Abe:2003zm}
Belle collaboration, K.~Abe {\em et~al.},
  \ifthenelse{\boolean{articletitles}}{\emph{{Study of $\Bm \to D^{**0} \pim \
  (D^{**0} \to D^{(*)+} \pim)$ decays}},
  }{}\href{http://dx.doi.org/10.1103/PhysRevD.69.112002}{Phys.\ Rev.\
  \textbf{D69} (2004) 112002}, \href{http://arxiv.org/abs/hep-ex/0307021}{{\tt
  arXiv:hep-ex/0307021}}\relax
\mciteBstWouldAddEndPuncttrue
\mciteSetBstMidEndSepPunct{\mcitedefaultmidpunct}
{\mcitedefaultendpunct}{\mcitedefaultseppunct}\relax
\EndOfBibitem
\bibitem{Aubert:2009wg}
\babar collaboration, B.~Aubert {\em et~al.},
  \ifthenelse{\boolean{articletitles}}{\emph{{Dalitz plot analysis of $\Bm \to
  \Dp\pim\pim$}}, }{}\href{http://dx.doi.org/10.1103/PhysRevD.79.112004}{Phys.\
  Rev.\  \textbf{D79} (2009) 112004},
  \href{http://arxiv.org/abs/0901.1291}{{\tt arXiv:0901.1291}}\relax
\mciteBstWouldAddEndPuncttrue
\mciteSetBstMidEndSepPunct{\mcitedefaultmidpunct}
{\mcitedefaultendpunct}{\mcitedefaultseppunct}\relax
\EndOfBibitem
\bibitem{Kuzmin:2006mw}
Belle collaboration, A.~Kuzmin {\em et~al.},
  \ifthenelse{\boolean{articletitles}}{\emph{{Study of $\bar{B}^0 \to D^0 \pi^+
  \pi^-$ decays}},
  }{}\href{http://dx.doi.org/10.1103/PhysRevD.76.012006}{Phys.\ Rev.\
  \textbf{D76} (2007) 012006}, \href{http://arxiv.org/abs/hep-ex/0611054}{{\tt
  arXiv:hep-ex/0611054}}\relax
\mciteBstWouldAddEndPuncttrue
\mciteSetBstMidEndSepPunct{\mcitedefaultmidpunct}
{\mcitedefaultendpunct}{\mcitedefaultseppunct}\relax
\EndOfBibitem
\bibitem{LHCb-PAPER-2014-035}
LHCb collaboration, R.~Aaij {\em et~al.},
  \ifthenelse{\boolean{articletitles}}{\emph{{Observation of overlapping
  spin-$1$ and spin-$3$ $\bar{D}^0 K^-$ resonances at mass $2.86$~GeV/$c^2$}},
  }{}\href{http://dx.doi.org/10.1103/PhysRevLett.113.162001}{Phys.\ Rev.\
  Lett.\  \textbf{113} (2014) 162001},
  \href{http://arxiv.org/abs/1407.7574}{{\tt arXiv:1407.7574}}\relax
\mciteBstWouldAddEndPuncttrue
\mciteSetBstMidEndSepPunct{\mcitedefaultmidpunct}
{\mcitedefaultendpunct}{\mcitedefaultseppunct}\relax
\EndOfBibitem
\bibitem{LHCb-PAPER-2014-036}
LHCb collaboration, R.~Aaij {\em et~al.},
  \ifthenelse{\boolean{articletitles}}{\emph{{Dalitz plot analysis of
  $B^0_s\to\bar{D}^0K^-\pi^+$ decays}},
  }{}\href{http://dx.doi.org/10.1103/PhysRevD.90.072003}{Phys.\ Rev.\
  \textbf{D90} (2014) 072003}, \href{http://arxiv.org/abs/1407.7712}{{\tt
  arXiv:1407.7712}}\relax
\mciteBstWouldAddEndPuncttrue
\mciteSetBstMidEndSepPunct{\mcitedefaultmidpunct}
{\mcitedefaultendpunct}{\mcitedefaultseppunct}\relax
\EndOfBibitem
\bibitem{Lees:2014abp}
\babar collaboration, J.~P. Lees {\em et~al.},
  \ifthenelse{\boolean{articletitles}}{\emph{{Dalitz plot analyses of $B^0 \to
  D^- D^0 K^+$ and $B^+ \to \overline{D}^0 D^0 K^+$ decays}},
  }{}\href{http://dx.doi.org/10.1103/PhysRevD.91.052002}{Phys.\ Rev.\
  \textbf{D91} (2015) 052002}, \href{http://arxiv.org/abs/1412.6751}{{\tt
  arXiv:1412.6751}}\relax
\mciteBstWouldAddEndPuncttrue
\mciteSetBstMidEndSepPunct{\mcitedefaultmidpunct}
{\mcitedefaultendpunct}{\mcitedefaultseppunct}\relax
\EndOfBibitem
\bibitem{delAmoSanchez:2010vq}
\babar collaboration, P.~del Amo~Sanchez {\em et~al.},
  \ifthenelse{\boolean{articletitles}}{\emph{{Observation of new resonances
  decaying to $D\pi$ and $D^*\pi$ in inclusive $e^+e^-$ collisions near
  $\sqrt{s}=$10.58 GeV}},
  }{}\href{http://dx.doi.org/10.1103/PhysRevD.82.111101}{Phys.\ Rev.\
  \textbf{D82} (2010) 111101}, \href{http://arxiv.org/abs/1009.2076}{{\tt
  arXiv:1009.2076}}\relax
\mciteBstWouldAddEndPuncttrue
\mciteSetBstMidEndSepPunct{\mcitedefaultmidpunct}
{\mcitedefaultendpunct}{\mcitedefaultseppunct}\relax
\EndOfBibitem
\bibitem{LHCb-PAPER-2013-026}
LHCb collaboration, R.~Aaij {\em et~al.},
  \ifthenelse{\boolean{articletitles}}{\emph{{Study of $D_J$ meson decays to
  $D^+\pi^-$, $D^0\pi^+$ and $D^{*+}\pi^-$ final states in $pp$ collisions}},
  }{}\href{http://dx.doi.org/10.1007/JHEP09(2013)145}{JHEP \textbf{09} (2013)
  145}, \href{http://arxiv.org/abs/1307.4556}{{\tt arXiv:1307.4556}}\relax
\mciteBstWouldAddEndPuncttrue
\mciteSetBstMidEndSepPunct{\mcitedefaultmidpunct}
{\mcitedefaultendpunct}{\mcitedefaultseppunct}\relax
\EndOfBibitem
\bibitem{PDG2014}
Particle Data Group, K.~A. Olive {\em et~al.},
  \ifthenelse{\boolean{articletitles}}{\emph{{\href{http://pdg.lbl.gov/}{Review
  of particle physics}}},
  }{}\href{http://dx.doi.org/10.1088/1674-1137/38/9/090001}{Chin.\ Phys.\
  \textbf{C38} (2014) 090001}\relax
\mciteBstWouldAddEndPuncttrue
\mciteSetBstMidEndSepPunct{\mcitedefaultmidpunct}
{\mcitedefaultendpunct}{\mcitedefaultseppunct}\relax
\EndOfBibitem
\bibitem{PhysRevLett.10.531}
N.~Cabibbo, \ifthenelse{\boolean{articletitles}}{\emph{{Unitary symmetry and
  leptonic decays}},
  }{}\href{http://dx.doi.org/10.1103/PhysRevLett.10.531}{Phys.\ Rev.\ Lett.\
  \textbf{10} (1963) 531}\relax
\mciteBstWouldAddEndPuncttrue
\mciteSetBstMidEndSepPunct{\mcitedefaultmidpunct}
{\mcitedefaultendpunct}{\mcitedefaultseppunct}\relax
\EndOfBibitem
\bibitem{PTP.49.652}
M.~Kobayashi and T.~Maskawa,
  \ifthenelse{\boolean{articletitles}}{\emph{{\CP-violation in the
  renormalizable theory of weak interaction}},
  }{}\href{http://dx.doi.org/10.1143/PTP.49.652}{Progress of Theoretical
  Physics \textbf{49} (1973) 652}\relax
\mciteBstWouldAddEndPuncttrue
\mciteSetBstMidEndSepPunct{\mcitedefaultmidpunct}
{\mcitedefaultendpunct}{\mcitedefaultseppunct}\relax
\EndOfBibitem
\bibitem{Gronau:1990ra}
M.~Gronau and D.~London, \ifthenelse{\boolean{articletitles}}{\emph{{How to
  determine all the angles of the unitarity triangle from $\Bd \to D \KS$ and
  $\Bs \to D\phi$ }},
  }{}\href{http://dx.doi.org/10.1016/0370-2693(91)91756-L}{Phys.\ Lett.\
  \textbf{B253} (1991) 483}\relax
\mciteBstWouldAddEndPuncttrue
\mciteSetBstMidEndSepPunct{\mcitedefaultmidpunct}
{\mcitedefaultendpunct}{\mcitedefaultseppunct}\relax
\EndOfBibitem
\bibitem{Gronau:1991dp}
M.~Gronau and D.~Wyler, \ifthenelse{\boolean{articletitles}}{\emph{{On
  determining a weak phase from charged \B decay asymmetries}},
  }{}\href{http://dx.doi.org/10.1016/0370-2693(91)90034-N}{Phys.\ Lett.\
  \textbf{B265} (1991) 172}\relax
\mciteBstWouldAddEndPuncttrue
\mciteSetBstMidEndSepPunct{\mcitedefaultmidpunct}
{\mcitedefaultendpunct}{\mcitedefaultseppunct}\relax
\EndOfBibitem
\bibitem{Sinha:2004ct}
N.~Sinha, \ifthenelse{\boolean{articletitles}}{\emph{{Determining $\gamma$
  using $B \to D^{**}K$}},
  }{}\href{http://dx.doi.org/10.1103/PhysRevD.70.097501}{Phys.\ Rev.\
  \textbf{D70} (2004) 097501}, \href{http://arxiv.org/abs/hep-ph/0405061}{{\tt
  arXiv:hep-ph/0405061}}\relax
\mciteBstWouldAddEndPuncttrue
\mciteSetBstMidEndSepPunct{\mcitedefaultmidpunct}
{\mcitedefaultendpunct}{\mcitedefaultseppunct}\relax
\EndOfBibitem
\bibitem{Alves:2008zz}
LHCb collaboration, A.~A. Alves~Jr.\ {\em et~al.},
  \ifthenelse{\boolean{articletitles}}{\emph{{The \lhcb detector at the LHC}},
  }{}\href{http://dx.doi.org/10.1088/1748-0221/3/08/S08005}{JINST \textbf{3}
  (2008) S08005}\relax
\mciteBstWouldAddEndPuncttrue
\mciteSetBstMidEndSepPunct{\mcitedefaultmidpunct}
{\mcitedefaultendpunct}{\mcitedefaultseppunct}\relax
\EndOfBibitem
\bibitem{LHCb-DP-2014-002}
LHCb collaboration, R.~Aaij {\em et~al.},
  \ifthenelse{\boolean{articletitles}}{\emph{{LHCb detector performance}},
  }{}\href{http://arxiv.org/abs/1412.6352}{{\tt arXiv:1412.6352}}, {to appear
  in Int. J. Mod. Phys. A}\relax
\mciteBstWouldAddEndPuncttrue
\mciteSetBstMidEndSepPunct{\mcitedefaultmidpunct}
{\mcitedefaultendpunct}{\mcitedefaultseppunct}\relax
\EndOfBibitem
\bibitem{LHCb-DP-2014-001}
R.~Aaij {\em et~al.}, \ifthenelse{\boolean{articletitles}}{\emph{{Performance
  of the LHCb Vertex Locator}},
  }{}\href{http://dx.doi.org/10.1088/1748-0221/9/09/P09007}{JINST \textbf{9}
  (2014) P09007}, \href{http://arxiv.org/abs/1405.7808}{{\tt
  arXiv:1405.7808}}\relax
\mciteBstWouldAddEndPuncttrue
\mciteSetBstMidEndSepPunct{\mcitedefaultmidpunct}
{\mcitedefaultendpunct}{\mcitedefaultseppunct}\relax
\EndOfBibitem
\bibitem{LHCb-DP-2013-003}
R.~Arink {\em et~al.}, \ifthenelse{\boolean{articletitles}}{\emph{{Performance
  of the LHCb Outer Tracker}},
  }{}\href{http://dx.doi.org/10.1088/1748-0221/9/01/P01002}{JINST \textbf{9}
  (2014) P01002}, \href{http://arxiv.org/abs/1311.3893}{{\tt
  arXiv:1311.3893}}\relax
\mciteBstWouldAddEndPuncttrue
\mciteSetBstMidEndSepPunct{\mcitedefaultmidpunct}
{\mcitedefaultendpunct}{\mcitedefaultseppunct}\relax
\EndOfBibitem
\bibitem{LHCb-DP-2012-003}
M.~Adinolfi {\em et~al.},
  \ifthenelse{\boolean{articletitles}}{\emph{{Performance of the \lhcb RICH
  detector at the LHC}},
  }{}\href{http://dx.doi.org/10.1140/epjc/s10052-013-2431-9}{Eur.\ Phys.\ J.\
  \textbf{C73} (2013) 2431}, \href{http://arxiv.org/abs/1211.6759}{{\tt
  arXiv:1211.6759}}\relax
\mciteBstWouldAddEndPuncttrue
\mciteSetBstMidEndSepPunct{\mcitedefaultmidpunct}
{\mcitedefaultendpunct}{\mcitedefaultseppunct}\relax
\EndOfBibitem
\bibitem{LHCb-DP-2012-002}
A.~A. Alves~Jr.\ {\em et~al.},
  \ifthenelse{\boolean{articletitles}}{\emph{{Performance of the LHCb muon
  system}}, }{}\href{http://dx.doi.org/10.1088/1748-0221/8/02/P02022}{JINST
  \textbf{8} (2013) P02022}, \href{http://arxiv.org/abs/1211.1346}{{\tt
  arXiv:1211.1346}}\relax
\mciteBstWouldAddEndPuncttrue
\mciteSetBstMidEndSepPunct{\mcitedefaultmidpunct}
{\mcitedefaultendpunct}{\mcitedefaultseppunct}\relax
\EndOfBibitem
\bibitem{LHCb-DP-2012-004}
R.~Aaij {\em et~al.}, \ifthenelse{\boolean{articletitles}}{\emph{{The \lhcb
  trigger and its performance in 2011}},
  }{}\href{http://dx.doi.org/10.1088/1748-0221/8/04/P04022}{JINST \textbf{8}
  (2013) P04022}, \href{http://arxiv.org/abs/1211.3055}{{\tt
  arXiv:1211.3055}}\relax
\mciteBstWouldAddEndPuncttrue
\mciteSetBstMidEndSepPunct{\mcitedefaultmidpunct}
{\mcitedefaultendpunct}{\mcitedefaultseppunct}\relax
\EndOfBibitem
\bibitem{BBDT}
V.~V. Gligorov and M.~Williams,
  \ifthenelse{\boolean{articletitles}}{\emph{{Efficient, reliable and fast
  high-level triggering using a bonsai boosted decision tree}},
  }{}\href{http://dx.doi.org/10.1088/1748-0221/8/02/P02013}{JINST \textbf{8}
  (2013) P02013}, \href{http://arxiv.org/abs/1210.6861}{{\tt
  arXiv:1210.6861}}\relax
\mciteBstWouldAddEndPuncttrue
\mciteSetBstMidEndSepPunct{\mcitedefaultmidpunct}
{\mcitedefaultendpunct}{\mcitedefaultseppunct}\relax
\EndOfBibitem
\bibitem{Sjostrand:2006za}
T.~Sj\"{o}strand, S.~Mrenna, and P.~Skands,
  \ifthenelse{\boolean{articletitles}}{\emph{{PYTHIA 6.4 physics and manual}},
  }{}\href{http://dx.doi.org/10.1088/1126-6708/2006/05/026}{JHEP \textbf{05}
  (2006) 026}, \href{http://arxiv.org/abs/hep-ph/0603175}{{\tt
  arXiv:hep-ph/0603175}}\relax
\mciteBstWouldAddEndPuncttrue
\mciteSetBstMidEndSepPunct{\mcitedefaultmidpunct}
{\mcitedefaultendpunct}{\mcitedefaultseppunct}\relax
\EndOfBibitem
\bibitem{Sjostrand:2007gs}
T.~Sj\"{o}strand, S.~Mrenna, and P.~Skands,
  \ifthenelse{\boolean{articletitles}}{\emph{{A brief introduction to PYTHIA
  8.1}}, }{}\href{http://dx.doi.org/10.1016/j.cpc.2008.01.036}{Comput.\ Phys.\
  Commun.\  \textbf{178} (2008) 852},
  \href{http://arxiv.org/abs/0710.3820}{{\tt arXiv:0710.3820}}\relax
\mciteBstWouldAddEndPuncttrue
\mciteSetBstMidEndSepPunct{\mcitedefaultmidpunct}
{\mcitedefaultendpunct}{\mcitedefaultseppunct}\relax
\EndOfBibitem
\bibitem{LHCb-PROC-2010-056}
I.~Belyaev {\em et~al.}, \ifthenelse{\boolean{articletitles}}{\emph{{Handling
  of the generation of primary events in Gauss, the LHCb simulation
  framework}}, }{}\href{http://dx.doi.org/10.1088/1742-6596/331/3/032047}{{J.\
  Phys.\ Conf.\ Ser.\ } \textbf{331} (2011) 032047}\relax
\mciteBstWouldAddEndPuncttrue
\mciteSetBstMidEndSepPunct{\mcitedefaultmidpunct}
{\mcitedefaultendpunct}{\mcitedefaultseppunct}\relax
\EndOfBibitem
\bibitem{Lange:2001uf}
D.~J. Lange, \ifthenelse{\boolean{articletitles}}{\emph{{The EvtGen particle
  decay simulation package}},
  }{}\href{http://dx.doi.org/10.1016/S0168-9002(01)00089-4}{Nucl.\ Instrum.\
  Meth.\  \textbf{A462} (2001) 152}\relax
\mciteBstWouldAddEndPuncttrue
\mciteSetBstMidEndSepPunct{\mcitedefaultmidpunct}
{\mcitedefaultendpunct}{\mcitedefaultseppunct}\relax
\EndOfBibitem
\bibitem{Golonka:2005pn}
P.~Golonka and Z.~Was, \ifthenelse{\boolean{articletitles}}{\emph{{PHOTOS Monte
  Carlo: A precision tool for QED corrections in $Z$ and $W$ decays}},
  }{}\href{http://dx.doi.org/10.1140/epjc/s2005-02396-4}{Eur.\ Phys.\ J.\
  \textbf{C45} (2006) 97}, \href{http://arxiv.org/abs/hep-ph/0506026}{{\tt
  arXiv:hep-ph/0506026}}\relax
\mciteBstWouldAddEndPuncttrue
\mciteSetBstMidEndSepPunct{\mcitedefaultmidpunct}
{\mcitedefaultendpunct}{\mcitedefaultseppunct}\relax
\EndOfBibitem
\bibitem{Allison:2006ve}
Geant4 collaboration, J.~Allison {\em et~al.},
  \ifthenelse{\boolean{articletitles}}{\emph{{Geant4 developments and
  applications}}, }{}\href{http://dx.doi.org/10.1109/TNS.2006.869826}{IEEE
  Trans.\ Nucl.\ Sci.\  \textbf{53} (2006) 270}\relax
\mciteBstWouldAddEndPuncttrue
\mciteSetBstMidEndSepPunct{\mcitedefaultmidpunct}
{\mcitedefaultendpunct}{\mcitedefaultseppunct}\relax
\EndOfBibitem
\bibitem{Agostinelli:2002hh}
Geant4 collaboration, S.~Agostinelli {\em et~al.},
  \ifthenelse{\boolean{articletitles}}{\emph{{Geant4: a simulation toolkit}},
  }{}\href{http://dx.doi.org/10.1016/S0168-9002(03)01368-8}{Nucl.\ Instrum.\
  Meth.\  \textbf{A506} (2003) 250}\relax
\mciteBstWouldAddEndPuncttrue
\mciteSetBstMidEndSepPunct{\mcitedefaultmidpunct}
{\mcitedefaultendpunct}{\mcitedefaultseppunct}\relax
\EndOfBibitem
\bibitem{LHCb-PROC-2011-006}
M.~Clemencic {\em et~al.}, \ifthenelse{\boolean{articletitles}}{\emph{{The
  \lhcb simulation application, Gauss: Design, evolution and experience}},
  }{}\href{http://dx.doi.org/10.1088/1742-6596/331/3/032023}{{J.\ Phys.\ Conf.\
  Ser.\ } \textbf{331} (2011) 032023}\relax
\mciteBstWouldAddEndPuncttrue
\mciteSetBstMidEndSepPunct{\mcitedefaultmidpunct}
{\mcitedefaultendpunct}{\mcitedefaultseppunct}\relax
\EndOfBibitem
\bibitem{Feindt2006190}
M.~Feindt and U.~Kerzel, \ifthenelse{\boolean{articletitles}}{\emph{{The
  NeuroBayes neural network package}},
  }{}\href{http://dx.doi.org/10.1016/j.nima.2005.11.166}{Nucl.\ Instrum.\
  Meth.\ A \textbf{559} (2006) 190}\relax
\mciteBstWouldAddEndPuncttrue
\mciteSetBstMidEndSepPunct{\mcitedefaultmidpunct}
{\mcitedefaultendpunct}{\mcitedefaultseppunct}\relax
\EndOfBibitem
\bibitem{Pivk:2004ty}
M.~Pivk and F.~R. Le~Diberder,
  \ifthenelse{\boolean{articletitles}}{\emph{{sPlot: A statistical tool to
  unfold data distributions}},
  }{}\href{http://dx.doi.org/10.1016/j.nima.2005.08.106}{Nucl.\ Instrum.\
  Meth.\  \textbf{A555} (2005) 356},
  \href{http://arxiv.org/abs/physics/0402083}{{\tt
  arXiv:physics/0402083}}\relax
\mciteBstWouldAddEndPuncttrue
\mciteSetBstMidEndSepPunct{\mcitedefaultmidpunct}
{\mcitedefaultendpunct}{\mcitedefaultseppunct}\relax
\EndOfBibitem
\bibitem{LHCb-PAPER-2012-001}
LHCb collaboration, R.~Aaij {\em et~al.},
  \ifthenelse{\boolean{articletitles}}{\emph{{Observation of $CP$ violation in
  $B^\pm \to D K^\pm$ decays}},
  }{}\href{http://dx.doi.org/10.1016/j.physletb.2012.04.060}{Phys.\ Lett.\
  \textbf{B712} (2012) 203}, Erratum
  \href{http://dx.doi.org/10.1016/j.physletb.2012.05.060}{ibid.\
  \textbf{B713} (2012) 351}, \href{http://arxiv.org/abs/1203.3662}{{\tt
  arXiv:1203.3662}}\relax
\mciteBstWouldAddEndPuncttrue
\mciteSetBstMidEndSepPunct{\mcitedefaultmidpunct}
{\mcitedefaultendpunct}{\mcitedefaultseppunct}\relax
\EndOfBibitem
\bibitem{LHCb-PAPER-2012-048}
LHCb collaboration, R.~Aaij {\em et~al.},
  \ifthenelse{\boolean{articletitles}}{\emph{{Measurements of the
  $\Lambda_b^0$, $\Xi_b^-$, and $\Omega_b^-$ baryon masses}},
  }{}\href{http://dx.doi.org/10.1103/PhysRevLett.110.182001}{Phys.\ Rev.\
  Lett.\  \textbf{110} (2013) 182001},
  \href{http://arxiv.org/abs/1302.1072}{{\tt arXiv:1302.1072}}\relax
\mciteBstWouldAddEndPuncttrue
\mciteSetBstMidEndSepPunct{\mcitedefaultmidpunct}
{\mcitedefaultendpunct}{\mcitedefaultseppunct}\relax
\EndOfBibitem
\bibitem{LHCb-PAPER-2013-011}
LHCb collaboration, R.~Aaij {\em et~al.},
  \ifthenelse{\boolean{articletitles}}{\emph{{Precision measurement of $D$
  meson mass differences}},
  }{}\href{http://dx.doi.org/10.1007/JHEP06(2013)065}{JHEP \textbf{06} (2013)
  065}, \href{http://arxiv.org/abs/1304.6865}{{\tt arXiv:1304.6865}}\relax
\mciteBstWouldAddEndPuncttrue
\mciteSetBstMidEndSepPunct{\mcitedefaultmidpunct}
{\mcitedefaultendpunct}{\mcitedefaultseppunct}\relax
\EndOfBibitem
\bibitem{Hulsbergen:2005pu}
W.~D. Hulsbergen, \ifthenelse{\boolean{articletitles}}{\emph{{Decay chain
  fitting with a Kalman filter}},
  }{}\href{http://dx.doi.org/10.1016/j.nima.2005.06.078}{Nucl.\ Instrum.\
  Meth.\  \textbf{A552} (2005) 566},
  \href{http://arxiv.org/abs/physics/0503191}{{\tt
  arXiv:physics/0503191}}\relax
\mciteBstWouldAddEndPuncttrue
\mciteSetBstMidEndSepPunct{\mcitedefaultmidpunct}
{\mcitedefaultendpunct}{\mcitedefaultseppunct}\relax
\EndOfBibitem
\bibitem{Skwarnicki:1986xj}
T.~Skwarnicki, {\em {A study of the radiative cascade transitions between the
  Upsilon-prime and Upsilon resonances}}, PhD thesis, Institute of Nuclear
  Physics, Krakow, 1986,
  {\href{http://inspirehep.net/record/230779/}{DESY-F31-86-02}}\relax
\mciteBstWouldAddEndPuncttrue
\mciteSetBstMidEndSepPunct{\mcitedefaultmidpunct}
{\mcitedefaultendpunct}{\mcitedefaultseppunct}\relax
\EndOfBibitem
\bibitem{LHCb-PAPER-2012-018}
LHCb collaboration, R.~Aaij {\em et~al.},
  \ifthenelse{\boolean{articletitles}}{\emph{{Observation of $B^0 \to \bar{D}^0
  K^+ K^-$ and evidence for $B^0_s \to \bar{D}^0 K^+ K^-$}},
  }{}\href{http://dx.doi.org/10.1103/PhysRevLett.109.131801}{Phys.\ Rev.\
  Lett.\  \textbf{109} (2012) 131801},
  \href{http://arxiv.org/abs/1207.5991}{{\tt arXiv:1207.5991}}\relax
\mciteBstWouldAddEndPuncttrue
\mciteSetBstMidEndSepPunct{\mcitedefaultmidpunct}
{\mcitedefaultendpunct}{\mcitedefaultseppunct}\relax
\EndOfBibitem
\bibitem{Dalitz:1953cp}
R.~H. Dalitz, \ifthenelse{\boolean{articletitles}}{\emph{{On the analysis of
  tau-meson data and the nature of the tau-meson}},
  }{}\href{http://dx.doi.org/10.1080/14786441008520365}{Phil.\ Mag.\
  \textbf{44} (1953) 1068}\relax
\mciteBstWouldAddEndPuncttrue
\mciteSetBstMidEndSepPunct{\mcitedefaultmidpunct}
{\mcitedefaultendpunct}{\mcitedefaultseppunct}\relax
\EndOfBibitem
\bibitem{Fleming:1964zz}
G.~N. Fleming, \ifthenelse{\boolean{articletitles}}{\emph{{Recoupling effects
  in the isobar model. 1. General formalism for three-pion scattering}},
  }{}\href{http://dx.doi.org/10.1103/PhysRev.135.B551}{Phys.\ Rev.\
  \textbf{135} (1964) B551}\relax
\mciteBstWouldAddEndPuncttrue
\mciteSetBstMidEndSepPunct{\mcitedefaultmidpunct}
{\mcitedefaultendpunct}{\mcitedefaultseppunct}\relax
\EndOfBibitem
\bibitem{Morgan:1968zza}
D.~Morgan, \ifthenelse{\boolean{articletitles}}{\emph{{Phenomenological
  analysis of $I=\frac{1}{2}$ single-pion production processes in the energy
  range 500 to 700 MeV}},
  }{}\href{http://dx.doi.org/10.1103/PhysRev.166.1731}{Phys.\ Rev.\
  \textbf{166} (1968) 1731}\relax
\mciteBstWouldAddEndPuncttrue
\mciteSetBstMidEndSepPunct{\mcitedefaultmidpunct}
{\mcitedefaultendpunct}{\mcitedefaultseppunct}\relax
\EndOfBibitem
\bibitem{Herndon:1973yn}
D.~Herndon, P.~Soding, and R.~J. Cashmore,
  \ifthenelse{\boolean{articletitles}}{\emph{{A generalised isobar model
  formalism}}, }{}\href{http://dx.doi.org/10.1103/PhysRevD.11.3165}{Phys.\
  Rev.\  \textbf{D11} (1975) 3165}\relax
\mciteBstWouldAddEndPuncttrue
\mciteSetBstMidEndSepPunct{\mcitedefaultmidpunct}
{\mcitedefaultendpunct}{\mcitedefaultseppunct}\relax
\EndOfBibitem
\bibitem{blatt-weisskopf}
J.~Blatt and V.~E. Weisskopf, {\em Theoretical nuclear physics}, J. Wiley (New
  York), 1952\relax
\mciteBstWouldAddEndPuncttrue
\mciteSetBstMidEndSepPunct{\mcitedefaultmidpunct}
{\mcitedefaultendpunct}{\mcitedefaultseppunct}\relax
\EndOfBibitem
\bibitem{Aubert:2005ce}
\babar collaboration, B.~Aubert {\em et~al.},
  \ifthenelse{\boolean{articletitles}}{\emph{{Dalitz-plot analysis of the
  decays $B^\pm \to K^\pm \pi^\mp \pi^\pm$}},
  }{}\href{http://dx.doi.org/10.1103/PhysRevD.72.072003}{Phys.\ Rev.\
  \textbf{D72} (2005) 072003}, Erratum
  \href{http://dx.doi.org/10.1103/PhysRevD.74.099903}{ibid.\   \textbf{D74}
  (2006) 099903}, \href{http://arxiv.org/abs/hep-ex/0507004}{{\tt
  arXiv:hep-ex/0507004}}\relax
\mciteBstWouldAddEndPuncttrue
\mciteSetBstMidEndSepPunct{\mcitedefaultmidpunct}
{\mcitedefaultendpunct}{\mcitedefaultseppunct}\relax
\EndOfBibitem
\bibitem{Zemach:1963bc}
C.~Zemach, \ifthenelse{\boolean{articletitles}}{\emph{{Three pion decays of
  unstable particles}},
  }{}\href{http://dx.doi.org/10.1103/PhysRev.133.B1201}{Phys.\ Rev.\
  \textbf{133} (1964) B1201}\relax
\mciteBstWouldAddEndPuncttrue
\mciteSetBstMidEndSepPunct{\mcitedefaultmidpunct}
{\mcitedefaultendpunct}{\mcitedefaultseppunct}\relax
\EndOfBibitem
\bibitem{Zemach:1968zz}
C.~Zemach, \ifthenelse{\boolean{articletitles}}{\emph{{Use of angular-momentum
  tensors}}, }{}\href{http://dx.doi.org/10.1103/PhysRev.140.B97}{Phys.\ Rev.\
  \textbf{140} (1965) B97}\relax
\mciteBstWouldAddEndPuncttrue
\mciteSetBstMidEndSepPunct{\mcitedefaultmidpunct}
{\mcitedefaultendpunct}{\mcitedefaultseppunct}\relax
\EndOfBibitem
\bibitem{Garmash:2004wa}
Belle collaboration, A.~Garmash {\em et~al.},
  \ifthenelse{\boolean{articletitles}}{\emph{{Dalitz analysis of the three-body
  charmless decays $\Bp\to \Kp\pip\pim$ and $\Bp \to \Kp\Kp\Km$}},
  }{}\href{http://dx.doi.org/10.1103/PhysRevD.71.092003}{Phys.\ Rev.\
  \textbf{D71} (2005) 092003}, \href{http://arxiv.org/abs/hep-ex/0412066}{{\tt
  arXiv:hep-ex/0412066}}\relax
\mciteBstWouldAddEndPuncttrue
\mciteSetBstMidEndSepPunct{\mcitedefaultmidpunct}
{\mcitedefaultendpunct}{\mcitedefaultseppunct}\relax
\EndOfBibitem
\bibitem{Laura++}
{{\tt Laura++} Dalitz plot fitting package, \url{http://laura.hepfo
  rge.org}}\relax
\mciteBstWouldAddEndPuncttrue
\mciteSetBstMidEndSepPunct{\mcitedefaultmidpunct}
{\mcitedefaultendpunct}{\mcitedefaultseppunct}\relax
\EndOfBibitem
\bibitem{Ben-Haim:2014afa}
E.~Ben-Haim, R.~Brun, B.~Echenard, and T.~E. Latham,
  \ifthenelse{\boolean{articletitles}}{\emph{{JFIT: a framework to obtain
  combined experimental results through joint fits}},
  }{}\href{http://arxiv.org/abs/1409.5080}{{\tt arXiv:1409.5080}}\relax
\mciteBstWouldAddEndPuncttrue
\mciteSetBstMidEndSepPunct{\mcitedefaultmidpunct}
{\mcitedefaultendpunct}{\mcitedefaultseppunct}\relax
\EndOfBibitem
\bibitem{Williams:2010vh}
M.~Williams, \ifthenelse{\boolean{articletitles}}{\emph{{How good are your
  fits? Unbinned multivariate goodness-of-fit tests in high energy physics}},
  }{}\href{http://dx.doi.org/10.1088/1748-0221/5/09/P09004}{JINST \textbf{5}
  (2010) P09004}, \href{http://arxiv.org/abs/1006.3019}{{\tt
  arXiv:1006.3019}}\relax
\mciteBstWouldAddEndPuncttrue
\mciteSetBstMidEndSepPunct{\mcitedefaultmidpunct}
{\mcitedefaultendpunct}{\mcitedefaultseppunct}\relax
\EndOfBibitem
\end{mcitethebibliography}

\ifx\mcitethebibliography\mciteundefinedmacro
\PackageError{LHCb.bst}{mciteplus.sty has not been loaded}
{This bibstyle requires the use of the mciteplus package.}\fi
\providecommand{\href}[2]{#2}

\clearpage

\appendix
\section{Secondary minima}
\label{app:minima}

\newcommand{\mc}{\multicolumn}

The results, in terms of fit fractions and complex coefficients, corresponding to the two secondary minima discussed in Sec.~\ref{sec:dalitz} are compared to those of the global minimum in Table~\ref{tab:minima}.
The main difference between the global and secondary minima is in the interference pattern in the $D\pi$ P-waves, while the third minimum exhibits a different interference pattern in the $D\pi$ S-wave than the global minimum and has a very large total fit fraction due to strong destructive interference.

\begin{table}[!hb]
\centering
\caption{\small
  Results for the fit fractions and complex coefficients for the secondary minima with $2{\rm NLL}$ values 2.8 and 3.3 units greater than that of the global minimum of the NLL function.  
}
\label{tab:minima}
\resizebox{\textwidth}{!}{ 
\begin{tabular}{l|ccc|ccc|ccc|ccc|ccc} 
\hline 
Resonance & \mc{3}{c|}{Fit fraction (\%)} & \mc{3}{c|}{Real part} & \mc{3}{c|}{Imaginary part} & \mc{3}{c|}{Magnitude} & \mc{3}{c}{Phase} \\ 
$2\Delta{\rm NLL}$ & 0 & 2.8 & 3.3 & 0 & 2.8 & 3.3 & 0 & 2.8 & 3.3 & 0 & 2.8 & 3.3 & 0 & 2.8 & 3.3\\
\hline 
$D^{*}_{0}(2400)^{0}$ & $ 8.3$ & $ 9.6$ & $84.4$ & $-0.04$ & $-0.03$ & $-1.38$ & $-0.51$ & $-0.55$ & $-0.72$ & $0.51$ & $0.55$ & $1.56$ & $-1.65$ & $-1.62$ & $-2.66$ \\ 
$D^{*}_{2}(2460)^{0}$ & $31.8$ & $31.5$ & $34.9$ & \mc{3}{c|}{$ 1.00$} & \mc{3}{c|}{$ 0.00$} & \mc{3}{c|}{$1.00$} & \mc{3}{c}{$ 0.00$} \\ 
$D^{*}_{1}(2760)^{0}$ & $ 4.9$ & $ 4.6$ & $ 5.5$ & $-0.32$ & $-0.30$ & $-0.30$ & $-0.23$ & $-0.24$ & $-0.26$ & $0.39$ & $0.38$ & $0.40$ & $-2.53$ & $-2.46$ & $-2.42$ \\ 
\hline
S-wave nonresonant  & $38.0$ & $36.2$ & $  4.6$ & $ 0.93$ & $ 0.89$ & $-0.33$ & $-0.58$ & $-0.60$ & $ 0.15$ & $1.09$ & $1.07$ & $-0.36$ & $-0.56$ & $-0.59$ & $ 2.71$ \\ 
P-wave nonresonant  & $23.8$ & $22.6$ & $-31.9$ & $-0.43$ & $ 0.83$ & $-0.84$ & $ 0.75$ & $ 0.15$ & $ 0.45$ & $0.87$ & $0.85$ & $ 0.96$ & $ 2.09$ & $ 2.96$ & $ 2.65$ \\ 
\hline
$D^{*}_{v}(2007)^{0}$ & $ 7.6$ & $ 7.1$ &  $11.9$ & $ 0.16$ & $-0.38$ & $-0.28$ & $ 0.46$ & $-0.29$ & $-0.51$ & $0.49$ & $0.48$ & $ 0.58$ & $ 1.24$ & $-2.49$ & $-2.07$ \\ 
$B^{*}_{v}$          & $ 3.6$ & $1.0 $ &  $25.0$ & $-0.07$ & $-0.16$ & $-0.31$ & $ 0.33$ & $ 0.09$ & $ 0.79$ & $0.34$ & $0.18$ & $ 0.85$ & $ 1.78$ & $ 2.61$ & $ 1.94$ \\ 
\hline
Total fit fraction & 118.1 & 112.6 & 198.3 \\
\hline 
\end{tabular} 
}
\end{table}

\section{Results for interference fit fractions}
\label{app:iffstat}

The central values and statistical errors for the interference fit fractions are shown in Table~\ref{tab:iffstat}. 
The experimental systematic and model uncertainties are given in Tables~\ref{tab:iffexp} and~\ref{tab:iffmodel}. 
The interference fit fractions are common to both trigger subsamples.

\begin{table}[!htb]
\caption{\small Interference fit fractions (\%) and statistical uncertainties. The amplitudes are: ($A_0$) $D^*_v(2007)^0$,  ($A_1$) $D^*_0(2400)^0$,  ($A_2$) $D^*_2(2460)^0$,  ($A_3$) $D^*_1(2760)^0$,  ($A_4$) $B^*_v$,  ($A_5$) nonresonant S-wave,  ($A_6$) nonresonant P-wave. The diagonal elements are the same as the conventional fit fractions.
}
\label{tab:iffstat}
\centering
\vspace{1ex}
\resizebox{\textwidth}{!}{
\begin{tabular}{lccccccc}
\hline
& $\phantom{-}A_0$ & $\phantom{-}A_1$ & $\phantom{-}A_2$ & $\phantom{-}A_3$ & $\phantom{-}A_4$ & $\phantom{-}A_5$ & $\phantom{-}A_6$ \\
\hline
$A_0$ & $\phantom{-}7.6 \pm 2.3 $ & $\phantom{-}0.0 \pm 0.0 $ & $\phantom{-1}0.0 \pm 0.0 $ & $\phantom{-}2.4\pm 0.9 $ & $\phantom{-}4.8 \pm 1.3$ & $\phantom{-1}0.0 \pm 0.0 $ & $-14.2 \pm 5.3$ \\
$A_1$ &  & $\phantom{-}8.3 \pm 2.6 $ & $\phantom{-1}0.0 \pm 0.0 $ & $\phantom{-}0.0 \pm 0.0 $ & $-1.6 \pm 0.7 $ & $\phantom{-}18.1 \pm 2.6 $ & $\phantom{-1}0.0 \pm 0.0$ \\
$A_2$ &  &  & $\phantom{-}31.8 \pm 1.5 $ & $\phantom{-}0.0 \pm 0.0 $ & $-2.3\pm 0.6 $ & $\phantom{-1}0.0 \pm 0.0 $ & $\phantom{-1}0.0 \pm 0.0$ \\
$A_3$ &  &  &  & $\phantom{-}4.9 \pm 1.2 $ & $\phantom{-}2.0 \pm 0.8 $ & $\phantom{-1}0.0 \pm 0.0 $ & \phantom{1}$-9.6 \pm 2.9$ \\
$A_4$ &  &  &  &  & $\phantom{-}3.6 \pm 1.9$ & \phantom{1}$-6.7\pm 2.3 $ & $-11.1 \pm 3.6 $ \\
$A_5$ &  &  &  &  &  & $\phantom{-}38.0 \pm 7.4 $ & $\phantom{-1}0.0 \pm 0.0$ \\
$A_6$ &  &  &  &  &  &  & $\phantom{-}23.8 \pm 5.6$ \\
\hline
\end{tabular}
}
\end{table}

\begin{table}[!htb]
\centering
\caption{\small
Experimental systematic uncertainies on the interference fit fractions (\%). The amplitudes are: ($A_0$) $D^*_v(2007)^0$,  ($A_1$) $D^*_0(2400)^0$,  ($A_2$) $D^*_2(2460)^0$,  ($A_3$) $D^*_1(2760)^0$,  ($A_4$) $B^*_v$,  ($A_5$) nonresonant S-wave,  ($A_6$) nonresonant P-wave. The diagonal elements are the same as the conventional fit fractions.}
\label{tab:iffexp}
\begin{tabular}{lccccccc} 
\hline 
& $A_0$ & $A_1$ & $A_2$ & $A_3$ & $A_4$ & $A_5$ & $A_6$ \\ 
\hline 
$A_0$ & 1.3 & 0.0 & 0.0 & 0.4 & 0.6 & 0.0 & 2.6 \\ 
$A_1$ & & 0.6 & 0.0 & 0.0 & 0.4 & 0.6 & 0.0 \\ 
$A_2$ & & & 0.9 & 0.0 & 0.3 & 0.0 & 0.0 \\ 
$A_3$ & & & & 0.4 & 0.2 & 0.0 & 0.7 \\ 
$A_4$ & & & & & 0.9 & 1.1 & 1.2 \\ 
$A_5$ & & & & & & 1.5 & 0.0 \\ 
$A_6$ & & & & & & & 2.1 \\ 
\hline 
\end{tabular} 
\end{table}

\begin{table}[!htb]
\centering
\caption{\small
Model systematic uncertainies on the interference fit fractions (\%). The amplitudes are: ($A_0$) $D^*_v(2007)^0$,  ($A_1$) $D^*_0(2400)^0$,  ($A_2$) $D^*_2(2460)^0$,  ($A_3$) $D^*_1(2760)^0$,  ($A_4$) $B^*_v$,  ($A_5$) nonresonant S-wave,  ($A_6$) nonresonant P-wave. The diagonal elements are the same as the conventional fit fractions.}
\label{tab:iffmodel}
\begin{tabular}{lccccccc} 
\hline 
& \phantom{1}$A_0$\phantom{1} & \phantom{1}$A_1$\phantom{1} & \phantom{1}$A_2$\phantom{1} & \phantom{1}$A_3$\phantom{1} & \phantom{1}$A_4$\phantom{1} & \phantom{1}$A_5$ & \phantom{1}$A_6$\phantom{1} \\ 
\hline 
$A_0$ & 1.5 & 0.0 & 0.0 & 0.1 & 1.4 & \phantom{1}0.0 & 1.1 \\ 
$A_1$ & & 1.9 & 0.0 & 0.0 & 1.7 & \phantom{1}4.8 & 0.0 \\ 
$A_2$ & & & 1.4 & 0.0 & 0.5 & \phantom{1}0.0 & 0.0 \\ 
$A_3$ & & & & 0.9 & 0.6 & \phantom{1}0.0 & 3.4 \\ 
$A_4$ & & & & & 1.6 & \phantom{1}2.8 & 0.4 \\ 
$A_5$ & & & & & & \phantom{1}10.8 & 0.0 \\ 
$A_6$ & & & & & & & 3.67 \\ 
\hline 
\end{tabular} 
\end{table}

\clearpage
\centerline{\large\bf LHCb collaboration}
\begin{flushleft}
\small
R.~Aaij$^{41}$, 
B.~Adeva$^{37}$, 
M.~Adinolfi$^{46}$, 
A.~Affolder$^{52}$, 
Z.~Ajaltouni$^{5}$, 
S.~Akar$^{6}$, 
J.~Albrecht$^{9}$, 
F.~Alessio$^{38}$, 
M.~Alexander$^{51}$, 
S.~Ali$^{41}$
% ...

{\footnotesize \it
% ...
$ ^{5}$Clermont Universit\'{e}, Universit\'{e} Blaise Pascal, CNRS/IN2P3, LPC, Clermont-Ferrand, France\\
$ ^{6}$CPPM, Aix-Marseille Universit\'{e}, CNRS/IN2P3, Marseille, France\\
% ...
$ ^{9}$Fakult\"{a}t Physik, Technische Universit\"{a}t Dortmund, Dortmund, Germany\\
% ...
$ ^{37}$Universidad de Santiago de Compostela, Santiago de Compostela, Spain\\
$ ^{38}$European Organization for Nuclear Research (CERN), Geneva, Switzerland\\
% ...
$ ^{41}$Nikhef National Institute for Subatomic Physics, Amsterdam, The Netherlands\\
% ...
$ ^{46}$H.H. Wills Physics Laboratory, University of Bristol, Bristol, United Kingdom\\
% ...
$ ^{51}$School of Physics and Astronomy, University of Glasgow, Glasgow, United Kingdom\\
$ ^{52}$Oliver Lodge Laboratory, University of Liverpool, Liverpool, United Kingdom\\
% ...
}
\end{flushleft}

\end{document}